\documentclass[12pt]{article}
\usepackage{graphicx}
\def\lr #1{\mathrel{#1\kern-.75em\raise1.75ex\hbox{$\leftrightarrow$}}}
\def\bi{\begin{itemize}}
\def\ei{\end{itemize}}
\def\benu{\begin{enumerate}}
\def\eenu{\end{enumerate}}
\def\be{\begin{equation}}
\def\ee{\end{equation}}
\def\ba{\begin{eqnarray}}
\def\ea{\end{eqnarray}}
\def\R{{\hbox{{\rm I}\kern-.2em\hbox{\rm R}}}}   
\def\H{{\hbox{{\rm I}\kern-.2em\hbox{\rm H}}}}   
\def\N{{\hbox{{\rm I}\kern-.2em\hbox{\rm N}}}}   
\def\C{{\ \hbox{{\rm I}\kern-.6em\hbox{\bf C}}}} 
\def\Z{{\hbox{{\rm Z}\kern-.4em\hbox{\rm Z}}}}   

\def\downarcfill{${\mathsurround=0pt }
\braceld\leaders\vrule\hfill\bracerd $}
\def\arc#1{\mathop{\vbox{\ialign{##\crcr\noalign{\kern3pt}
\downarcfill\crcr\noalign{\kern3pt\nointerlineskip}
$\hfil\displaystyle{#1}\hfil$\crcr}}}\limits}

\def\mettresous#1\sous#2{\mathrel{\mathop{\kern0pt #2}\limits_{#1}}}
\def\sqr#1#2{{\vcenter{\vbox{\hrule height.#2pt
          \hbox{\vrule width.#2pt height#1pt \kern#1pt
           \vrule width.#2pt}
           \hrule height.#2pt}}}}

\def\ket#1{|#1\rangle}      
\def\bra#1{\langle #1|}     
\def\kvac {|0\rangle}                   
\def\braket#1#2{\mathrel{\langle #1|#2\rangle}}   

\newlength{\fleche}
\newcounter{FLECHE}

\newcounter{DEMIFLECHE}

\def\lrpartial{\mathrel{\partial\kern-.75em\raise1.75ex\hbox{$\leftrightarro
w$}}}
\def\lrD{\mathrel{{\cal D}\kern-.75em\raise1.75ex\hbox{$\leftrightarrow$}}}
\def\mettresous#1\sous#2{\mathrel{\mathop{\kern0pt #2}\limits_{#1}}}
\def\zentier{\ \hbox{{\rm Z}\kern-.4em\hbox{\rm Z}}}
\def\reel{\ \hbox{{\rm R}\kern-1em\hbox{{\rm I}}}}
\def\bfGamma{\ \hbox{{$\Gamma$}\kern-.5em\hbox{{\rm I}}}\,}
\def\Iden{\ \hbox{{1}\kern-.25em\hbox{{\rm I}}}\,}

\def\1op{\ \hbox{{\rm 1}\kern -0.23em\hbox{{\rm I}}}}

\newcommand{\figdef}[4]
{\begin{figure} \vglue10pt
\vspace{#1}
\special{illustration #2 scaled 800}
\caption{{\sl #3}}
\vspace{3mm}
\label{#4}
\end{figure}
}
\catcode`\@=11 \@addtoreset{equation}{section}
\def\theequation{\arabic{section}.\arabic{equation}}
\def\appendix{\renewcommand{\thesection}{\Alph{section}}\setcounter{section}{0}
              \renewcommand{\theequation}
            {\mbox{\Alph{section}.\arabic{equation}}}\setcounter{equation}{0}}

\def\ket#1{|#1>}
\def\bra#1{<#1|}
\def\braket#1#2{<#1|#2>}
\def\sgn{\mathop{\rm sgn}\nolimits}

\def\tr{\mathop{\rm tr}\nolimits}
\def\cF{{\cal F}}

\def\cMp{{\cal M}^+}

\def\cWpe{{\cal W}^+_\epsilon}
\def\cWme{{\cal W}^-_\epsilon}
\def\cMpe{{\cal M}^+_\epsilon}
\def\cMme{{\cal M}^-_\epsilon}
\def\bwp{{\mbox{\bf W}}^+}
\def\bwm{{\mbox{\bf W}}^-}
\def\cUwi#1{{\cal U}^\omega _{in,#1}}
\def\cUwo#1{{\cal U}^\omega _{out,#1}}
\def\cVwi#1{{\cal V}^\omega _{in,#1}}
\def\cVwo#1{{\cal V}^\omega _{out,#1}}
\def\F{{\mbox{\bf F}}}
\def\P{{\mbox{\bf P}}}
\def\L{{\mbox{\bf L}}}
\def\R{{\mbox{\bf R}}}
\def\TF{{\bf F\ }}
\def\TP{{\bf P\ }}
\def\TR{{\bf R\ }}
\def\TL{{\bf L\ }}
\def\tF{{\bf F}}
\def\tP{{\bf P}}
\def\tR{{\bf R}}
\def\tL{{\bf L}}

\def\NUwi#1{{\cal N}({\cal U}^\omega _{in,#1})}
\def\NVwi#1{{\cal N}({\cal V}^\omega _{in,#1})}
\def\NUwo#1{{\cal N}({\cal U}^\omega _{out,#1})}
\def\NVwo#1{{\cal N}({\cal V}^\omega _{out,#1})}
\def\QUwi#1{{Q}({\cal U}^\omega _{in,#1})}
\def\QVwi#1{{Q}({\cal V}^\omega _{in,#1})}

\def\QUiIpR#1{{Q}_{{\cal I}^+_R}({\cal U}^\omega _{in,#1})}
\def\QUoIpR#1{{Q}_{{\cal I}^+_R}({\cal U}^\omega _{out,#1})}

\def\QUoImR#1{{Q}_{{\cal I}^-_R}({\cal U}^\omega _{out,#1})}
\def\QViIpR#1{{Q}_{{\cal I}^+_R}({\cal V}^\omega _{in,#1})}

\def\QViImR#1{{Q}_{{\cal I}^-_R}({\cal V}^\omega _{in,#1})}

\def\QUiHpR#1{{Q}_{{\cal H}^+_R}({\cal U}^\omega _{in,#1})}

\def\QUiHmR#1{{Q}_{{\cal H}^-_R}({\cal U}^\omega _{in,#1})}
\def\QUoHmR#1{{Q}_{{\cal H}^-_R}({\cal U}^\omega _{out,#1})}
\def\QViHpR#1{{Q}_{{\cal H}^+_R}({\cal V}^\omega _{in,#1})}

\def\QViHmR#1{{Q}_{{\cal H}^-_R}({\cal V}^\omega _{in,#1})}

\def\QUiImL#1{{Q}_{{\cal I}^-_L}({\cal U}^\omega _{in,#1})}

\def\QUiHmL#1{{Q}_{{\cal H}^-_L}({\cal U}^\omega _{in,#1})}

\def\QViHmL#1{{Q}_{{\cal H}^-_L}({\cal V}^\omega _{in,#1})}

\def\Scrip{{{\cal I}^{+}}}
\def\Scrim{{{\cal I}^-}}
\def\ScripL{{{\cal I}^+_L}}
\def\ScripLF{{{\cal I}^+_{L,F}}}
\def\ScripFL{{{\cal I}^+_{F,L}}}
\def\ScripR{{{\cal I}^+_R}}
\def\ScripRF{{{\cal I}^+_{R,F}}}
\def\ScripFR{{{\cal I}^+_{F,R}}}
\def\ScrimL{{{\cal I}^-_L}}
\def\ScrimLP{{{\cal I}^-_{L,P}}}
\def\ScrimPL{{{\cal I}^-_{P,L}}}
\def\ScrimR{{{\cal I}^-_R}}
\def\ScrimRP{{{\cal I}^-_{R,P}}}
\def\ScrimPR{{{\cal I}^-_{P,R}}}
\def\sw{\sgn(\omega)\;}
\def\pxe{{\mbox{\rm e}}}
\def\chwm2{\cosh\left[\pi\left(\frac{\omega}a-\frac{m^2}{2qE}\right)\right]}

\def\emm2{\pxe^{-\pi\frac{m^2}{2qE}}}
\def\ewm2{\pxe^{\pi\left(\frac{\omega}a-\frac{m^2}{2qE}\right)}}
\def\emwm2{\pxe^{-\pi\left(\frac{\omega}a-\frac{m^2}{2qE}\right)}}

\def\epm2{\pxe^{\pi\frac{m^2}{2qE}}}

\def\HorpL{{{\cal H}^+_L}}
\def\HorpR{{{\cal H}^+_R}}
\def\HormL{{{\cal H}^-_L}}
\def\HormR{{{\cal H}^-_R}}

\def\m{{\mbox{m}}}

\def\Ar {{\mbox{\bf A}}(k \vert k^{\prime }, k^{{\prime }{\prime }})}

\def\Br {{\mbox{\bf B}}(k^\prime \vert k, -k^{{\prime }{\prime }})}

\def\fin{\end{document}}
\def\lrD{\mathrel{{\cal D}\kern-1.em\raise1.75ex\hbox{$\leftrightarrow$}}}
\def\lr #1{\mathrel{#1\kern-1.25em\raise1.75ex\hbox{$\leftrightarrow$}}}
\def\lrpartial{\mathrel
{\partial\kern-.75em\raise1.75ex\hbox{$\leftrightarrow$}}}
\def\kvac {{\vert 0 \rangle}}

\def\bV {{\mbox{\bf V}}}
\begin{document}
\title{
{\bf Quantum charged fields in (1+1) Rindler space.}}
\author{Cl. Gabriel\footnote{Aspirant du F.N.R.S.}\,\,\,\footnote{e-mail :
gabriel@sun1.umh.ac.be}
 and Ph. Spindel\footnote{e-mail : spindel@umh.ac.be}}
\date{{\em M\'ecanique et Gravitation,
Universit\'e de Mons-Hainaut}, \\ {\em 6, avenue du Champ de Mars, B-7000 Mons, Belgium}}
\maketitle
\begin{abstract}
We study, using Rindler coordinates,  the quantization of
a charged scalar field
interacting with a constant (Poincar\'e invariant), external, electric field in (1+1) dimensionnal flatspace: our main motivation is pedagogy. We illustrate in this framework the equivalence between various approaches to field quantization commonly used in the framework of curved backgrounds. First we establish
the expression of
the Schwinger vacuum decay rate, using the operator formalism. Then we
rederive it in the framework of the
Feynman path integral method. Our analysis reinforces the
conjecture which identifies the zero
winding sector of the Minkowski propagator with the Rindler propagator.
Moreover we compute the expression
of the Unruh's modes that allow to make connection between
Minkowskian and Rindlerian quantization
scheme by purely algebraic relations. We use these modes to study the
physics of a charged two level
detector moving in  an electric field whose transitions are due to the
exchange of charged quanta. In the
limit where the Schwinger pair production mechanism of the exchanged quanta
becomes negligible we recover
the Boltzman equilibrium ratio for the population of the levels of the
detector. Finally we explicitly
show how the detector can be taken as the large mass and charge limit of an
interacting fields system.
\end{abstract}
\vspace*{3cm}
\pagebreak
\section{Introduction} Shortly after Hawking's discovery of black hole
evaporation \cite{Haw}, Unruh \cite{Unr} showed that a uniformly
accelerated detector moving in flat space perceives the Minkowski vacuum
to be thermally populated at temperature
$T_U=a/2\pi$. On the other hand, Heisenberg and Euler \cite{HeEu} showed
in 1936 that the vacuum state of a charged quantum field interacting with
a static electric field is unstable and decays into pairs. In 1951
Schwinger \cite{Schw}gave the expression of this decay rate, using the
technique known today as the Schwinger proper time representation of the functional integral. One can use  the same
formalism to describe  these two phenomena, wherein the second occurs in
the secure framework of usual quantum field in flat space. In this way we have
a physically relevant model whose interpretation is unambiguous and which
allows to exemplify several formal developments and check their validity.
In this paper we mainly study, using Rindler coordinates on the (1+1)
dimensional Minkowski space, the
quantization of a charged field interacting with a background, constant,
electric field $E$.The main motivation of this work is
pedagogical:
 we hope to illustrate (and
check) in the framework of an exactly (but non trivial) solvable model several
conjectures that often are taken for granted in quantum field theory on curved spaces. In the framework of a charged quantum field interacting with an
external
constant electric field, we  explicitly show that,  the formal
evaluation of vacuum decay rate based on an expression of
$\langle 0,out\vert 0,in\rangle$ as a functional integral on quantum fields
leads to results in accord with standard calculations,even in Rindler
coordinates, where the integrals are no more gaussian. This supports the use
of a similar approach for more complicated problems, for which no
exact solution is known. We also illustrate several aspects of physics that we
believe
to be important; for instance: 1) the occurrence of a boundary effect, whose
analog in the framework of the physics of the Boulware vacuum in Schwarzschild
geometry erases the Hawking radiation; 2) the validity of the thermodynamical
equilibrium relation in (and only in) the limit of heavy systems; 3) the
significance of the "rindlerian energy" balance 4) the contribution of the
Schwinger-like and Unruh-like
vacuum fluctuations on the transitions of the detector, similar to the
Schwinger
and Hawking mechanisms of
charged black-holes evaporation.
\\
\noindent Our paper is organized as follows. Section {\bf 1} is just a summary of the main body
of the text.
In section {\bf 2} we recall the definition of Rindler coordinates and
classify
the classical trajectories of charged particles in a constant electric
field. The results of this analysis are intensively exploited when we
 interpret the subsequent
results. In section {\bf 3} we recall the derivation of the Schwinger effect
in the framework of
the usual quantization  (using the creation and annihilation
operators formalism in Lorentzian coordinates) of a charged scalar field
interacting with an external constant electric field. Then, in section
{\bf 4}, we perform the same work in Rindler coordinates. The results differ
from those obtained using global coordinates by boundary terms. We interpret
this difference as the manifestation of the same mechanism leading
to the difference between the Boulware's and Unruh's vacua in the
framework of black hole physics. Then, in section {\bf 5}, we  reconsider
the problem in the light of the functional path integral method, and
illustrate the validity of the standard approximation scheme (saddle point
-- W.K.B. approximation) used to evaluate functional integrals, by comparing
their predictions to the results obtained previously;  we show that both
methods (fortunately) agree up to boundary terms. In the same way, we also
reinforce the interpretation of the usual minkowskian propagator as a sum over
winding Rindlerian propagators \cite{BPS}. Indeed using an expression for
what we believed to be the exact Rindler propagator to compute the rate of
particle creation from the vacuum, we shall recover at one and the same time
its leading term and the boundary correction terms. We also compute, in
section {\bf 6}, the Rindlerian particle content of the Minkowskian vacua. In
section {\bf 7}, we study the behavior of accelerated charged detectors.
We show that the detector transition (in the limit where the charge of the
exchanged quanta vanishes) are
dominated by the induced transition of the detector with preexisting vacuum
fluctuations that maximally overlap
the detector world line. We also establish precisely how to interpret a two
level charged detector  as  a limiting case of large mass $M$ and charge $Q$,
but of
finite acceleration $Q\,E/M$, of an interacting field \cite{PaMa,GMPS}
system.
Finally, some mathematical appendices provide  the technicalities
underlying the
results of the main text.\\

\section{Classical trajectories in Rindler coordinates}
 We recall that Rindler coordinates divide Minkowski space into four
patches,
 denoted hereafter by the labels
\underline{\tR}(ight), \underline{\tL}(eft), \underline{\tP}(ast) and
\underline{\tF}(uture). On each of these patches the coordinate
transformations
 between Minkowski $(t,z)$ and Rindler $(\tau,\xi)$ coordinates are given
by
\par\noindent \footnotesize
\begin{eqnarray}  &
\begin{array}{lcl}
\R \left\{\begin{array}{lcl}
         t & = &\  a^{-1} e^{a\, \xi_R} \sinh a\tau_R \; \\
           z & = &\  a^{-1} e^{a\, \xi_R} \cosh a\tau_R \;
           \end{array}\right. (z>0, \vert t\vert <z)& &
\L \left\{\begin{array}{lcl}
           t & = & - a^{-1} e^{a\, \xi_L} \sinh a\tau_L \;  \\
           z & = & - a^{-1} e^{a\, \xi_L} \cosh a\tau_L  \;
           \end{array}\right.(z<0,  \vert t\vert <z)
\\ \nonumber &&\\
\P \left\{\begin{array}{lcl}
         t & = & - a^{-1} e^{a\, \xi_P} \cosh a\tau_P \; \\
           z & = & - a^{-1} e^{a\, \xi_P} \sinh a\tau_P \;
           \end{array}\right. (t<0,  \vert z\vert <\vert t\vert)& &
\F \left\{\begin{array}{lcl}
           t & = &\  a^{-1} e^{a\, \xi_F} \cosh a\tau_F \;  \\
           z & = &\  a^{-1} e^{a\, \xi_F} \sinh a\tau_F \;
           \end{array}\right.(t>0,  \vert z\vert < t )
\end{array}
\\
\end{eqnarray}
\normalsize
\par\noindent On each of these quadrants, the respective Rindler coordinates run from $-\infty$ to $\infty$. In the following, from time to time, we shall also make use of another
standard Rindler coordinate~:
$$
\rho = a^{-1}e^{[2\,a\,\xi]}\qquad ,
$$
which is obviously always positive.\\
\par\noindent Let us emphasize that on \TR and \TL the vector field
$\partial_\tau$ is  timelike, pointing respectively to the future and the
past while on \TF and \TP it is
$\partial_\xi$ that provides us timelike directions, pointing toward the
future on
\TF and the past on \tP. In the following, we will be led to consider the
various components of the boundaries of these patches. In two dimensions,
the boundary of the full Minkowski space consists into four points: the
future and past timelike infinities, denoted by $i^+ and \ i^-$,  the
spacelike left and right infinities, $i^0_L$ and $i^0_R$, and the four
components of the null infinity: $\ScripR,\ \ScripL,\
\ScrimR$ and $\ScrimL$. Moreover, the light cone issuing  from the origin
$O$ defines the acceleration horizons of stationary Rindler's  observers,
whose trajectory equations are $\xi=C^{te}$ on
\tR , \TL and $\tau=C^{te}$ on \TP and \TF . It splits into four branches
(constituted by the null rays emerging from its vertex $O$ or ending on
it):
$\HorpR,\ \HormR,\ \HorpL$ and $\HormL$. They constitute parts of the
boundary of the different  Rindler's patches. The other components of
their ``boundaries" {\footnote {Between quotation marks because they are not really boundaries of regions in Minlowski space but boundaries of the conformal Carter-Penrose diagram. In the rest of the paper, we shall not make explicit this distinction.}}are the points
$i^+,\ i^-,\ i^0_L,\ i^0_R$ and the eight pieces of the null infinity
components denoted by
$\ScripRF$ (the future null infinity boundary of the \TR quadrant),
$\ScripFR,\
\ScripLF,\
\dots ,\
\ScrimPR \;$. All these geometrical considerations are summarized on
fig. ({\bf 1})
which represents the well-known conformal Carter-Penrose diagram of (1+1)
dimensional Minkowski space-time.

A constant electric field, which in flat coordinates is given by $ {\cal
F}
 =  E d z \wedge d t$ with $E=Cst$, reads as \footnote{ When no confusion arises we
do not make explicit on which patch the Rindler's coordinates are
considered.}
${\cal F } =  \epsilon E e^{2a\, \xi} d \xi \wedge  d \tau$ with $\epsilon =
+1 $ on
\tR, \TL and
$\epsilon = - 1$ on \TP and \tF. This field is invariant with respect to the Poincar\'e group, acting on the (1+1) Minkowski space. It can be derived from the
potential:
\begin{equation} A = {E \over 2} ( z d t - t d z) = \epsilon {E \over 2}
e^{2 a\, \xi} a^{-1} \label{rindgauge} d
\tau\qquad.
\end{equation}  Classically, it accelerates positively charged particles
towards the right ($z>0$), negatively charged particles towards the left
($z<0$). The Hamiltonian describing these classical motions is
\begin{eqnarray}  H_{cl} = - {\epsilon \pxe^{-2 a\, \xi}  \over 2 m} \left[
\left( p_\tau - {\epsilon q E \pxe^{2a\, \xi} \over 2 a} \right)^2 -
p^2_\xi \right] \qquad , \label{ham}
\end{eqnarray} the momentum $ p_\tau$ is a constant of motion:
\begin{eqnarray}   p_\tau = \epsilon m \pxe^{2 a\, \xi} \left( {q E \over 2
m a} -
\dot{\tau}
\right) = - \omega \label{ptau}
\end{eqnarray} and, once we impose the evolution parameter to be the
proper time\footnote{A dot indicating a derivative with respect to it.}
along the trajectory, the mass shell
value  of the Hamiltonian is fixed~:
$H_{cl}=-m/2$, which determines
$p_\xi$: \begin{eqnarray}   p_\xi^2 = - \epsilon m^2 e^{2 a\, \xi} + \left(
\omega + \epsilon {q E
\over 2 a} e^{2 a\, \xi}
 \right)^2\quad.
\end{eqnarray} On \TR and \tL, where $ \epsilon = 1$, $\dot\tau$ is of
constant sign on timelike trajectories
$(\dot\tau^2=\dot \xi^2+ \pxe^{-2a\xi})$, while $p_\xi (= + m \dot \xi
e^{2 a\, \xi})$   vanishes at the turning points $ \xi _+$ and $ \xi _-$:
\begin{eqnarray}
 \xi_\pm = {1\over a}\log \left\{ \left( ma  \over qE \right)
\left[ 1\pm (1 - 2
\Omega)^{1/2} \right ] \right\}
\end{eqnarray} when $ \Omega \equiv {\omega qE / m^2 a}$ is less than ${1
\over 2}$. At these turning points:
\begin{eqnarray}
\dot \tau_\pm = {qE \over 2a m\Omega} \left[ 1 \mp (1 - 2 \Omega)^{1/2}
\right]\qquad ;
\end{eqnarray} showing that, for  $\Omega<0$,  $\dot \tau_-$ is negative
on the  trajectories for which
$\xi<\xi_-$.The interpretation of these latter is obtained by noticing
that particles moving forward in
 time (with respect to the global time orientation of Minkowski space),
move with
$\dot \tau >0$ on
\TR and $\dot \tau < 0$ on \tL, while anti-particles go in the opposite
way.\\ Finally, the  acceleration at the turning points is:
\begin{eqnarray}
\ddot \xi=\pm a e^{-2 a\, \xi_{\pm}}(1-2\Omega)^{1/2}.
\end{eqnarray} Classical trajectories are branches of hyperbola (see fig. ({\bf 2})). The
geometrical significance of the constant $\Omega$ is obtained by noticing
that~:
\begin{equation}
\Delta^2=\left(\frac{m}{qE}\right)^2\left[1-2\,\Omega\right]
\label{Omegacentre}
\end{equation}
is the squared invariant distance from the origin $O$
to the center of the hyperbolic trajectory (see fig. ({\bf 2})).\\

\par\noindent On \TR and \tL,  according the values of $\Omega$, we
distinguish three classes of trajectories.
\begin{itemize}
\item Those, on which
$\Omega > {1 \over 2}$. They have no turning points and correspond to
motion with
$\xi$ varying between from $  - \infty$ to $  + \infty$, with $ \dot \xi >
0$. The centers of these hyperbolas lie in the sector $\tP$ or $\tF$.
Moreover, $\dot
\tau>0$, so these trajectories describe particles in the $\tR$ sector and
anti-particles in $\tL$; more precisely, they describe:
\begin{itemize}
\item On \tR, particles [1] (the numbers between brackets refer to the
trajectories of fig. ({\bf
1})) entering from
\TP (with
$\dot
\xi>0$), across
    the past horizon  $\HormR$ with $\dot \xi>0$ and running towards the
null infinity component $\ScripRF$
 or [2] coming from $\ScrimRP$ (with $\dot \xi<0$) and leaving \TR via
$\HorpR$,
\item On \tL, anti-particles (particles going backward in time) moving
between the  infinity and horizon components
$\ScripLF$ and $\HormL$ with $\dot \xi<0$, or between  $\ScrimLP$
and$\HorpL$ with
$\dot \xi<0$.
\end{itemize}
\item When $ {1 \over 2} > \Omega > 0$ the
 trajectories present a turning point but $\dot \tau$ is still positive,
thus they describe motions of particles on \TR  and anti-particles on
\TL. The center of the hyperbolas are now located in the $\tR$ or $\tL$
sectors, but are "inside" the hyperbolic trajectory $\Delta^2=\left(\frac m{qE}\right)^2$. In
the $\tR$ sector, the trajectories [3] with turning point $\xi_-$ describe
particles, which enter in $\tR$ by crossing
$\HormR$ with $\dot \xi>0$, pass through maximum $\xi$ at the turning
point, and quit $\tR$ by crossing $\HorpR$. The trajectories [4] with turning
point $\xi_+$ connect
$\ScrimRP$ and
$\ScripRF$, and pass through a minimum $\xi$ at the turning.
\item  When $ \Omega < 0$, the center of the hyperbolas are located
"outside" the limit hyperbola $\Delta^2=\left(\frac m{qE}\right)^2$. Trajectories [4']
corresponding to the turning points $\xi_+$ describe also particles
(because $\dot \tau>0) $ connecting
$\ScrimRP$ and $\ScripRF$. They differ from those with ${1 \over 2} >
\Omega > 0$ by the fact that the "partner" trajectories [5] (the other
branches of the hyperbolas) intersect the $\tR$ sector. Finally,
trajectories associated to the turning points
$\xi_-$ describe anti-particles entering \TR by crossing $\HormR$ and
leaving by
$\HorpR$.
\end{itemize} On the quadrants \TP and \tF , $(\epsilon = - 1)$, the sign
of the momentum
$p_\xi = - m \dot \xi e^{2 a\, \xi}$ remains  constant along the
trajectories and allows us to distinguish particle trajectories from
anti-particle ones. Particles move with $\dot \xi >0$ on \TF and $\dot
\xi < 0$ on \tP. At null infinity
$ {\dot
\tau}_{\infty} =\pm \dot \xi_{\infty}= {qE/2m a} > 0$ in accordance with
the fact that particles [6] enter into \TP from the right past
null infinity
$\ScrimRP$ (with $ d
\tau > 0$, $ d \lambda > 0$ and $\dot \xi_{\infty}<0)$ while
anti-particles [7] enter from the left past null infinity $\ScrimLP$ (with $
d
\tau < 0$, $ d \lambda < 0)$ and $\dot \xi_{\infty}>0)$. Similarly, on
\tF,  particles [8] go asymptotically to  the right component of the null
infinity
$(\ScripFR )$ (with
   $ d \tau $ and $ d \lambda$ positive) while anti-particles [9] go near $
\ScripFL$ (with $ d \tau$ and $ d \lambda$ negative and $d\xi>0$). On the
horizons,
$\dot \tau$ and $\Omega$ have opposite signs. So, the variable $\tau$
varies monotonically when
$\Omega <0$ but when $\Omega>0$,  $\dot \tau$ vanishes at
$\xi=\frac{1}{2a}\log [\frac{2a\omega}{qE}]$. When $\Omega >0$, on the
\TP sector,  the corresponding (particle) trajectories [6] connect $\ScrimPR$
to $\HormR$ and the (anti-particle)  trajectories $\ScrimPL$
 to $\HormL$. On the \TF sector, they describe  anti-particles connecting
$\ScripFL$ to
$\HorpL$ or particles between
$\HorpR$ and $\ScripFR$.  The world lines of typical trajectories are depicted
on fig. ({\bf 2}). Note that, due to the conformal character of this
picture, the
neighborhood of infinity is contracted into a finite region and the asymptotic
tangency of trajectories with their
asymptotic null
rays is no more explicitly apparent, excepted in the fact that both reach the same points
at infinity.
\section{Field quantization and Schwinger effect}
\protect\label{RindQuant} The quantization of a charged scalar field in
an external background electric field is straightforward. It consists
into three main steps:
\begin{itemize}
\item[1)] Introduce a covariant derivative $ {\cal D}_\mu = \partial_\mu
- i q A_\mu$  and determine a complete set of solutions of the wave
equation
\begin{eqnarray}   {\cal D}_\mu {\cal D}^\mu \phi = m^2 \phi \qquad
,\label{fweq}
\end{eqnarray} \item[2)] Separate the modes into two classes $\{ \phi_A\}$
associated to particles and $ \{\psi_{\tilde A}^* \}$ associated to
anti-particle, normalize them using the scalar product built from the usual bilinear current
\ba J_\mu(\phi_A,\phi_{A^\prime})=-i\phi_A^*\; \lrD_\mu\;
\phi_{A^\prime}\label{jform}
\ea
integrated on appropriated Cauchy's surface $\Sigma$:
\ba
\langle \phi_A,\phi_{A^\prime}\rangle=\int_{\Sigma}J_\mu(\phi_A,\phi_{A^\prime})dx^\mu\quad, \label{scalaire}
\ea
and define the quantum field
\begin{eqnarray}
\hat \Phi = \sum _{A} a(A) \phi_A +{\sum _{\tilde A}} b^+ (\tilde A^)
\psi_{\tilde A}^* \qquad ,
\end{eqnarray} whose amplitudes satisfy usual commutation relations :
\begin{equation} \begin{array}{lcl} [ a (A), a^+ (A^\prime)]= \delta_{A
A^\prime}&\qquad&[ b(\tilde A), b^+ (\tilde A^\prime)]=\delta_{\tilde A
\tilde A^\prime} \\
\strut [ a(A),b(\tilde A)]=0 & &\mbox{\rm etc ...}
\end{array}
\label{ccc}
\end{equation}
 \item[3)] Build the Fock space from the vacuum state $ \kvac $ defined by
\begin{equation}  a (A) \kvac = 0 = b (\tilde A) \kvac \quad .
\end{equation}
 \end{itemize} The interesting physics, the well known Schwinger
phenomenon \cite{Schw}, results from the  scattering of the waves on the
external field. In the framework of second quantized field theory, a mode
describing an incoming particle evolves into a mode associated to the
superposition of outgoing particles and anti-particles \cite{BMPS}. Let
us briefly recall how this mechanism works in Minkowskian coordinates. In
the gauge (\ref{rindgauge}), equation (\ref{fweq}) reads:
\begin{equation} \left[(\partial_z +
\frac{iqEt}{2})^2-(\partial_t-\frac{iqEz}{2})^2
\right] \phi= m^2 \phi\qquad ,
\label{feqxt}
\end{equation}
 but as in the gauge $A=E zdt$, the Dalembertian operator  and
$\partial_t$ commute, the solutions of equation (\ref{feqxt}) can be
expressed as superposition of the modes
$\pxe^{-i\sigma t}\pxe^{i\frac{qE}{2}tz}
\varphi_\sigma (z)$ where the functions  $\varphi_\sigma (z)$ obey the
equation:
\begin{equation} \left[\partial^2_z+(\sigma+qEz)^2 \right] \varphi_\sigma
(z)= m^2
\varphi_\sigma (z).  \end{equation}
 whose general solution  can be expressed as a linear combination of
parabolic
 cylinder functions $D_\nu[\zeta]$; (see for instance  refs \cite{BPS,BMPS} for a detailed
discussion of the solutions of this equation and their physical
interpretation). The modes describing incoming (anti-)particles are expressed in terms of parabolic cylinder functions (whose integral representation will be given in section {\bf 6}).
\begin{eqnarray}
\phi_\sigma^{p \, in}&=&\frac 1M D_{i\frac{m^2}{2qE}-\frac 12}\left[\zeta \right] \pxe^{-i\sigma
t}\pxe^{i\frac{qE}{2}tz}\qquad ,\qquad
\phi_\sigma^{a \, in *}=\frac {e^{\frac {3i\pi} 4}}{M}
D_{-i\frac{m^2}{2qE}-\frac 12}\left[ -\zeta^*\right] \pxe^{-i\sigma
t}\pxe^{i\frac{qE}{2}tz}
\qquad , \label{Dmodein}
\end{eqnarray}
with $\zeta=
e^{-\frac {3 i
\pi} 4} \sqrt{2qE}(z+\frac \sigma {qE})$.
Equivalently,  modes corresponding to outgoing
(anti)-particles are given by $\phi_\sigma^{p \, out}[t,z]=\left(\phi_\sigma^{p \, in}[-t,z]\right)^*
 \quad {\mbox{\rm and}}\quad \phi_\sigma^{a \, out}[t,z]=\left(\phi_\sigma^{a \, in}[-t,z]\right)^*$. The constant $M$ has been fixed in order that these modes are normalized as
follows:
\begin{eqnarray}
\langle\phi_\sigma^{p \, in},\phi_{\sigma '} ^{p \, in} \rangle
&=&\langle\phi_\sigma^{p \, out},\phi_{\sigma '} ^{p \, out} \rangle =
\delta(\sigma-\sigma ')\qquad ,\nonumber \\
\langle\phi_\sigma^{a \, in},\phi_{\sigma '} ^{a \, in}\rangle
&=&\langle\phi_\sigma^{a \, out},\phi_{\sigma '} ^{a \, out} \rangle =
-\delta(\sigma-\sigma ') \qquad ,\nonumber \\
\langle\phi_\sigma^{p \, in},\varphi_{\sigma '} ^{a \, in *} \rangle
&=&\langle\phi_\sigma^{p \, out},\phi_{\sigma '} ^{a
\, out *} \rangle = 0\qquad . \label{normphi}
\end{eqnarray} These two basis are related by the Bogoljubov
transformation:
\begin{eqnarray}
\phi_\sigma^{p \, out}&=&\gamma \;\phi_\sigma^{p \, in}-\delta^*
\phi_\sigma^{a\, in *}\qquad , \nonumber \\
 \phi_\sigma^{a \, out *}&=&\gamma ^* \phi_\sigma^{a \, in *}-\delta
\;\phi_\sigma^{p
\, in }
\label{bogomink}
\end{eqnarray} with:
\begin{eqnarray}
\gamma &=&\frac {\sqrt{2 \pi}}{\Gamma[\frac 12 +i\frac{m^2}{2qE}]}
e^{-\frac {\pi m^2}{4 qE}}e^{i\frac \pi 2} \nonumber\quad ,\quad
\delta =e^{-\frac {\pi m^2}{2 qE}}e^{i\frac \pi 2}\qquad ,
\label{gamdel}
\end{eqnarray} coefficients that verify the charge conservation relation $\vert \gamma \vert^2-\vert \delta \vert^2 = 1$.\\
The $in$ and $out$ vacua are related by:
\begin{equation} \vert 0, Mink,in \rangle=N^{-1/2} e^{\frac
{\delta^*}{\gamma}\sum_\sigma a^{\dag \, out}_\sigma b^{\dag \,
out}_\sigma}
\vert 0, Mink, out\rangle \label{videoutin}
\end{equation}
 where we have fixed an arbitrary phase by setting $N=\left(\prod_\sigma
\vert
\gamma
\vert^2 \right)$. Equation (\ref{videoutin}) show explicitly that the
$in$-vacuum is filled with $out$-particles. This is the content of the
Schwinger effect. Computing the probability of persistence of the vacuum,
we obtain:
\begin{eqnarray}
\vert\braket
{0,out}{0,in}\vert^2&=&\exp\left[-\sum_{\sigma}\ln\left(1+\vert\delta\vert^2
\right)\right]\nonumber
\\ &=&\pxe^{-\mbox{\rm  L\,T$\;\Gamma$}} \label{Drate}
\end{eqnarray} i.e., in the limit of large space-time volume $L\times T$,
the rate:
\begin{equation} \Gamma=\frac{qE}{2\;\pi}\ln \;(1+\pxe^{-\frac{\pi m^2}{qE}})
\end{equation}
 of pair creation by unit of space-time volume. The link between the
 number of modes $\sum_\sigma$
 and the factor ${\mbox{\rm L\, T ${qE}/{2\;\pi}$}}$ is obtained as
 follows \cite{BMPPS,BMPS}. Classically, the trajectories of charged
 particles in a constant electric field are hyperbolas. Each is
 characterized by the location of its center  and its radius $= m/qE$. Quantum
 mechanically, it appears that it is
 in the neighborhood of the classical center, in a region of
 space-time extension of the order of $(m/qE)\times (m/qE)$ that a
  wave packet describing an incoming particle gives birth to an
  anti-particle. Accordingly, the number of modes relevant to the
  pair creation mechanism in a given
  space-time domain, of dimensions T$\times$L, are those who centers belong
to this domain. The
  density of modes of frequency $\sigma$ in the time-$t$ interval T is
  $(\mbox{\rm T}/2\,\pi)\ d\sigma$. The centers of the relevant
  modes are located at the points $z=-\sigma/qE$ belonging to the
  space interval L=$z_{1}-z_{2}$. Therefore the values of $\sigma$,
which contribute to the sum,
  are those include in the interval $[-qE\, z_{1},\; -qE\,z_{2}]$ so
  that in the limit of large space-time volume (in order that the
  boundary effects that are not taken into account in this counting of
  modes become negligible) we obtain $\sum_{\sigma} = (T/2\,\pi)
  \int^{qEz_{2}}_{qEz_{1}}  d\sigma= T\,L\,(qE/2\,\pi)$.
\section{Rindlerian field quantization} In Rindler coordinates, the wave
equation (\ref{fweq}) becomes:
\begin{eqnarray}
\left[ -  e^{-2 a\, \xi} (\partial_\tau^2 - \partial_\xi^2) + i {\epsilon qE
\over a} \partial_\tau + \left( {q E \over 2} \right)^2 a^{-2} e^{2 a\, \xi}
\right]
\phi = \epsilon m^2 \phi \qquad .
\label{Rweq}
\end{eqnarray} As $\partial_\tau$ commutes with the Dalembertian
operator,  the general solution of the wave equation can be expressed as
a superposition of  modes
$\phi_{\omega}= (e^{- i \omega
\tau}/\sqrt{2\pi})  {\cal {F}}_\omega (\xi)$. The functions $  {\cal
F}_\omega (\xi)$ obey the second order differential equation:
\begin{equation} \left[\frac{d^2}{d\xi^2}+\left(\omega + \epsilon
\frac{qE}{2a}\pxe^{2a\xi}\right)^2 -\epsilon
m^2\pxe^{2a\xi}\right]{\cal{F}}_{\omega}(\xi)=0\qquad .\label{FFweq}
\end{equation}
 The asymptotic behaviors of the solutions of this equation\footnote{The
solution of this equation will be expressed in terms of Whittaker's
functions in the next section.}
 are easily obtained by the standard W.K.B. technique.  The modes
$\phi_{\omega}$\footnote{In principle, we would have had to indicate, for
each mode, the patches on which it is considered, and index the various
coefficients
$C$ and
$D$ that it defines by labels $R,L,P$ or $F$. As we found the notations
 cumbersome enough as they are, we shall omit these precisions, hoping
that the context will always be clear enough to avoid ambiguities.} can
be expressed as:
\begin{eqnarray}
\phi_{\omega}&\approx& \left\{ C_{+}^{\omega}\exp+i\left(\epsilon
\frac{qE}{4a^2}\pxe^{2a\xi}+(\omega
-\frac{m^2a}{qE})\xi\right)\right.\nonumber\\
&&+\left.C_{-}^{\omega}\exp-i\left(\epsilon
\frac{qE}{4a^2}\pxe^{2a\xi}+(\omega
-\frac{m^2a}{qE})\xi\right)\right\}\left(\frac{2a}{qE}\right)^{1/2}\pxe^{-a\xi}
\frac{\pxe^{-i\omega\tau}}{\sqrt{2\pi}}\label{approxC}
\end{eqnarray} when $\xi \approx +\infty$ i.e. near $\Scrip$ and
$\Scrim$, and
\begin{eqnarray}
\phi_{\omega}&\approx& \frac{1}{\sqrt{2\vert\omega\vert}}\left\{
D_+^{\omega}\exp i\left[\omega\,
\xi\right]+D_-^{\omega}\exp-i\left[\omega\, \xi\right]\right\}
\frac{\pxe^{-i\omega
\tau}}{\sqrt{2\pi}}\label{approxD}
\end{eqnarray} when $\xi \approx -\infty$ i.e. near the horizons. More
precisely, near $\ScripR$ and
$\ScrimL$, whose points are labeled by $u=\tau - \xi$ coordinates, the
relevant part of a mode is given by the weight of its $\exp(-i\omega u)$
component, i.e. its
$C_{+}^{\omega}$ coefficient. Similarly, near $\ScrimR$ and $\ScripL$, it
is its
$\exp(-i\omega v)$ part, i.e. its
$C_{-}^{\omega}$ coefficient that governs it. Near the horizons, there
are the components $D_{-}^{\omega}\exp(-i\omega v)$ and
$D_{+}^{\omega}\exp(-i\omega u)$ that become relevant according to whether we
are close to $\HorpR$ or
$\HormL$ on one hand or near $\HormR$ or $\HorpL$ on the other.
 Of course all these different weights
($C_{+}^{\omega},\ C_{-}^{\omega},\ D_-^{\omega},\ D_+^{\omega}$)  are not
independent. They
are related by charge conservation. The scalar product (\ref{jform}, \ref{scalaire}) is defined on the Rindler patches $\tR$ and $\tL$ by choosing as Cauchy's surfaces sections $\tau=C^{te}$. On
such sections the scalar product of the modes we are considering reads as:
\begin{eqnarray}
<\phi_\omega,\phi_{\omega'}>_{R,L}&=&\theta_{R,L}\int_{-\infty}^{+\infty}(
\omega+\omega'+{qE\over a}\pxe^{2a\xi}){\cal F}_{\omega'}^* {\cal F}_\omega
\ d\xi \nonumber \\ &\equiv&Q^{R,L}_\omega \delta(\omega-\omega')\qquad.
\end{eqnarray}
The prefactor $\theta_{R,L}$ is given by the relative sign
of the orientation of the vector field $\partial_\tau$ with respect to
the global time orientation:
$\theta_{R}=1,\theta_{L}=-1$. Using the asymptotic expressions (\ref{approxC}, \ref{approxD}) of the
modes
 we obtain
\begin{eqnarray}
  Q^{R,L}_\omega &=&
\theta_{R,L}\left[|C_{+}^{\omega}|^2+\sw|D_{-}^{\omega}|^2\right]\nonumber
\\
&=&\theta_{R,L}\left[|C_{-}^{\omega}|^2+\sw|D_{+}^{\omega}|^2\right]\qquad
,\label{QRL}
\end{eqnarray} in accordance with the wronskian theorem. By
taking as surfaces of integration
$\tau=\pm \infty$, we easily see that each term of these sums  represents a
fraction of the total charge of the mode that comes from  past or goes
to  future components of the boundaries
 of the patches $\tR$ and
$\tL$. More precisely we have:
\begin{equation} \begin{array}{lcl}
\begin{array}{ccc} Q^R_\omega&=&Q^{\ScripRF}_\omega +Q^{\HorpR}\\
\strut &=&Q^{\ScrimRP}_\omega +Q^{\HormR}
\end{array} &\quad&
\begin{array}{ccc} Q^L_\omega&=&Q^{\ScripLF}_\omega +Q^{\HorpL}\\
\strut&=&Q^{\ScrimLP}_\omega +Q^{\HormL}
\end{array}
\end{array}\qquad ,\label{Qsplit}
\end{equation}
 with the various charges associated to the mode $\phi_{\omega}$ given by
the coefficients appearing in its asymptotic expansion:
$Q^{\ScripRF}_\omega=+|C_{+}^{\omega}|^2$, $Q^{\HorpR}=+\sw |D_{-}^{\omega}|^2$...\\ On the quadrant $\tP$ and $\tF$, the scalar product is
given by the integration of the current (\ref{jform})
 over $\xi =C^{te}$ surfaces. It expresses itself in terms of the Wronskian of the solutions $\cF_\omega^*$ and $\cF_\omega
$ of eq. (\ref{FFweq}) as:
\begin{eqnarray}
  <\phi_\omega,\phi_{\omega'}>_{P,F}&=&\theta_{P,F}\; i\;
\cF_\omega^*\lrpartial_\xi\cF_\omega
\delta(\omega-\omega')\nonumber \\ &\equiv&Q^{P,F}_\omega
\delta(\omega-\omega')\label{QFP}
\end{eqnarray} where, due to the time orientation of the vector field
$\partial_\xi$ we have
$\theta_{F}=1,\theta_{P}=-1$.\\
 Here the evaluation of the charge is simpler. The Wronskian theorem tells
us that:
\begin{eqnarray}
  Q^{P,F}_\omega &=&
\theta_{P,F}\sw\left[|D_-^{\omega}|^2-|D_{+}^{\omega}|^2\right]\nonumber\\
&=&\theta_{P,F}\left[|C_{+}^{\omega}|^2-|C_{-}^{\omega}|^2\right]\qquad
\end{eqnarray}
 As in eq. (\ref{Qsplit}), the total charge $Q^{P,F}_\omega$ can be split in terms
of the amounts of charge crossing the different components of the
boundaries of the Rindler patches. We obtain~:
\begin{equation}
\begin{array}{lcl}
\begin{array}{ccc} Q^F_\omega&=& Q^{\HorpL}+Q^{\HorpR}\\
\strut&=&Q^{\ScripFL}_\omega +Q^{\ScripFR}_\omega
\end{array} &\quad&
\begin{array}{ccc} Q^{P}_\omega &=& Q^{\HormR}_\omega +
Q^{\HormL}_\omega\\

                                &=& Q^{\ScrimPL}_\omega +
Q^{\ScrimPR}_\omega
\end{array}
\end{array}\qquad ,
\end{equation}
 with $Q^{\ScripFR}_\omega=\ +|C_{+}^{\omega}|^2$, $Q^{\ScrimPR}=\
+|C_{-}^{\omega}|^2$...\\
The particle assignment of these modes is obtained by
considering the asymptotic behaviors of wave packets of almost fixed
value of $\omega$ built on these modes, for instance slightly spread
around a value $\bar \omega$ with a Gaussian weight. Indeed,
asymptotically, they are expressed as superpositions of W.K.B.
solutions
$ \exp i S (\tau , \xi ,\omega )$ where $ S (\tau, \xi, \omega)$ is the
Maupertuis action for the Hamiltonian (\ref{ham}). By evaluating these
superpositions in the saddle point approximation we see that these
packets are localized on classical trajectories $
\left.{\partial S / \partial \omega}\right|_{\bar \omega} = C^{te}$. This
allows to interpret them in terms of particles, in the light of the analysis of the previous section. Moreover, having this particle interpretation of the modes, we will obtain, in the subsequent subsections, a precise description of the Schwinger mechanism of pair creation in the Rindler vacua; fig. ({\bf 3}) summarizes the results.
\bigskip
\subsection{Quantization on the right quadrant (\tR)} In Appendix
{\bf\ref{A}}\ we discuss the solutions
 of eq. (\ref{Rweq}) in terms of Whittaker's functions and, using integral
representations of them, we give the expression of the various
coefficients needed to determine the charges carried by the modes.  These
considerations lead us to define on the right quadrant $in$- and
$out$-modes as follows:\\

$\bullet$ Particles $in$-modes are given by the functions:
\begin{eqnarray}   {\cal U}_{in,R} ^\omega= \NUwi{R} \frac{e^{- i \omega
\tau}}{\sqrt{2\pi}}  e^{- a\, \xi} W_{i({\omega\over 2a} - {m^2
\over 2 qE}), i{\omega\over 2a}}
\left[ - i {qE \over 2a^2} e^{2 a\, \xi} \right]\label{uin}
\end{eqnarray} whose supports (in terms of wave packet) are on $ \HormR,\
\HorpR
$ and $ \ScripR$ and
\begin{eqnarray}   {\cal V}_{in,R} ^\omega=  \NVwi{R} \frac{e^{- i \omega
\tau}}{\sqrt{2\pi}}  e^{- a\, \xi} M_{-i({\omega\over 2a} - {m^2
\over 2 qE}), -i{\omega\over 2a}}
\left[ + i {qE \over 2a^2} e^{2 a\, \xi} \right] \label{vin}
\end{eqnarray} whose supports are on $ \ScrimR, \ScripR$ and ${\cal H}_R
^+$. Obviously $ {\cal V}_{in}^\omega$ and $ {\cal U}_{in}^\omega$ modes
are orthogonal; their supports do not overlap on the Cauchy surface $\HormR
\cup \ScrimR$. They are also orthogonal among themselves:
\begin{eqnarray}   \langle{\cal U}_{in,R}^\omega,  {\cal
U}_{in,R}^{\omega^\prime}\rangle &=& \vert \NUwi{R} \vert^2
 \left(\frac{qE}{a}\right) \frac{\pxe^{\frac{\pi \omega}{2a}}} {\sinh (\pi
\omega/ a )} {\cosh \left[\pi \left( {\omega\over a} - {m^2 \over 2 qE}
\right)\right]}\;
\delta (\omega- \omega^\prime) \nonumber \\
 &\equiv&  \QUwi{R} \delta (\omega-
\omega^\prime) \qquad, \nonumber \\  \langle {\cal V}_{in,R}^\omega, {\cal V}_{in,R}
^{\omega^\prime}\rangle  &=& \vert \NVwi{R} \vert ^2
\left(\frac{\omega/a}{\sinh(\pi\omega/a)}\right)
\left(\frac{qE}{a}\right)
\pxe^{\frac{\pi}2 ({{\omega}\over{2a}}+{{m^2}\over{qE}})}
{\cosh\left[\pi\left(\frac{\omega}a-\frac{m^2}{2qE}\right)\right]}
\delta (\omega- \omega^\prime)  \nonumber \\
 &\equiv&  \QVwi{R} \delta (\omega-
\omega^\prime)\qquad. \label{normuv}
\end{eqnarray} \noindent The modes $\cUwi{R}$ enter into \TR via $\HormR$
while the modes
$\cVwi{R}$ emerge from $\ScrimR$. They are normalized such that
$\QUwi{R} = \sw=\pm 1$ and $\QVwi{R} = + 1$ represent their total charge.
From our previous discussion we obtain :
\begin{eqnarray}
\QUiHpR{R} \  =&\sw
\frac{\pxe^{-\frac{\pi\omega}{a}}\;{\cosh\left[\pi\frac{m^2}{2qE}\right]}}
{{\cosh\left[\pi\left(\frac{\omega}a-\frac{m^2}{2qE}\right)\right]}}
&\equiv \ \sw q_1\nonumber\qquad,\\ &&\nonumber\\
\QUiIpR{R}\ =&
\frac{\pxe^{-\frac{\pi
m^2}{2qE}}\;\left\vert\sinh\left[\pi\frac{\omega}a\right]
\right\vert}
{{\cosh\left[\pi\left(\frac{\omega}a-\frac{m^2}{2qE}\right)\right]}}
&\equiv\  q_2 \qquad. \label{QUin}
\end{eqnarray} In the same way we obtain:
\begin{eqnarray}
\QViIpR{R}= q_1\quad ,\quad \QViHpR{R}  = \sw q_2 &&\quad , \nonumber \\
\QVwi{R} = \QViImR{R}=\QViHpR{R}+\QViIpR{R}=+1&&\quad .\label{QVin}
\end{eqnarray} The interpretation of these equations is obvious. The modes
$ {\cal U}_{in,R}^\omega$ describe incoming particles when $ \omega> 0$
and incoming anti-particles when $\omega < 0$. The modes $ {\cal
V}_{in,R}^\omega$ always describe incoming particles. A schematic drawing
of wave packets built with these modes are depicted on fig. ({\bf 3}).  During
their voyage in
\tR, each incoming mode splits into two outgoing  branches,
one ending on $ {\cal I}_R^+$ and the other crossing
${\cal H}_R^+$ when they leave the quadrant. Charge conservation implies
that:
\begin{eqnarray}   Q  & = Q _{{\cal H}_R^+}  + Q_{{\cal I}_R^+} \qquad .
\nonumber
\end{eqnarray} Moreover, in a constant electric field only positively
(resp. negatively) charged particles can arrive from $
\ScrimR$ (resp. $\ScrimL$) and go up to $ \ScripR$  (resp.
$\ScrimR$). This is reflected by the positivity of $Q_{\ScrimR} ({\cal
V}_{in}^\omega)$ and $
\QUiIpR{R}$. The charged scalar quantum field operator can be represented
on \TR as
\begin{eqnarray}
\hat\phi_R&=& \int_{- \infty} ^{+ \infty} d \omega[ a_{{\cal V}_R} ^{in}
(\omega)
\cVwi{R}+
\theta (\omega) a_{{\cal U}_R}^{in} (\omega) \cUwi{R}\nonumber\\ & &+
\theta (-
\omega) b_{{\cal U}_R}^{+\; in} (\omega)
\cUwi{R}]\qquad, \label{QFORi}
\end{eqnarray} the various field amplitude operators being defined as
usual by:
\begin{eqnarray}  a_{\cal U}^{in} (\omega)\kvac_{{\cal
U}^{\omega}_{in}}^{\R}=0\qquad\omega>0\qquad,\nonumber \qquad b_{\cal U}^{in}
(\omega)\kvac_{{\cal U}^{\omega}_{in}}^{\R}=0\qquad \omega<0\qquad,\qquad
a_{\cal V} ^{in} (\omega)\kvac_{{\cal V}^{\omega}_{in}}^{\R}=0\qquad,
\end{eqnarray} \noindent and the \TR $in$-vacuum state\footnote {This
vacuum state is not the end  of the story. Indeed the true vacuum state
must be defined on a Cauchy surface of the whole space, i.e. it must
include a factor representing the \tL\, part of the vacuum.}\   as the
tensorial product
\begin{eqnarray}
 \kvac_{in}^{\R} = \prod_{\omega=-\infty}^{+\infty}\left( \kvac_{{\cal
U}^{\omega}_{in}}^{\R}\otimes
\kvac_{{\cal V}^{\omega}_{in}}^{\R}\right).
\end{eqnarray} $\bullet$ Similarly we may consider modes associated to
outgoing particles. They are given by:
\begin{eqnarray}   {\cal V}_{out,R}^\omega= \NVwo{R} \frac{e^{-i
\omega\tau}}{\sqrt{2\pi}} e^{-a\, \xi}  W_{- i({\omega\over 2a} - {m^2
\over 2 q E}), i {\omega\over 2 a}}
\left[ +i {q E \over 2 a} e^{2 a\, \xi} \right]\qquad ,\label{VoR}
\end{eqnarray}  with $|\NVwo{R}|=|\NUwi{R}|$. They correspond to modes
entering into \TR from $ \HormR$ and $ \ScrimR$ and leaving it via
$\HorpR$. They carry the asymptotic charges
\begin{equation}  Q ({\cal V}_{out,R}^\omega) = Q_{\HorpR} ({\cal
V}_{out,R}^\omega)=
\sw \quad ,\quad Q_{\HormR} ({\cal V}_{out,R}^\omega) =
\sw q_1\quad ,\quad Q_{\ScrimR} ({\cal V}_{out,R}^\omega) = q_2\quad
.\label{QVoutR}
\end{equation} \noindent These modes are also orthogonal. They describe
outgoing particles $(\omega> 0)$ or anti-particles $(\omega< 0)$ and of
course obey the charge conservation rule:
\begin{eqnarray}   Q ({\cal V}_{out,R}^\omega) = Q_{\HormR} ({\cal
V}_{out,R}^\omega) + Q_{\ScrimR} ({\cal V}_{out,R}^\omega)=\sw
\end{eqnarray} \noindent with $ Q_{\ScrimR} ({\cal V}_{out,R}^\omega)$
always positive. As for the
$in$-modes, to obtain a complete set on
\tR, we have to  add to the set of functions (\ref{VoR}) the modes
\begin{eqnarray}   {\cal U}_{out,R}^\omega&=& \NUwo{R} \frac{e^{-i
\omega\tau}}{\sqrt{2\pi}}e^{-a\xi}
 M_{i({\omega\over 2 a} - {m^2 \over 2 qE}), i {\omega\over 2a}} \left[-
i{ qE \over 2a^2} e^{2 a\, \xi} \right] \qquad,
\label{UoR}
\end{eqnarray} which are also orthogonal among themselves and with respect
to the set $\{ {\cal V}_{out,R} ^\omega\}$. These modes come from $
\HormR$ and $\ScrimR$ and end on $ {\ScripR}$ and  $
\ScripR$. They charge content is
\begin{equation} \QUoIpR{R} = 1 \quad ,\quad
\QUoImR{R} =q_1 \quad ,\quad
\QUoHmR{R} = \sw q_2  \quad .\label{QUoutR}
\end{equation}
 These charge are such that:
\begin{eqnarray}  Q ({\cal U}_{out}^\omega) = Q_{\HormR} ({\cal
U}_{out}^\omega) + Q_{\ScrimR} ({\cal U}_{out}^\omega)=1.
\end{eqnarray} With these modes we may express the quantum field operator
(\ref{QFORi}) as
\begin{eqnarray}
\hat\phi_R &=& \int_{- \infty} ^{+ \infty} d \omega[ a_{{\cal U}_R} ^{out}
(\omega)
\cUwo{R}+
\theta (\omega) a_{{\cal V}_R}^{out} (\omega) \cVwo{R}\nonumber\\ & & +
\theta (-
\omega) b_{{\cal V}_R}^{+\; out} (\omega) \cVwo{R}]\label{QFORo}\qquad.
\end{eqnarray} Of course $out$-  and $in$-modes are related by a
Bogoljubov transformation. Using formulas recalled in Appendix
{\bf\ref{A}}, we easily obtain~:
\begin{eqnarray}
 {\cal
U}_{out,R}^{\omega}&=&\theta(\omega)\alpha^{R}_{{\cal{U}}{\cal{U}}}(\omega){
\cal
U}_{in,R}^{\omega}+\alpha^{R}_{{\cal{U}}{\cal{V}}}(\omega){\cal{V}}_{in,
R}^{\omega}+\theta(-\omega)\beta^{R}_{{\cal{U}}{\cal{U}}}(\omega){\cal
U}_{in,R}^{\omega}\qquad,\\   {\cal  V}_{out,R}^{\omega}
&=&\theta(\omega)\left\{\alpha^{R}_{{\cal{V}}{\cal{U}}}(\omega){\cal{U}}_{in,
R}^{\omega}+\alpha^{R}_{{\cal{V}}{\cal{V}}}(\omega){\cal{V}}_{in,
R}^{\omega}\right\}+\theta(-\omega)\left\{\gamma^{R}_{{\cal{V}}{\cal{V}}}(
\omega) {\cal{V}}_{in,
R}^{\omega}+\epsilon^{R}_{{\cal{V}}{\cal{U}}}(\omega){\cal
U}_{in,R}^{\omega}\right\}\qquad,\label{bogoRindR}
\end{eqnarray} whose  coefficients, fully displayed in Appendix
{\bf {B}}, are such that:
\begin{eqnarray}
\vert \alpha^{R}_{{\cal{U}}{\cal{V}}}(\omega)
\vert^2=\vert\epsilon^{R}_{{\cal{V}}{\cal{U}}}(\omega)\vert^2=\vert
\alpha^{R}_{{\cal{V}}{\cal{U}}}(\omega) \vert^2=q_1&&\qquad ,\nonumber\\
\vert \alpha^{R}_{{\cal{U}}{\cal{U}}}(\omega)\vert^2=\vert
\beta^{R}_{{\cal{U}}{\cal{U}}}(\omega)\vert^2=\vert
\gamma^{R}_{{\cal{V}}{\cal{V}}}(\omega)\vert^2=q_2 &&\qquad ,
\end{eqnarray} and (remember that $q_1$ and $q_2$ are function of $\omega$):
\begin{eqnarray}
\vert \alpha^{R}_{{\cal{U}}{\cal{V}}}(\omega<0) \vert^2-\vert
\beta^{R}_{{\cal{U}}{\cal{U}}}(\omega<0)\vert^2 \equiv q_1-q_2=1&&\qquad
,\nonumber \\
\vert \alpha^{R}_{{\cal{U}}{\cal{V}}}(\omega>0)\vert^2+\vert
\alpha^{R}_{{\cal{U}}{\cal{U}}}(\omega>0)\vert^2 \equiv q_1+q_2=1&&\qquad
.
\end{eqnarray} The last relation shows that the Bogoljubov transformation
between the $in$- and $out$- vacuum state factors is non trivial only
for $\omega<0$, in which case we obtain
\begin{equation} \kvac_{{\cal U}^{\omega}_{in}}^{\R}
\kvac_{{\cal V}^{\omega}_{in}}^{\R}=|\alpha_{{\cal U}{\cal V}}
(\omega)|^{-1}
\pxe^{-\frac {\gamma_{{\cal V}{\cal V}} (\omega)}{\alpha_{{\cal U}{\cal
V}} (\omega)}a^{\dagger\, out}_{{\cal U}}(\omega) b^{\dagger\, out}_{{\cal
V}}(\omega) }
\kvac_{{\cal U}^{\omega}_{out}}^{\R}
\kvac_{{\cal V}^{\omega}_{out}}^{\R} \qquad (\forall \omega<0),
\end{equation}
 reflecting the instability of the $in$-vacuum and the pair production
predicted by the Schwinger mechanism. The fact that only the $\omega <0$ vacuum
factors appears to be unstable is in accord with formulas
(\ref{QUin}),(\ref{QVin}) which indicate that it is only for these values
of $\omega$ that an $in$-mode of a given charge generate
$out$-branches with opposite charges. Note also that this instability can
be interpreted as a tunneling between classical trajectory living in \TR
and its ``partner" when this latter visits the \TR quadrant.

The probability of vacuum persistence is given by:
\begin{eqnarray}
\vert\strut^{\R}\hspace{-1mm}\braket{0,in}{0,out}^{\R}\vert^2
&=&\prod_{\omega<0}|\alpha_{{\cal U}{\cal V}} (\omega)|^{-2}\nonumber\\
&=&\exp\left[-\sum_{\omega <0}\ln\left(1+\vert\beta_{{\cal U}{\cal
U}}^R\vert^2\right)\right]\nonumber \\ &=&\exp\left[-\sum_{\omega
<0}\ln\left(\frac{1+{\pxe^{-\pi\frac{m^2}{qE}}}}{1+{\pxe^{-\pi\frac{m^2}{qE}}}
{\pxe^{2\frac{\pi\omega}{a}}}}\right)\right]\qquad .\label{inoutR}
\end{eqnarray} The significance of a symbol as $\sum_{\omega<0}$ is
analyzed in ref. \cite{BMPPS,BMPS} and recalled at the end of section
{\bf 3}.
The density of modes of frequency
$\omega$, in an interval of width ${\cal T}$ in the $\tau$ variable, is
${\cal T}/2\pi\
d\omega $. The sum over $\omega <0 $ corresponds to consider essentially
the particle trajectories whose mirror trajectories also visit the \TR
quadrant. The spacelike coordinate
$\xi_c$ of the center of contributing trajectories are obtained
immediately from equation (\ref{Omegacentre}) which gives
$d\omega=\frac{a m^2}{qE}d\Omega=- a qE\rho_c d\rho_c$ and thus
${\cal T}\,d  |\omega| =qE\,dV$.  As a consequence we obtain:
\begin{equation}
\vert\strut^{\R}\hspace{-1mm}\braket{0,in}{0,out}^{\R}\vert^2=
\exp\left[-\ln\left(1+{\pxe^{-\pi\frac{m^2}{qE}}}\right)
\frac{qE}{2\pi}\bV +\mbox{\rm ${\cal T} \times$ finite terms}\right]\qquad
,\label{vacioR}
\end{equation}
 where in the limit of large space-time volume $\bV$ (i.e. large interval
$[-\omega_1,-\omega_2]$ of length $\Delta
\omega \gg a $~), we have
\begin{equation}
\bV=\frac {\cal T}{qE}{\Delta\omega }=\frac{{\cal T}}{2\,
a}\left(\exp[2a\xi_c(-\omega
_2)] -\exp[2a\xi_c(-\omega _1)]\right)\equiv T\ R\qquad ,
\end{equation}
where we have denoted the difference of Rindler radii by
$R=\rho_{c}(\omega_2)-\rho_{c1}(\omega_1)$ and introduced
a mean proper time $T={\cal
T}(\rho_{c}[\omega_2]+\rho_{c1}(\omega_1)]/(2\,a)\equiv {\cal T}\, \bar R$.
>From eq. (\ref{vacioR}) we see that the rate of Rindler pair creation
from Rindler right quadrant vacuum consists in two pieces. The first
one, as expected, is proportional to the measure of the volume multiplied
by the
Schwinger rate. To this dominant (in the large volume limit) term, a
surface
correction, which can be expressed in term of dilogarithm (Spence) function
$Li_2$,
has to be subtracted:
\begin{eqnarray}
\sum_{\omega <0}{1+{\pxe^{-\pi\frac{m^2}{qE}}}
{\pxe^{2\frac{\pi\omega}{a}}}}&=&\frac{{\cal
T}}{2\,\pi}\int_{-\omega_{2}}^{-\omega_{1}}
\left({1+{\pxe^{-\pi\frac{m^2}{qE}}}
{\pxe^{-2\frac{\pi\omega}{a}}}}\right)\,d\omega \nonumber \\
&=& \frac{a\,{\cal T}}{4\, \pi^2}\left[Li_2(-e^{-{\pi \,qE}\, \rho_{c2}^2})-
Li_2(-e^{-{\pi \,qE}\, \rho_{c1}^2})\right]\label{surfacedilog}\nonumber
\qquad .
\end{eqnarray}

  As emphasized by Robert Brout \cite{RB}, it is instructive to compare the
formula
for the persistence of Rindler vacuum, eq. (\ref{inoutR}), with a similar
situation which arises in the theory of Hawking black hole radiation.
In particular, our result stands in strong analogy to the analysis
of $s$-wave emission by a Schwarzschild black hole, and even more so
when this latter is reduced to a pseudo ($1+1$) dimensional problem,
when one neglects the effects of the $s$-wave `'centrifugal" barrier
(See ref. \cite{BD}). It will be noted that the rate eq. (\ref{inoutR})
defines is a
difference between a volume term, proportional to R, and a surface
term wherein R is replaced by a characteristic length $a/(qE\, {\bar R})$.
Each of these
terms is proportional to T, hence giving rise to a rate per unit
volume and rate per unit surface respectively. For the Schwinger
case, the object of our present study, R is macroscopic and the
surface term is negligible. But for the case of $s$-wave black hole
evaporation, the steady state production comes from a region of
${\cal O}(M)$ of the horizon where $2\,M$ is the Schwarzschild
radius. Recall the formula for the rate of particle emission
\begin{equation}
\frac{\langle T_{o}^{r}\rangle}{k_B\,T_{\mbox{\rm{\scriptsize Hawk.}}}}=
        \frac{\pi}{12}k_B T_{\mbox{\rm{\scriptsize Hawk.}}}\qquad
.\label{Hawkflux}
\end{equation}
There is a prefactor $M$ in the above, the volume of the
``skin"; this replaces $R$ in our formula.

Equation (\ref{Hawkflux}) contains the vacuum expectation value of the
energy flux
in Unruh vacuum, the outgoing mode piece of the Hartle-Hawking vacuum.
In ref. \cite{BD} there is displayed the same calculation in Boulware
vacuum, the analog of what we have called Rindler vacuum. It is shown
that in this case vacuum polarization effects cancel against the
emission effects and $\langle T_{o}^{r}\rangle_{\mbox{\rm \scriptsize
Boulware}}=0$. In our eq. (\ref{vacioR}), this subtraction comes up in
analogous fashion, due to the particular structure of the Rindler
modes, and hence of Rindler vacuum, in the vicinity of the horizons.
The difference between the two cases is that in the black hole case
 the cancellation of
$\langle T_{or}\rangle_{Boulware}$ must be total. This is a consequence
 of the static character of the
Schwarzschild metric which underlies the construction of Boulware vacuum.
 In the Schwinger case the problem
is actually intrinsically time dependent. Indeed the electric field has
to be (adiabatically)
 switched on and off in order to prepare the $in$-vacuum initial state (or
to observe the
$out$-one).
 \subsection{Quantization on the left quadrant (\tL)} Formally the field
equations and their solutions are the same on the left and right quadrants.
Their only difference results that on \TL the vector field
$\partial_\tau$ points  near the past. This implies that the signs of the
charges and the $in$ and $out$ labels have to be interchange with
respect to their values on the \TR quadrant. The analytical expression of
the
$in$- and
$out$-modes on the
\tL-quadrant can be deduced immediately from the drawing (see fig. ({\bf 3})) of
the supports of the modes built on the
\tR-quadrant. Adding a extra subscript ($R$, $L$, etc \dots) to make the
distinction between the modes defined on the various quadrants, we obtain
for the
$in$-mode the expression:
\begin{eqnarray}   {\cal U}_{in,L}^\omega(\xi_L, \tau_L) &=& \NUwi{L} \frac{e^{-i \omega\tau_L}}{\sqrt{2\pi}}  e^{- a\, \xi_L}
M_{i({\omega\over 2a} - {m^2 \over 2 qE}), i {\omega\over 2a}} \left[ - i
{q E \over 2a^2} e^{2a\xi_L} \right] \nonumber
\\ &\propto& {\cal
U}_{out,R}^\omega(\xi_L,
\tau_L)
\end{eqnarray} and for their charge content:
\begin{eqnarray}   Q({\cal U}_{in,L}^\omega) &=& Q_{{\cal I}_L^-} ({\cal
U}_{in,L}^\omega) = - 1 = - Q_{{\cal I}_R^+} ({\cal U}_{out,R}^\omega)
\nonumber \qquad,\\ Q_{{\cal H}_L^+} ({\cal U}_{in,L}^\omega) &=& - Q_{{\cal
H}_R^-} ({\cal U}_{out,R}^\omega)=-\sw q_2
\nonumber \qquad,\\ Q_{{\cal I}_L^+} ({\cal U}_{in,L}^\omega) &=& - Q_{{\cal
I}_L^-} ({\cal U}_{out,R}^\omega)=-q_1\qquad .
\end{eqnarray}  Similarly
\begin{eqnarray}   &{\cal V}_{in, L}^\omega(\xi_L, \tau_L) \propto {\cal
V}_{out,R}^\omega(\xi_L,
\tau_L) &\quad, \\
&{\cal U}_{out,L}^\omega(\xi_L,
\tau_L)\propto {\cal U}_{in, R}^\omega(\xi_L, \tau_L) &\quad  , \\  &{\cal V}_{out,L}^\omega(\xi_L, \tau_L) \propto {\cal V}_{in ,
R}^\omega(\xi_L,
\tau_L) &
.\nonumber
\end{eqnarray} Here again, we may define $in$- and $out$- vacua, Fock
spaces, etc \dots and the amplitude of vacuum persistence is given by an
expression similar to (\ref{vacioR}).
\subsection{Quantization on the past quadrant (\tP)} On the past quadrant,
it is the vector field $ - \partial_{\xi_p}$ that defines the future
direction. The various modes can again be expressed again in terms of
Whittaker's functions: with their charges  given by eqs
(\ref{QFP}). More precisely we may choose as particle incoming modes
\begin{eqnarray}   {\cal V}_{in,P}^\omega(\xi_P, \tau_P) = \NVwi{P}
\frac{e^{-i \omega
\tau_P}}{\sqrt{2\pi}}
 e^{- a\, \xi_P} W_{-i ({\omega\over 2a} - {m^2
\over 2 q E} )\; , i {\omega\over 2a}} \left[ - i {q E \over 2a^2} e^{2a\,
\xi_P} \right]\qquad.
\end{eqnarray} \noindent They came out from $ \ScrimR$, carrying a charge
\begin{eqnarray}
\QViImR{P}  = \vert \NVwi{P} \vert ^2 \left(\frac{qE}{a}\right ) e^{-{\pi
\over 2}({\omega\over a} - {m^2 \over q E})} = + 1
\end{eqnarray} and leave \TP via $ \HormR$ and $ \HormL$, taking away the
charges
\begin{eqnarray}
\QViHmR{P} = \pxe^{\pi\left(\frac\omega
a-\frac{m^2}{2qE}\right)}\;\frac{{\cosh\left[\pi\frac{m^2}{2qE}\right]}}
{{\sinh\left[\frac{\pi\omega}a\right]}}\equiv \sw q_3\qquad,
\end{eqnarray} \noindent and
\begin{eqnarray}
\QViHmL{P} = -\pxe^{-\frac{\pi
m^2}{2qE}}\;\frac{{\cosh\left[\pi\left(\frac{\omega}a-\frac{m^2}{2qE}\right)
\right]}}{{\sinh\left[\frac{\pi\omega}a\right]}}\equiv -\sw q_4
\qquad .
\end{eqnarray} The complex conjugates of the anti-particle incoming modes
are given by
\begin{eqnarray}
\cUwi{P}(\xi_P,\tau_P)&=& \NUwi{P} \frac{e^{- i
\omega\tau_P}}{\sqrt{2\pi}}
  e^{- a\, \xi_P} W_{i ({\omega\over 2a} - {m^2 \over 2 q E}), i
{\omega\over 2a}}
\left[ i {q E \over 2a^2} e^{2 a\, \xi_P} \right]\\
&\propto&\left[\cVwi{P}(\xi_P,-\tau_P)\right]^*\qquad.
\end{eqnarray} Their asymptotic charge content, when they appears on
$\ScrimL$, is:
\begin{eqnarray}
\QUiImL{P}=- 1 \qquad .
\end{eqnarray} This charge splits into
\begin{eqnarray}
\QUiHmR{P} &=& \sw q_4 \qquad,\nonumber \\
\QUiHmL{P} &=& -\sw q_3\qquad ,
\end{eqnarray} with $\sw (q_3-q_4)=1$. \\  Outgoing $\cal U$-modes on \TP
have to vanish on
$\HormL$ i.e. their coefficient $D_-$ must be zero. They are given by
Whittaker's M-functions:
\begin{eqnarray}   {\cal U}_{out,P}^\omega (\xi_P,\tau_P)&=& \NUwo{P}
\frac{e^{-i
\omega\tau_P}}{\sqrt{2\pi}}  \pxe^{-a\xi_P}
 M_{i({\omega\over 2a} - {m^2 \over 2q E} ),  i {\omega\over 2a}} \left[
i {q E
\over 2a^2} e^{2a\, \xi_P} \right]
\end{eqnarray} with
\begin{equation} \vert\NUwo{P}\vert^{-2}=\vert D_+(\cMp _-)\vert^{2}=\vert
{\omega}\vert\left(\frac {qE}{2a^2}\right)\pxe^{-\frac{\pi\omega}{2a}}
\end{equation}
 whose charge content is
\begin{eqnarray}  Q({\cal U}_{out,P}^\omega)  &\equiv& Q_{{\cal H}_R^-}
({\cal U}_{out,P}^\omega) =+\sw
\qquad ,\nonumber \\ Q_{{\cal I}_L^-}({\cal U}_{out,P}^\omega) &=& -
q_4\qquad <0
\qquad ,\nonumber \\ Q_{{\cal I}_R^-}({\cal U}_{out,P}^\omega) &=& \
q_3\qquad >0\qquad .
\end{eqnarray} Accordingly, modes with positive (resp. negative) values of
$\omega$ have to be associated to particles (resp. antiparticles) and, as
expected, the sum of the two last charges reproduces the total charge of
the mode. Finally, we have also to consider the modes
\begin{eqnarray}   {\cal V}_{out, p}^\omega (\xi_P,\tau_P) &=& \NVwo{P}
\frac{e^{-i
\omega\tau_P}}{\sqrt{2\pi}}  e^{- a\xi_P}
 M_{-i({\omega\over 2a} - {m^2 \over 2 q E}), -i {\omega\over 2a}} \left[-
i {q E
\over 2a^2} e^{2a\, \xi_P} \right]\\
&\propto&\left[\cUwi{P}(\xi_P,-\tau_P)\right]^*
\end{eqnarray} \noindent that emerges from $ \Scrim$ and leaves \TP via
$\HormR$, with the charges
\begin{eqnarray}   Q ({\cal V}_{out, P}^\omega)  &\equiv& Q_{{\cal H}_L^-}
({\cal V}_{out, P}^\omega) = -\sw \nonumber \qquad,\\  Q_{{\cal I}_L^-}({\cal
V}^{\omega}_{out,P}) &=& - q_3
\nonumber \qquad,\\  Q_{{\cal I}_R^-}({\cal V}^{\omega}_{out,P})&=& \ q_4\qquad .
\end{eqnarray} Here also the quantum field operator can be expressed in
two equivalent forms, using the $in$- and  $out$- modes:
\begin{eqnarray}
\hat\phi_P &=& \int_{- \infty} ^{+ \infty} d \omega\left[ a_{{\cal V}_P}
^{out} (\omega)
\cVwi{P}+ b_{{\cal U}_P}^{+\;in} (\omega) \cUwi{P}\right]\label{QFOPi}\\
 &=& \int_{- \infty} ^{+ \infty} d \omega\left[\theta (\omega)\left(
a_{{\cal U}_P} ^{out} (\omega)
\cUwo{P}+b_{{\cal V}_P}^{+\; out} (\omega) \cVwo{P}\right)
\right.\nonumber\\ &&\ \ \
\ +\left.\theta (-\omega)\left( a_{{\cal V}_P}^{out} (\omega) \cVwo{P}
 +  b_{{\cal U}_P}^{+\; out} (\omega)
\cUwo{P}\right)\right]\label{QFOPo}\qquad .
\end{eqnarray} Using the relations given in Appendix {\bf\ref{A}},  we
obtain easily the Bogoljubov transformation relating
$in$- and $out$-modes:
\begin{eqnarray}
\cUwo{P}&=&
\theta(\omega)\left\{\alpha^{P}
_{{\cal{U}}{\cal{V}}}(\omega)\cVwi{P}+\beta^{P}_{{\cal{U}}{\cal{U}}}(\omega)
\cUwi{P}
\right\}+\theta(-\omega)\left\{\gamma^{P}_{{\cal{U}}{\cal{V}}}(\omega)\cVwi{P}+
\epsilon^{P}_{{\cal{U}}{\cal{U}}}(\omega)
\cUwi{P}\right\}\qquad, \nonumber\\
\cVwo{P}&=&
\theta(-\omega)\left\{\alpha^{P}_{{\cal{V}}{\cal{V}}}(\omega)\cVwi{P}+\beta^
{P}_{{\cal{V}}{\cal{U}}}(\omega)
\cUwi{P}
\right\}+\theta(\omega)\left\{\gamma^{P}_{{\cal{V}}{\cal{V}}}(\omega)\cVwi{P}+
\epsilon^{P}_{{\cal{V}}{\cal{U}}}(\omega)
\cUwi{P}\right\}\qquad,\label{UVoiP}
\end{eqnarray} whose coefficients (see Appendix {\bf{B}}) satisfy
charge conservation relations:
\begin{eqnarray}
\vert\alpha^P_{{\cal{U}}{\cal{V}}}(\omega>0)\vert^2-\vert\beta^P_{{\cal{U}}{\cal
{U}}} (\omega>0)\vert^2&=&1\qquad ,
\end{eqnarray} and:
\begin{eqnarray}
\vert \gamma^P_{{\cal{U}}{\cal{V}}}(\omega<0)\vert^2-\vert
\epsilon^P_{{\cal{U}}{\cal{U}}}(\omega<0)\vert^2&=&-1\qquad .
\end{eqnarray}
From these equations we may repeat the steps leading to
the probability (\ref{vacioR}) of no pair creation  and obtain:
\begin{eqnarray}
\vert\strut^{\P}\hspace{-1mm}\braket{0,in}{0,out}^{\P}\vert^2&=&
\exp\left\{-\left[\sum_{\omega
>0}\ln\left(1+\vert\beta^P_{{\cal{U}}{\cal{U}}}\vert^2\right)+
\sum_{\omega
<0}\ln\left(1+\vert\gamma^P_{{\cal{U}}{\cal{V}}}\vert^2\right)\right]\right\}
\nonumber
\\ &=&\exp\left[-\sum_{\omega
>0}\ln\left(1+{\pxe^{-\pi\frac{m^2}{2qE}}}\frac{\cosh\pi(\frac \omega
a-\frac{m^2}{2qE})}{{\sinh\left[\frac{\pi\omega}a\right]}}\right)\nonumber
\right. \\ &&\left.
\left(1+\pxe^{-\pi\left(\frac{\omega}a+\frac{m^2}{2qE}\right)}\frac{{\cosh
\left[\pi\frac{m^2}{2qE}\right]}}{{\sinh\left[\frac{\pi\omega}a\right]}}\right
)\right]
 \qquad .\label{vacioP}
\end{eqnarray}
\subsection{Quantization on the future quadrant (\tF)} The modes on the
\TF quadrant are obtained from those defined on \TP in the same way we
pass from
\TR modes to \TL modes~:
\begin{eqnarray}
\cVwi{F}(\xi_F,\tau_F)
 &\propto&\cVwo{P}(\xi_F,\tau_F) \qquad , \nonumber\\
\cUwi{F}(\xi_F,\tau_F)&\propto&\cUwo{P}(\xi_F,\tau_F) \qquad , \nonumber\\
\cVwo{F}(\xi_F,\tau_F)
 &\propto&\cVwi{P}(\xi_F,\tau_F) \qquad , \nonumber\\
\cUwo{F}(\xi_F,\tau_F)
&\propto&\cUwi{P}(\xi_F,\tau_F)\qquad , \label{future}
\end{eqnarray} the proportionality factors reflecting the arbitrary
phases appearing in the normalization factors. Note also that, as
expected, these modes satisfy charge conjugation relations
\begin{eqnarray}
\cUwi{F}(\xi_F,\tau_F)\propto\left[\cUwi{F}(\xi_F,-\tau_F)\right]^*
\quad , \quad
\cUwo{F}(\xi_F,\tau_F)\propto\left[\cVwo{F}(\xi_F,-\tau_F)\right]^*
\qquad .
\end{eqnarray} Figure {\bf  3} summarizes all the
discussion of this section. We have schematically represented, for the
four Rindler quadrants, the behavior of typical wavepackets built with the
different modes discussed here above and the charge content that each of
them carries when it crosses the horizon or reaches infinity.

\section{Vacuum decay rates revisited} In this section we shall reobtain
the rates of vacuum decay by using functional methods. Our motivation is
that we hope in this way  to obtain an alternative picture of the physics
underlying the pair creation. In addition, as such methods often
necessitate the use of approximations, it is useful to use them as a
test in examples for which the exact answer is known since there
sometimes arise physical situations in which they are the only tool
available.  First we
shall briefly recall the general formalism and illustrate it in the
context of our scalar field interacting with a constant electric field, the
calculation being performed with respect to an inertial $t,z$ coordinate
system, i.e. with respect to the vacuum states defined by inertial
observers. Here, the interaction being quadratic, all the integrations are
gaussian and the calculations can be performed analytically. Then we
shall consider the same problem with respect to accelerated observers,
living in the \TR quadrant. We shall first perform our calculations using
the expression of the Feynman propagator as sum of modes and recover the
results obtained in the previous sections. Then we shall consider the
Schwinger propagator which expres itself as a  kernel integrated
with respect to a time variable (Schwinger proper time\footnote {Also called
fifth time in the framework of quantum field theory in 3+1 dimension.}).
Thanks to the  invariance with respect to
translations both in time $t$ and $\tau$  (boost) we shall be able to
reduce the expression of the Schwinger propagator to the evaluation of
path integrals of ordinary one-dimensional quantum mechanical problems.
We shall then evaluate these path integrals at one-loop approximation
(which is exact for quadratic potentials) and compare the results
 to those previously obtained. Finally we shall also compute the $in$-$out$
vacuum amplitude, using an expression for the Feynman propagator obtained
by unwinding the inertial propagator, and show  that it does  coincide
with the amplitude calculated in the secure framework of the mode
analysis.

The previous vacuum persistence amplitudes obtained from  ``mode"
calculations, may also be derived from the Green functions of the
quantized field. Indeed, as  is well known (see for instance the book
\cite{BD}), the vacuum persistence amplitude can be expressed as
 a functional integral~:
 \begin{eqnarray}  \langle 0,out\vert
0,in\rangle_{J=J^*=0}={\cal{Z}}\left[0,0\right]\equiv e^{i{\cal{W}}}
\end{eqnarray} where ${\cal{Z}}[J,J^*]$ is defined as~: \begin{eqnarray}
{\cal{Z}}\left[J,J^*\right]&=&\int{\cal{D}}\phi{\cal{D}}\phi^*\,
\exp\left\{iS\left[\phi,\phi^*\right]+i\int J^*\phi d^d x+i\int J \phi^*
d^dx \right\}
\end{eqnarray} in terms of the action $S\left[\phi,\phi^*\right]$ of the
charged scalar field, minimally coupled to the electric field:
 \begin{eqnarray}  S\left[\phi,\phi^*\right]&=&-\frac12\int d^dx
\left\{({\cal{D}}_\mu\phi) ({\cal{D}}^\mu\phi)^*+({\cal{D}}_\mu\phi)^*
({\cal{D}}^\mu\phi)+(m^2-i\epsilon)\phi
\phi^*+(m^2+i\epsilon)\phi^*\phi)\right\}\nonumber\\ &=&-\int dx
dx'\phi^*(x) {\cal{H}}_{xx'}\phi(x')\qquad.
 \end{eqnarray}
 (Here, and in the following, we denote collectively by $x$, $x'$, the
coordinate variables of the field; the context indicating if they are
Rindlerian or minkowskian.) \\
 A standard computation allows to give a sense to the expression of
$\cal{W}$~:
\begin{eqnarray}  {\cal{W}}=-i{\tr}\ln(-{\cal{G}}) =\int_{m^2}^\infty
dm^2 \int d^dx \, G_F(x,x)
\end{eqnarray}  where ${\cal{G}}=-{\cal{H}}^{-1}$ is related to the
Feynman propagator
\begin{equation} G_F(x,y)=-i {\langle 0,out\vert{\cal{T}}
(\phi(x)\phi^\dagger(y))\vert 0,in
\rangle}/{\langle 0,out\vert 0,in\rangle}
\end{equation}
 by:
\begin{eqnarray} {\cal{G}}(x,y)=\frac12 \left\{
G_F(x,y)+G_F^*(y,x)\right\}\qquad.
\end{eqnarray}
 Thus, the imaginary part (the one which encodes the vacuum instability)
of
 $\cal W$ reads:
\begin{eqnarray}  {\cal{I}}m{\cal{W}}={\cal{R}}e\int_{m^2}^\infty dm^2
\int d^dx \, {\cal{G}}(x,x)={\cal{R}}e\int_{m^2}^\infty dm^2 \int d^dx \,
G_F(x,x)\qquad.
\end{eqnarray}  Our purpose now is to evaluate this expression for an
inertial observer and an accelerated one, using different schemes of
calculation.
\subsection{Mode representation of the Feynman propagator}
\label{exactPersist} If the asymptotic vacua are those of an inertial
observer, using $(t,z)$ coordinates, it is
  straightforward to obtain the Feynman propagator \cite{PaBr} as
superposition of modes (eqs \ref{Dmodein} and followings in the text)~:
\begin{eqnarray}  iG_F(x,x^\prime) &=&\frac {\delta} {\gamma} \int d\sigma
\,\varphi_\sigma^{p\,out}(x)\varphi_\sigma^{a\,out}(x^\prime)\nonumber \\
&&+\theta(t-t')\int
\, d\sigma \varphi_\sigma^{p\, out}(x)\varphi_\sigma^{p\,out\,*}(x^\prime)
+\theta(t'-t)\int \, d\sigma \varphi_\sigma^{a\,
out\,*}(x)\varphi_\sigma^{a\,out}(x^\prime)\qquad. \end{eqnarray}  Therefore we
get~:
\begin{eqnarray}  {\cal{G}}(x,x^\prime)&=&-\frac i {2}\frac {\delta}
{\gamma}
\int d\sigma
\left(\varphi_\sigma^{p\,out}(x)\varphi_\sigma^{a\,out}(x^\prime)-
\varphi_\sigma^{p\,out\,*}(x^\prime)\varphi_\sigma^{a\,out\,*}(x)\right)
\nonumber
\\ &&-\frac{i\epsilon(t-t')}{2}\int \, d\sigma \left(\varphi_\sigma^{p\,
out}(x)\varphi_\sigma^{p\,out\,*}(x^\prime)- \varphi_\sigma^{a\,
out\,*}(x)\varphi_\sigma^{a\,out}(x^\prime)\right)\qquad.
\end{eqnarray}  The first term in this last expression is traceless, so
we obtain~:
\begin{equation}  {\cal{R}}e \, \tr\,{\cal{G}}={\cal{I}}m \frac {\delta}
{\gamma} \int d^2x \int d\sigma
\,\varphi_\sigma^{p\,out}(x)\varphi_\sigma^{a\,out}(x)\qquad.
\end{equation}

$\bullet$ Using the integral representation {\it \`a la Schwinger}
for the product of wave functions derived in the Appendix {\bf{C}},
we can rewrite this as:
\begin{equation} {\cal{R}}e \, \tr\,{\cal{G}}=-{\cal{I}}m
\frac{e^{-\frac{\pi m^2}{qE}}}{2\pi}\frac{\sqrt{qE}}{2i\pi}\int d^2x\int
d\sigma\int_0^{\infty} \frac{ds}{(\sinh 2qEs)^{\frac
12}}e^{-im^2s}e^{iqE(z+\frac
\sigma{qE})^2 \tanh qEs}
\end{equation} and the imaginary part of the $\cal W$ amplitude reads:
\begin{equation} {\cal{I}}m{\cal {W}}={\cal{I}}m \frac
1{2\pi}\frac{\sqrt{qE}}{2i\pi}\int d^2x\int d\sigma\int_0^{\infty}
\frac{ds}{(s-i\frac \pi{qE})}\frac{e^{-im^2s}}{(\sinh 2qEs)^{\frac
12}}e^{iqE(z+\frac
\sigma{qE})^2 \tanh qEs}\qquad.
\end{equation} After the $t$ and $z$ integrations we get, in accord with
eq.(\ref{Drate}):
\begin{eqnarray}  {\cal{I}}m{\cal {W}}&=&-\frac T {8\pi}\int d\sigma\int
_{-\infty-i\pi}^{\infty-i\pi}\frac {d\theta}{\theta \sinh
\theta}e^{-i\frac {m^2}{qE}\theta}=\frac
T{4\pi}\ln(1+e^{-\pi
\frac{m^2}{qE}})\qquad .
\end{eqnarray}

$\bullet$ In Rindler coordinates $(\tau,\rho)\equiv (\tau,a^{-1} e^{a\,\xi})$ on
\tR, the $in$-$out$
 Feynman propagator, expressed as a mode superposition, reads
as~:
\begin{eqnarray}
iG_F^R(x,x^\prime)&=&\theta(\tau-\tau^\prime)\left\{\int_0^\infty
d\omega\left\{{\cal{U}}_\omega^{out}(x){\cal{U}}_\omega^{out\,*}(x^\prime)+{
\cal{V}}_\omega^{out}(x){\cal{V}}_\omega^{out\,*}(x^\prime)\right\}+\nonumber
\right.\\ &&\left. \int_{-\infty}^0 d\omega
\frac{1}{\alpha_{{\cal{U}}{\cal{V}}}}{\cal{U}}_\omega^{out}(x)
{\cal{V}}_\omega^{in\,*}(x^\prime)\right\}\nonumber
\\ &+&\theta(\tau^\prime-\tau)\int_{-\infty}^0
d\omega\frac{1}{\alpha_{{\cal{U}}{\cal{V}}}}{\cal{U}}_\omega^{in}(x)
{\cal{V}}_\omega^{out\,*}(x^\prime)
\end{eqnarray} and the part which encodes the pairs production is given
by~:
\begin{eqnarray} {\cal{R}}e \, \tr\,{\cal{G}}(x,x)&=&{\cal{I}}m\int d^2x
\int_{-\infty}^0 d\omega \frac 1
{\alpha_{{\cal{U}}{\cal{V}}}}{\cal{U}}_\omega^{out}(x){\cal{V}}_\omega^{in\,
*}(x)\qquad.
\end{eqnarray} Using the integral representation (given in Appendix
{\bf{C}}) of the product of modes appearing in this last equation  we
get, once the space-time volume element is reexpressed as $d^2x=a\rho
d\rho d\tau$:
\begin{eqnarray}  {\cal{R}}e \, \tr\,{\cal{G}}(x,x)&=&{\cal{I}}m\int
d\tau \int_{-\infty}^0 d\omega\frac 1{4\pi }e^{-\frac{\pi
m^2}{2qE}}e^{\frac{\pi
\omega}{2a}}\int_0^\infty
\rho d\rho e^{-\epsilon \rho^2}\nonumber \\ &&\int_{-\infty}^{\infty}\frac
{du }{\cosh u}e^{i(\frac \omega a-\frac{m^2}{qE}) u}e^{i\frac{qE}2\rho^2
\tanh u}I_{i\frac \omega a}(\frac{qE\rho^2}{2\cosh u})\qquad.
\end{eqnarray}  This representation allows us to  perform the integration on $m^2$ and we get for the imaginary part of the functional $\cal
W$:
\begin{eqnarray}  {\cal{I}}m{\cal{W}}=-\frac{qE}{4\pi }{\cal{I}}m\int
d\tau\int_{-\infty}^0 d\omega\int_{-\infty-i\frac \pi 2}^{\infty -i\frac
\pi 2}\frac {d\theta}{\theta
\sinh\theta}e^{-i\frac{m^2}{qE}\theta}e^{i\frac \omega a
\theta}\int_0^\infty \rho d\rho e^{-\epsilon \rho^2}
e^{i\frac{qE}2\rho^2\coth \theta}e^{\frac{\pi
\omega}{2a}}J_{i\frac \omega a}(\frac{qE\rho^2}{2\sinh
\theta})\qquad.\nonumber\\&&
\end{eqnarray}
 where we have introduced the new variable  $\theta=u-i\frac \pi 2$ and
rewrite an Bessel-$I$ function as a $J$ function. The next step consists now to
carry out the
$\rho^2$ integration which, thanks to the formula (6.611.1) of
\cite{GR}, leads us to~:
\begin{eqnarray} {\cal{I}}m{\cal{W}}&=&{\cal{I}}m \,\frac{e^{-i{\frac \pi
2}}}{{4\pi }}\int d\tau\int_{-\infty}^0 d\omega\int_{-\infty-i\frac \pi
2}^{\infty-i\frac \pi 2}d\theta\frac {{
\sgn ({\cal{R}}e\theta)}}{\theta
\sinh\theta}e^{-i\frac{m^2}{qE}\theta}e^{i\frac \omega a
\theta}{\left\{\sgn ({\cal{R}}e\,\theta)\,\sinh \theta
+\cosh\theta\right\}^{i\frac \omega a}}\nonumber\quad .
\end{eqnarray} Finally, by splitting the $\theta$  integral according to
the sign of ${\cal{R}}e\,\theta$ we obtain:
\begin{eqnarray}  {\cal{I}}m{\cal{W}}&=&\frac 1{8\pi}\int
d\tau\int_{-\infty}^{\infty}\frac{d\theta}\theta\int_{-\infty}^0
d\omega\frac{e^{-i\frac{m^2}{qE}\theta}e^{2 i\frac \omega a \theta}}{\sinh
\theta}-\frac 1{8\pi}\int
d\tau\int_{-\infty}^{\infty}\frac{d\theta}\theta\int_{-\infty}^0
d\omega\frac{e^{-i\frac{m^2}{qE}\theta}}{\sinh \theta}\nonumber\\ &=&\frac
T{4\pi}\int_{-\infty}^0 d\omega\, \ln(1+e^{-\frac {\pi m^2}{qE}})-\frac
T{4\pi}\int_{-\infty}^0 d\omega\,
\ln(1+e^{-\frac {\pi m^2}{qE}}e^{2\pi \frac \omega a})\label{ImWRind}
\end{eqnarray} in accord with our result (\ref{inoutR}) established by
using the Bogoljubov transformation coefficient between $in$- and $out$-modes.
\subsection{Schwinger representation of the Feynman propagator}
For the sake
of completeness, let us briefly recall the so-called Schwinger
representation of the Feynman propagator. Formally,  the Feynman
propagator $G_F$ is the inverse of the kinetic operator $K=\frac
{\delta^d(x-y)}{\sqrt{-g(x)}}\left\{-{\cal D}_\mu {\cal D}^\mu
+(m^2-i\epsilon)\right\}$ and can thus be computed as $G_F=-K^{-1}=-i \int_0^{\infty}ds e^{-i K s }$.
To express it, Schwinger \cite{Schw} introduces  the
kernel  defined as
$K(x,y;s)=\langle x^\prime\vert e^{-i K s} \vert x\rangle
$ which  obeys a Schr\"odinger  equation~:
\begin{eqnarray}
({\cal D}_\mu {\cal D}^\mu
-m^2+i\frac{\partial}{\partial s})K(x,y;s)=0
\end{eqnarray} with the boundary condition $K(x,y;s\rightarrow
0^+)=\delta^d(x-y)$.
 Using this proper time representation of the Feynman propagator
 \begin{equation}
 G_F(x,x')=\int_0^\infty ds\,K(x,x';s)\qquad ,\label{Schwprop}
 \end{equation} the vacuum persistence amplitude can be written as:
\begin{equation}  {\cal{I}}m{\cal{W}}={\cal{I}}m\int_{m^2}^\infty dm^2
\int d^dx
\int_0^\infty ds\, K(x,x;s) \qquad .\label{WKint}
\end{equation} The kernel $K(x,x^\prime;s)$, defined as the matrix elements
of the evolution operator
$K$, can be expressed as a path integral:
\begin{equation}  K(x,x^\prime;s)=\int
{\cal{D}}X^{\mu}(s^\prime)e^{iS(x,x^\prime;X^\mu(s^\prime);s)}\label{Kkernel}
\end{equation} where the domain of integration covers all the paths that
connect the point $x$ to the point
$x^\prime$ in a (rescaled) time\footnote{This time variable is
proportional to the proper time measured along the natural trajectory
that connects $x$ to $x'$, but, in general, does not coincide with it.}\  $s$,
and $S(x,x^\prime;x^\mu(s^\prime);s)$ is the classical action along these
trajectories, given by: \begin{eqnarray}
S(x,x^\prime;X^\mu(s^\prime);s)=\int_0^s ds\left\{\frac{\dot X^2}4+q\dot
X^\mu A_\mu(X[s]) -m^2\right\}\qquad.
\end{eqnarray}
Moreover, let us emphasized
that if it is often supposed that the Feynman propagator obtained from modes coincides with the
Schwinger representation (\ref{Schwprop}) displayed here above, it is only
in a few cases that their equivalence has been proved (See for example
\cite{ThP,KhLa}).
\subsubsection*{Reduction of functional integrals}  The Schwinger
propagator introduced above via a functional integral is a formal
object that has to be defined more accurately. To this end we remind the
Feynman prescription \cite{FeHi}. We divide the time interval $s$ into
$N$ subintervals of length $\epsilon$,  such that $s=N\epsilon$, and
introduce
 $N$ times the closure relation in the  expression of
$K(x,x^\prime;s)$ taken as the matrix element:
\begin{eqnarray} K(x,x^\prime;s)&=&\langle t^\prime,z^\prime\vert
e^{-i\hat Hs}\vert t,z\rangle\nonumber\\  &=&\int \left(\prod_{i=1}^N
dz_i dt_i \right) \langle x^\prime\vert e^{-iH\epsilon}\vert x_N\rangle
\dots\langle x_1\vert e^{-i\hat H\epsilon}\vert x\rangle\nonumber\\
&=&\int \left(\prod_{i=1}^N dz_i
dt_i\right)\prod_{j=0}^N\left\{\delta(t_{j+1}-t_j)\delta(z_{j+1}-z_j)-i
\epsilon \langle x_{j+1}\vert \hat H\vert x_j\rangle
\right\},\label{Kmatrep}  \end{eqnarray} before taking the limit
$N\mapsto \infty$ at the end of the calculation.

Let us first consider the problem for an inertial observer. Here the
hamiltonian operator reads: \begin{equation}
\hat H= \left\{(\partial_t-iqA_t)^2-\partial_z^2+m^2\right\}=\left\{-(\hat
p_t-qA_t)^2+\hat p_z^2+m^2\right\}.
\end{equation} We compute its matrix elements in position representation,
with $\vert x
\rangle=\vert z\rangle \otimes \vert t \rangle$. Using the orthonormalized
eigenvectors of the momentum operators $\hat p_z\vert k\rangle=k\vert
k\rangle$ and
$\hat p_t\vert \omega \rangle=-\omega \vert \omega \rangle$ which connect
to the position eigenvector by $\langle k\vert z\rangle=\frac{e^{-i k
z}}{\sqrt{2\pi}}$ and
$\langle t\vert \omega\rangle=\frac{e^{-i\omega t}}{\sqrt{2\pi}}$. Thus,
\begin{eqnarray}
\langle x_{j+1}\vert \hat H\vert x_j\rangle&=&-\delta(z_{j+1}-z_j)\int
d\omega \langle t_{j+1}\vert
\omega\rangle\langle\omega\vert(\hat p_t-qA_t(z_j))^2\vert
t_j\rangle\nonumber\\ &+&\delta(t_{j+1}-t_j)\int dk\langle z_{j+1}\vert
k\rangle\langle k \vert
\hat p_z^2
\vert z_j\rangle+\delta(z_{j+1}-z_j)\delta(t_{j+1}-t_j)m^2\nonumber \\
&=&-\delta(z_{j+1}-z_j)\int\frac{ d\omega}{2\pi}
(\omega+qA_t(z_j))^2e^{-i\omega(t_{j+1}-t_j)}\nonumber\\
&+&\delta(t_{j+1}-t_j)\int
\frac{dk}{2\pi}k^2e^{ik(z_{j+1}-z_j)}+\delta(z_{j+1}-z_j)\delta(t_{j+1}-t_j)
m^2\nonumber\\ &=&\int dk \, d\omega
\frac{e^{ik(z_{j+1}-z_j)}}{2\pi}\frac{e^{-i\omega(t_{j+1}-t_j)}}{2\pi}H_{cl}
(k,\omega,z_j)
\end{eqnarray} where we have introduced the classical Routh function
$H_{cl}(k,\omega,z_j)=-(\omega+qA_t(z_j))^2+k^2+m^2$.\\ Inserting this in
eq. (\ref{Kmatrep}), the kernel $K(x,x^\prime; s)$ reads as:
\begin{eqnarray} K(x,x^\prime; s)&=&\int (\prod_{j=1}^N dz_j
dt_j)\prod_{j=0}^N\int dk_j
\,d\omega_j\frac{e^{ik_j(z_{j+1}-z_j)}}{2\pi}\frac{e^{-i\omega_j(t_{j+1}-t_j
)}}{2\pi}e^{-i\epsilon H_{cl}(k_j,\omega_j,z_j)}\nonumber\\
 &=&\int (\prod_{j=0}^N\frac{dk_j}{2\pi})\int
\frac{d\omega_0}{2\pi}e^{-i\omega_0(t^\prime-t)}\int\prod_{j=1}^N dz_j
e^{ik_j
\dot z_j \epsilon}e^{-i\epsilon H_{cl}(k_j,\omega_0,z_j)}\nonumber\\
&=&\int\prod_{i=1}^N dz_i\int
\frac{d\omega_0}{2\pi}e^{-i\omega_0(t^\prime-t)}(4\pi
i\epsilon)^{-\frac{N+1}2}e^{i\epsilon {\dot
z_j^2}/4}e^{i\epsilon(\omega+qEz_j)^2}e^{-i\epsilon m^2}\nonumber\\
&=&\int
\frac{d\omega}{2\pi}e^{-i\omega(t^\prime-t)}K_\omega(z,z';s)\label{FeynmanMink}
\end{eqnarray}
where the last equality defines the Fourier transform of
the kernel
$K(x,x^\prime; s)$ as the path integral
\begin{equation} K_\omega(z,z';s)=\int {\cal{D}}z(s) e^{i\int_0^s
{\cal{L}}_\omega(z,\dot z)ds^\prime},\label{funcintMin}
\end{equation}
with respect to  the (classical) Lagrangian
${\cal{L}}_\omega(z,\dot z)=\dot z^2/4+(\omega+qEz)^2-m^2$.

A similar calculation can be done in Rindler coordinates ($\tau,\,
\rho=a^{-1}\exp(a\xi)$),
 where we use the
$\tau$ independence of the hamiltonian to perform the reduction of the
path integral to a one dimension quantum mechanical problem. However, due
to the explicit coordinate dependence of the metric, the calculation do
not reduce just to a strict rewriting of the previous one. The Schwinger
kernel is now given by~:
\begin{eqnarray}  K(x,x^\prime;s)&=&\langle\rho^\prime,\tau^\prime\vert
e^{-i\hat Hs}\vert
\rho,\tau\rangle\nonumber\\  &=&\frac
1{\sqrt{a^2 \rho\rho^\prime}}\int\left(\prod_{i=1}^N d\rho_i
d\tau_i\right) \prod_{j=0}^N
\left\{\delta(\rho_{j+1}-\rho_{j})\delta(\tau_{j+1}-\tau_{j})-i\epsilon\langle
x_{j+1}\vert (a\hat \rho)^{\frac 12} \hat H (a\hat \rho)^{\frac 12}\vert
x_{j}\rangle\right\}\nonumber\qquad.\label{Krind}
\end{eqnarray} The appearance of extra $\rho^{\frac 12}$ factors results
from the coordinate dependence of the volume element. Here the
hamiltonian reads as~:
\begin{equation}
\hat H=\hat p_\rho^2-(\hat p_\tau-\frac{aqE}2 \rho^2)^2 \frac 1
{a^2\rho^2}-\frac 1{4\rho^2}+m^2 \label{RindHam}
\end{equation} where the momentum operators are $\hat p_\tau=\frac 1 i
\partial_\tau$ and $\hat p_\rho=\frac 1 i (\partial_\rho
+\frac1{2\rho})$. Compared to the classical hamiltonian, an extra quantum term $1/4\rho^2$ appears, just as for the plane rotator \cite{KhLa}. This is dictated both by the
requirement of hermiticity with respect to the measure
$\rho\,d\rho$ and the classical commutation relations (see refs
\cite{PaDW}).  The orthonormalized eigenstates of these operators and the
connectors between them and the position eigenvectors
$\left(\vert x\rangle =\vert \rho \rangle \otimes \vert \tau \rangle
\right.$, with $\left.\langle
\rho '\vert \rho \rangle = (a \rho)^{-1} \delta (\rho'-\rho) \right)$ are:
\begin{eqnarray} \hat p_\rho \vert k\rangle=k\vert k\rangle&\qquad,\qquad
&\hat p_\tau \vert \omega \rangle=-\omega \vert\omega\rangle
\qquad,\nonumber\\
\qquad\langle
k\vert\rho\rangle=(a\rho)^{-\frac12}\frac{e^{-ik\rho}}{\sqrt{2\pi}}
&\qquad,\qquad&\langle\tau\vert
\omega\rangle=\frac{e^{-i\omega\tau}}{\sqrt{2\pi}}\qquad . \end{eqnarray}
So we obtain the matrix elements:
\begin{eqnarray}
\langle x_{j+1}\vert (a\rho)^{\frac 12} \hat H (a\rho)^{\frac 12}\vert
x_{j}\rangle&=&(a^2\rho_j\rho_{j+1})^{\frac 12}\langle x_{j+1}\vert \hat
H \vert x_{j}\rangle\nonumber\\ &=&-\frac
1{a^2\rho_j^2}\delta(\rho_{j+1}-\rho_j)\langle
\tau_{j+1}\vert (\hat p_\tau-qE A_\tau(\xi))^2\vert
\tau_j\rangle\nonumber\\ &+&\delta(\tau_{j+1}-\tau_j)\int dk
\frac{e^{ik(\rho_{j+1}-\rho_j)}}{2\pi}k^2+(m^2-\frac
1{4\rho^2_j})\delta(\tau_{j+1}-\tau_j)\delta(\rho_{j+1}-\rho_j)
\nonumber \\ &=&\int dk d\omega \frac {e^{i
k(\rho_{j+1}-\rho_j)}}{2\pi}\frac{e^{-i\omega(\tau_{j+1}-\tau_j)}}{2\pi}H_{c
l}(k,\omega,\rho_j)
\label{Hmatrind}
\end{eqnarray}  where we have set $H_{cl}(k,\omega,\rho)=\left\{-\frac
1{a^2\rho^2}(\omega+\frac{aqE\rho^2}2)^2+k^2-\frac 1{4\rho^2}\right\}+m^2$.
Inserting, the matrix elements (\ref{Hmatrind}) into  the expression of
the kernel (\ref{Krind}) we obtain~:
\begin{eqnarray}  K(x,x^\prime;s)&=&(a^2\rho \rho^\prime)^{-\frac
12}\int\prod_{i=1}^N d\rho_i d\tau_i \int \prod_{j=0}^N dk_j
d\omega_j\frac {e^{i
k_j(\rho_{j+1}-\rho_j)}}{2\pi}\frac{e^{-i\omega_j(\tau_{j+1}-\tau_j)}}{2\pi}
e^{-i\epsilon H_{cl}(k_j,\omega_j,\rho_j)}\nonumber \\ &=&(a^2\rho \rho^\prime)^{-\frac
12}\int\prod_{i=1}^N d\rho_i\int
\frac{d\omega_0}{2\pi} e^{-i\omega_0(\tau^\prime-\tau)}\prod_{j=0}^N
\frac{dk_j}{2\pi}e^{ik_j\dot
\rho_j\epsilon}e^{-i\epsilon H_{cl}(k_j,\omega_0,\rho_j)}\qquad.
\end{eqnarray}  As in the calculation done in minkowskian coordinates,
the independence of the hamiltonian with respect to the time variable
makes its conjugate momentum conserved. Each $\tau_j$ integration
generates a delta function of $\omega_j-\omega_{j-1}$, making all
$\omega_j$ integration trivial but the first one on $\omega_0$. Moreover
the integrals on the $k_j$ are gaussian, and we easily obtain:
\begin{eqnarray}  K(x,x^\prime;s)&=&\frac1{(a^2\rho
\rho^\prime)^{\frac 12}}(4\pi i
\epsilon)^{-\frac{N+1}2}\int\frac{d\omega_0}{2\pi}
e^{-i\omega_0(\tau^\prime-\tau)} \int\prod_{i=1}^N d\rho_i
\prod_{j=0}^Ne^{i\epsilon \dot
\rho_j^2 /4}e^{i\epsilon \frac
1{\rho_j^2}(\omega_{0}+qEA_\tau(\rho_j))^2}e^{-i(m^2-\frac
1{4\rho_j^2})\epsilon}\nonumber \\  &\equiv&\frac1{(a^2\rho\rho^\prime)^{\frac
12}}\int\frac{d\omega}{2\pi}
e^{-i\omega(\tau^\prime-\tau)}K_\omega(\rho,\rho';s)\label{RinSchwinker}
\end{eqnarray} where, as for the inertial propagator, we have introduced
a Fourier transform with respect to the Rindler $\tau$ time as the path
integral~:
\begin{equation} K^{Rin}_\omega(\rho,\rho';s)=\int {\cal{D}}\rho \exp
i\int_0^s{\cal{L}}_\omega(\rho,\dot
\rho)ds^\prime\qquad ;\label{funcintRin}
\end{equation}  the lagrangian being now given by~:
\begin{equation} {\cal{L}}_\omega(\rho,\dot \rho)=\dot \rho^2/4+\frac
1{a^2\rho^2}(\omega+q\frac E2 a\rho^2)^2+\frac 1{4\rho^2}-m^2\qquad
.\label{lagRin}
\end{equation}

To pursue our analysis we have to evaluate the two (unidimensional)
functional integrals
(\ref{funcintMin}) and (\ref{funcintRin}). The first one can be evaluated
explicitly, but we did not succeed  in expressing the second in a close
unambiguous form. To go ahead, we shall  evaluate them at one-loop
approximation, which actually is exact for gaussian integrals and thus
gives the correct answer for the  functional integral referring to
minkowskian coordinates (\ref{funcintMin}). In this approximation we just
have (in principle) to compute the classical action $S_{cl}$, as the well
known Pauli--Van Vleck formula \cite{PVV,CDeW} gives~:
\begin{eqnarray}  &K({\bf q},t;{\bf q^\prime},t^\prime)=\int {\cal{D}}{\bf
q}(t)\exp iS({\bf q}(t))=\left(\frac 1{2\pi i} \det
\frac{\partial^2 S_{cl}}{\partial {\bf q^\prime}\partial {\bf
q}}\right)^{\frac 12}\exp i S_{cl}\qquad.&
\end{eqnarray}  In principle this formula necessitates the evaluation of
the hessian matrix $\partial^2_{q\,q'}S_{cl}$. At first sight this
implies the knowledge of the general two point expression of the
action. But actually it suffices to know the general classical solution
of the equation of motion to obtain it. Indeed it can be shown (see for
instance \cite{CDD}) that the hessian matrix is the inverse of the matrix
built on Jacobi fields vanishing at point ${\bf q}$. These are simply
obtained by varying the general solution of the equation of motion with
respect to the initial velocity components\footnote{The reader will
easily recognize in this method the n-dimensional generalization of the
method advocated by S. Coleman to compute functional determinant
\protect\cite{Col}.}\cite{CDeW,Spi}. \\ Let us apply this method to
compute the  kernels (\ref{funcintMin} and \ref{funcintRin}) relevant to the evaluation  of trace of the Green
function and rate of vacuum decays.
Here we have only to take into account closed paths
in the functional
integrals, starting and ending from the same value of the $z$
or the $\rho$ coordinate.\\
\noindent In $z$-coordinate the calculation is straightforward. The
classical motion equation is: \begin{eqnarray}
\frac{d^2z}{ds'^2}-4 q^2E^2(z+\frac\omega{qE})=0
\end{eqnarray} whose particular solution such that  $z(s'=\mp\frac s
2)=z$ is given by~:
\begin{eqnarray}
\bar z(s')=-\frac \omega {qE}+(z+\frac \omega {qE})\frac{\cosh
2qEs'}{\cosh qEs }\qquad.
\end{eqnarray} The value of the classical action computed along this
closed trajectory reads:
\begin{eqnarray}  S_{cl}&=&\int_{-\frac s 2}^{\frac s
2}ds'\,{\cal{L}}_\omega(z,\dot z)\vert _{z=\bar z(s')}\nonumber\\
&=&qE(z+\frac \omega {qE})^2\frac{\cosh 2qEs -1}{\sinh 2 qEs}-m^2 s\qquad
.
\end{eqnarray} Here the computation of Jacobi fields is immediate,
because for quadratic potential it only consists in solving a harmonic
oscillator problem~:
\begin{equation}
\left\{\frac{d^2}{d^2s'}-4q^2 E^2\right\}h(s')=0\qquad.
\end{equation} whose solution obeying the Cauchy conditions $h(s'=-\frac
s 2)=0$ and $\frac \partial {\partial s'}h(s')\vert_{s'=-\frac s 2}=2$ (because the unusual normalization $1/4$ of the kinetic part of the
lagrangian) is:
\begin{eqnarray} h(s')=\frac1 {qE}\sinh 2qE(s'+\frac s 2)\qquad .
\end{eqnarray} Thus we get:
\begin{eqnarray} K_\omega(z,z;s)=(\frac {qE}{2\pi i\sinh
2qEs})^{\frac12}e^{iqE(z+\frac \omega {qE})^2\frac{\cosh 2qEs -1}{\sinh 2
qEs}-im^2s}\qquad.\label{intquad}
\end{eqnarray} Inserting this expression in eq. (\ref{WKint}), the
vacuum amplitude persistence, after $t$ and
$m^2$ integration $(T=\int dt)$, amount to~:
\begin{eqnarray} {\cal {I}}m \,{\cal{W}}&=&{\cal{I}}m \frac 1 i(\frac
{qE}{2i\pi})^{\frac 12}\frac T{2\pi}\int {d\omega}\int_0^\infty
\frac{ds}s e^{-im^2s}\int \frac{dz}{(\sinh 2qEs)^{\frac
12}}e^{iqE\frac{\sinh qEs }{\cosh  qEs}(z+\frac \omega {qE})^2}\nonumber
\\ &=&\frac{T}{4\pi}\int
{d\omega}\ln(1+e^{-\pi \frac {m^2}{qE}})\qquad.\label{tauxmink}
\end{eqnarray} This one loop computation gives the exact result, as
expected, because here the lagrangian is quadratic and the one-loop
approximation (\ref{intquad}) is actually exact.\\ Let us turn now to the
same problem formulated in Rindler coordinates, for the accelerated
observer. We  have to evaluate the functional integral (\ref{funcintRin})
on closed paths starting from and ending on $\rho(s'=-\frac s
2)=\rho(s'=\frac s 2)=\rho$. The equation of motion, derived from the
lagrangian (\ref{lagRin}) is:
\begin{equation}
\frac 12 \frac {d^2\rho}{ds'^2} +
\frac{2}{\rho^2}(\frac{\omega^2}{a^2}+\frac14)-\frac 12 q^2E^2\rho=0
\label{equrho}
\end{equation} which, by use of the energy theorem, reduces to
\begin{eqnarray}
\frac 1 {q^2E^2}\left(\frac {d\rho}{ds'}\right)^2=2\beta +(\rho^2+\frac
{{\tilde\Omega}^2}{\rho^2 })\label{drho}
\end{eqnarray} where $\beta$ is an integration constant and
\begin{equation}
\tilde\Omega=\frac{2}{qE}\sqrt{\frac{\omega^2}{a^2}+\frac {1}{4}}>0\qquad ,
\end{equation} the correspondence between this quantum number and the
classical constant of motion being
$\tilde\Omega \approx (m/qE)^2
\Omega$. When $\beta>0$, obviously there is no turning points in $\rho$;
the motion goes from
$\rho=0$ to $\rho=\infty$, and the trajectory never crosses twice the same
point. Actually, to get a trajectory with a bounce on the potential
barrier, $\beta$ must be less than $-\tilde\Omega$. Hereafter we
restrict ourselves to this class of trajectories. Integrating equation
(\ref{drho}), we obtain as solution of equation (\ref{equrho})~:
\begin{eqnarray}
\rho^2(s')+\beta=\varepsilon (\rho_0^2+\beta)\frac {\cosh\left\{
2qE(s'-s_0)+\varphi\right\}}{\cosh
\varphi}\label{genrho} \end{eqnarray} in terms of arbitrary constants
$\rho_0$, $\beta$ and $s_0$ and where $\varepsilon=\pm1$ and we have set,
for further notational convenience, $\cosh
\varphi=(\rho_0^2+\beta)/\sqrt{\beta^2-\tilde\Omega^2}$. The solution
which bounces in time $s$, such that $\rho(s'=-\frac s 2)=\rho(s'=\frac s
2)=\rho$ is:
\begin{equation}
\bar\rho^2(s')+\bar \beta=(\rho^2+\bar \beta)\frac{\cosh 2qEs'}{\cosh \bar
\varphi}
\end{equation} where the value of $\beta$ is now fixed to $\bar
\beta=\left\{ \rho^2-\cosh (qEs)\sqrt{\rho^4+\tilde\Omega^2 \sinh^2
(qEs)}\right\}/\sinh^2 (qEs)$, which implies that $\bar \varphi=-(qEs)$.
Obviously $\bar \beta$ is negative and
$\rho^2+\bar \beta$ will be negative or positive according to $\rho^2$ is
less or greater than  $\tilde\Omega$ i.e. according to the position of
$\rho$  with respect to  ($\sqrt{\tilde \Omega}$), where the minimum of the
potential
$\rho^2+\tilde\Omega^2/\rho^2$ occurs. The computation of the value of
the classical action along this trajectory is straightforward. We
obtain, in terms of the variable
$\psi$, defined through the relation
$\sinh\psi =\tilde\Omega\sinh qEs/\rho^2$:
\begin{equation}  S^{\,\prime}_\omega(\rho,s)=\frac{
\sqrt{\omega^2+a^2}}a\left(\frac {\cosh qEs-\cosh\psi
}{\sinh\psi}+\psi\right)\qquad ,
\end{equation} on which one it is easy to check that:
\begin{equation}  \frac{\partial S^{\prime}_\omega (\rho,s)}{\partial
s}=-\frac{q^2E^2\beta}2\qquad. \end{equation} It remains now to compute the
Jacobi field
$h(s')$ obeying the Cauchy conditions $h(s'=-\frac s 2)=0$ and
$\partial_{s'} h(s')\vert_{s'=-\frac s 2}=2$. Varying the general solution
(\ref{genrho})  of the equation of motion with respect to the ``energy"
parameter $\beta$ we get a Jacobi field vanishing for
$s'=-\frac s 2$. Once normalized such that $\left.\partial_s' \tilde h(s')\right|_{s'=-\frac s 2}
=2$, its value for
 $s'= s/2$ becomes:
 \begin{equation}
  h(\frac s 2)=\frac 2{qE}{\sinh qEs}\cosh\psi
\end{equation}
and gives the value of the Van Vleck determinant. Collecting all these results, we obtain
for the path integral (\ref{funcintRin}) with
$\rho=\rho'$, the approximated expression~:
\begin{equation}   K_\omega^{Rin}(\rho,\rho;s)\approx
\frac{
      e^{-im^2s+i\frac {qE \omega} a
s+i\frac{\sqrt{\omega^2+a^2}}{a}
\left(\frac {\cosh qEs-\cosh\psi}{\sinh\psi}+\psi\right) }}
{\sqrt{\frac{4i\pi}{qE}{\sinh qEs}\cosh\psi}}\qquad ,
\end{equation}  and for the vacuum persistence amplitude~:
\begin{eqnarray}
 {\cal {I}}m \,{\cal{W}}&=&{\cal{I}}m\frac T{2\pi i}\int {d\omega}
\int_0^\infty \frac{ds}s\,e^{-im^2s+i\frac {qE \omega} a s}
\sqrt{\frac{qE\tilde\Omega}{16i\pi}}\int_0^\infty d\psi
\frac{(\cosh\psi)^{\frac 12}}{(\sinh\psi)^{\frac 32}}e^{i\frac{\sqrt
{\omega^2+a^2}}a\left(\frac
{\cosh\theta-\cosh\psi}{\sinh\psi}+\psi\right)}\qquad.\nonumber\\
\end{eqnarray}  Let us evaluate the $\psi$ integral by the saddle point
method. Indeed, given the approximations used to evaluate
$K^{Rin}_\omega(\rho,\rho;s)$,
it would be inconsistent to endeavor to go beyond this approximation.
The phase reaches its maximum (with respect
to $\psi$) at
$\psi=qEs$, corresponding to a saddle point of  width $(\sqrt
{\omega^2+a^2}/a)\coth qEs$. This leads to the gaussian approximation~:
\begin{equation}
\int_0^\infty d\psi
\frac{(\cosh\psi)^{\frac 12}}{(\sinh\psi)^{\frac 32}}e^{i\frac{\sqrt
{\omega^2+a^2}}a\left(\frac
{\cosh\theta-\cosh\psi}{\sinh\psi}+\psi\right)}\approx
\sqrt{\frac{2i\pi a}{\omega^2+a^2}}
\frac{e^{i\frac{\sqrt{
\omega^2+a^2}}aqEs}}{\sinh qEs}
\end{equation}
 and the vacuum persistence amplitude reads:
\begin{eqnarray}   {\cal {I}}m \,{\cal{W}}&\approx&{\cal{I}}m\frac T{4\pi
i}\int {d\omega}
\int_0^\infty
\frac{ds}{s\sinh qEs}\,e^{-im^2s}e^{i\frac { \omega} a
qEs}e^{i\frac{\sqrt{\omega^2+a^2}} a qEs}\label{tauxrind}\\
&\approx&{\cal{I}}m\frac T{4\pi i}\int {d\omega}
\int_0^{\infty}
\frac{ds}{s\sinh
 qEs}\,\left(\theta(\omega)e^{-im^2s+i2\frac\omega a qEs}
 +\theta(-\omega)e^{-im^2s}\right)\qquad .\label{ImWRindapprox}
\end{eqnarray}
Comparing eq. (\ref{ImWRindapprox}) with eq. (\ref{ImWRind})
 we see that their difference reduces to
\begin{equation}
{\cal R}e\int_{-\infty}^{+\infty} d\omega\int_{-\infty}^{+\infty}
\frac{e^{i(2\omega\,\theta-\mu^2\theta)}}{\theta\,\sinh
\theta}\,d\theta=-\frac{1}{2\,\epsilon^2}
\end{equation}
once we take into account the $i\,\epsilon$ prescription.
This divergent quantity is physically meaningless. It is
independent of the physical parameters of the problem and must
be subtracted in order to recover the stability of the
Rindler vacuum in the absence of the electric field.
  Of course all the
finite corrections we have mentioned before, and that we recover here above,
are mathematically meaningless; the
approximations we used for the Schwinger kernel making them all irrelevant.
Our motivation for having
discussed them is just to show that no spurious term are introduced in
the dominant contribution to ${\cal {I}}m
\,{\cal{W}}$. Actually this leading order can be obtained directly by
ignoring all subtleties but just using a quadratic approximation of the
potential~:
\begin{eqnarray} V(\rho)&=&\frac1{\rho^2}(\frac\omega
a+\frac{qE}2\rho^2)^2+\frac1{4\rho^2}-m^2 \nonumber\\
&\approx&(\frac{2\omega qE}a-m^2)+q^2E^2(\rho-\rho_0)^2\label{Vquad}
\end{eqnarray} around the minimum
$\rho_0^2=\tilde\Omega\approx\frac{2}{aqE}\omega$.
 The computation is similar to the one discussed for the inertial observer
once we have substituted $m^2-\frac {\tilde
\omega qE}a-\frac{\omega qE}a$ to $m^2$. As for the evaluation of
(\ref{intquad}) we obtain now~:
\begin{equation} K^{Rin}_\omega (\rho,\rho;s)\approx \sqrt{\frac {qE}{2i\pi
\sinh 2qEs}}e^{i(\frac {\sqrt{\omega^2+a^2}+\omega qE}a
-m^2)s}e^{iqE(\rho-\rho_0)^2\frac{\cosh 2qEs -1}{\sinh 2 qEs}}\label{Vquadbis}
\end{equation}  whose trace gives expression (\ref{tauxrind}), analog to
(\ref{tauxmink}).
\subsubsection{Is the zero winding propagator the Rindler propagator?} In
the appendix of ref. \cite{BPS}, the Fourier transform of the inertial
Feynman propagator was expressed as a sum of winding
terms, similar to those given in ref. \cite{ToVD}. Each term of this sum
corresponds the Fourier transform of
 a path integration restricted to the subclass of paths
joining their ends after having performed a fixed number of
 winding around the common vertex of the four Rindler
quadrants (around the origin $O$, see fig. ({\bf 1})). It was conjectured that the kernel $\tilde
K^{w=0}_\omega$ obtained from the zero
 winding sector, provides the kernel leading to the
Rindler propagator~: $\tilde K^{w=0}_\omega=K^{Rin}_\omega$. Hereafter we
reinforce this conjecture
by showing that the rate calculated from this zero winding kernel~:
\begin{equation}
\tilde K_\omega^{w=0}=-\frac 1{4\pi}\frac{e^{-i(m^2-\frac
\omega a)Es}}{\sinh qEs}e^{i\frac{qE}4(\rho^2+\rho^{\prime\,2})\coth
qEs}I_{-i\frac{\vert\omega \vert}a}
\left(-i\frac{qE\rho\rho^\prime}{2\sinh qEs}\right)\label{Kzw}
\end{equation}
leads to our previous results. To this end we substitute
in eq. (\ref{WKint}) the expression of $K^{Rin}(x,x,s)$ obtained by using
the expression (\ref{Kzw}) in eq. (\ref{RinSchwinker}). Reexpressing
the Bessel-$I$ function as a sum of Bessel-$K$ functions and integrating
over the $\rho$ variable, we obtain~:
\begin{equation}
\begin{array}{lll}
&{\cal R}e \frac{\cal
T}{2\,\pi}\int_0^{\infty}\rho\,d\rho\int_{-\infty}^{\infty}d\omega
\int_0^{\infty}\tilde K^{w=0}_\omega(\rho,\rho;s)\frac{ds}{s}&\\
&\ \ ={\cal R}e \frac{\cal
T}{2\,\pi}\int_{-\infty}^{\infty}d\omega
\int_0^{\infty} e^{-i(m^2s+\frac{\omega}{a}qe\,s)}\frac{ds}{s}
\int_0^{\infty}\rho\,d\rho&
\\
&\ \ \ \ \ \ \frac{qE\,\rho\,e^{i\frac{qE}{2}\rho^2 \coth
qEs}}{2\,\pi\,\sinh(qE\,s)}
\left\{K_{i\frac\omega a}\left(i\frac{qe\,\rho^2}{2\,\sinh
(qE\,s)}\right) +\left( 2\,\theta(\omega)\sinh(\pi\frac \omega
a)-e^{\pi
\frac \omega a}\right)\,K_{i\frac\omega
a}\left(-i\frac{qe\,\rho^2}{2\,\sinh (qE\,s)}\right) \right\}\qquad .&
\end{array}
\end{equation}
Formula  ({\bf 6.611.3}) of ref. \cite{GR} provides us the result of the $\rho$
integration. So we obtain  for the
vacuum persistence amplitude, computed using (\ref{Kzw}):
\begin{equation}
{\cal {I}}m {\cal{W}}={\cal{R}}e\frac {\cal T}{4\pi
}\int_{-\infty}^{\infty}\frac {d\omega}{\sinh(\pi\frac{\omega}{a})}
\int_0^\infty \frac{ds}{s\sinh (qE\,s)}\,e^{-im^2s}
\left\{ \theta(\omega) \sinh(\pi\frac \omega a)\,
e^{i2\frac { \omega} a qE\,s}-\theta(-\omega) \sinh(\pi\frac \omega a)
\right\}
\end{equation}
i.e. expression (\ref{ImWRindapprox}) which is equivalent to eq.
(\ref{ImWRind}).

Let us note that the Fourier transform of the Schwinger kernel has to
satisfy the equation
\begin{equation}
\left[i\partial_s-m^2+\partial_\rho^2 +\frac{(qE)^2}{4}\rho^2+\frac{1}
{\rho^2}\left(\frac{\omega^2}{a^2}+\frac 14\right)\right]\tilde
K_\omega(\rho,\rho';s)=0
\end{equation}
and the initial condition on the \TR quadrant
\begin{equation}
\lim_{s\rightarrow 0}\tilde
K_\omega(\rho,\rho';s)=\delta (\rho-\rho')\qquad.
\end{equation}
The expression (\ref{Kzw}) satisfies this equation and the
initial condition in the following sense:
\begin{equation}
\lim_{s\rightarrow 0}\tilde
K_\omega^{w=0}(\rho,\rho';s)=\delta (\rho-\rho') +
[2\,\theta(\omega) \sinh(\pi\frac \omega a)-e^{\pi \frac{\omega} a}]
\delta (\rho + \rho') \qquad .
\end{equation}
This extra term is not a surprise. It is localized on the boundary of
the \TR patch and reflects the correlations between the left and right
quadrants  that appear when the vacuum decay and that pair
creation occurs. Actually the existence on the horizon of a singular
contribution to the Rindler (uncharged) scalar field  has been put into
evidence some time ago by Parentani \cite {PaCQG} who has shown that
such terms are essential to the validity of the theorem stating that
the usual Poincar\'e invariant vacuum state is
the state of minimal energy.
\section{Construction of Unruh modes}
In this section we shall establish connections between Minkowskian and
Rindlerian modes.  Instead of working with integral transforms, we find
more convenient to diagonalize the Bogoljubov  transformation. This is
achieved by introducing Unruh modes. They correspond to  new  basis of
the Fock spaces, which share the same quantum numbers as the Rindler
modes but that do not  mix positive and negative frequencies  of the
Minkowskian modes. The purpose of this section is to build these modes
which allows  to easily describes algebraically the Minkowskian particle content of
the Rindler vacua.
Completeness of Minkowskian and Rindler sets of modes insures  each
element of one set can be expressed as superposition  of members of the
other one. To obtain the relation between them we shall, instead of
computing directly the overlapping integrals, made use of integral
representations of the parabolic cylinder functions (see ref.\cite{GR},
eq. ({\bf 9.241.2}) for the mathematics and
\cite{BMPS} for their physical meaning) . Accordingly, we may express the
$in$- and
$out$-modes (eqs \ref{Dmodein} and followings in the text) as~:
\begin{eqnarray}
\phi^{p \, in}_\sigma (U,V)&=&\frac {1}{\Gamma[\frac 12
-i\frac{m^2}{2qE}]}\frac {e^{-\frac {\pi
m^2}{4qE}}e^{-\frac {i\pi} 4}}{\sqrt{2\pi}(2qE)^{1/4}} e^{-i\sigma V}e^{-i\frac {qE}4
V^2}e^{i\frac {qE}4UV}e^{-i\frac {\sigma^2}{2qE}}\label{Direp}\nonumber\\ &&\int_0^\infty dx
e^{ix\sqrt{2qE}(\frac {V-U}2 +\frac \sigma {qE}-i\frac {x^2}2})x^{-i\frac
{m^2}{2qE}-\frac 12} \\
\phi^{a \, in}_\sigma (U,V)&=&\frac
{1}{\Gamma[\frac 12
-i\frac{m^2}{2qE}]}\frac {e^{-\frac {i\pi}4}e^{-\frac {\pi
m^2}{4qE}}} {\sqrt{2\pi}(2qE)^{1/4}} e^{i\sigma
U}e^{-i\frac {qE}4 U^2}e^{i\frac {qE}4UV}e^{-i\frac {\sigma^2}{2qE}}\nonumber\\ &&\int_0^\infty dx e^{-ix\sqrt{2qE}(\frac {V-U}2
+\frac \sigma {qE}-i\frac {x^2}2})x^{-i\frac {m^2}{2qE}-\frac 12}
\end{eqnarray} where we have used the ingoing and outgoing light-like
coordinates $U=t-z$ and
$V=t+z$. Using the reflexion operation $U\mapsto -V$ and
$V\mapsto -U$, we obtain~:
\begin{eqnarray}
\nonumber
\phi^{p \, out}_\sigma (U,V)&=&(\phi^{p \, in}_\sigma (-V,-U))^*\qquad,\\
\nonumber
\phi^{a \, out}_\sigma (U,V)&=&(\phi^{a \, in}_\sigma (-V,-U))^* \qquad .
\end{eqnarray}
\\ On the other hand, from the integral representations
(\ref{Wintrep1})  of the Whittaker's function, we get the two
representations of the Rindler modes (see eqs. {\ref{RindFunW}}):
\begin{eqnarray}  {\cal{W}}^{-}_{\epsilon} (U,V)&=&\left(\frac{qE}{4 \pi
a^2}\right)^{1/2}e^{i\frac{qE}4 UV}(\frac {qEV^2}2)^{-i\frac \omega
{2a}}\frac{e^{i\epsilon \frac \pi 4}e^{\epsilon \frac {\pi \omega}{4a}}}
{\Gamma(-\frac {i m^2}{2qE}+\frac 12)} \\ \nonumber &&\int_0^\infty dt
e^{i\frac {qE}2 UVt}t^{-i\frac {m^2}{qE}-\frac 12}(1+t)^{-i\frac \omega a
+i\frac {m^2}{2qE}-\frac 12} \\  &=&\left(\frac{qE}{4 \pi
a^2}\right)^{1/2}e^{i\frac{qE}4 UV}(\frac {qEU^2}2)^{i\frac \omega
{2a}}\frac{e^{i\epsilon \frac \pi 4}e^{-\epsilon
\frac {\pi \omega}{4a}}} {\Gamma(i\frac \omega a -\frac {i m^2}{2qE}+\frac
12)}\\
\nonumber &&\int_0^\infty dt e^{i\frac {qE}2 UVt}t^{i\frac \omega a
-i\frac {m^2}{qE}-\frac 12}(1+t)^{i\frac {m^2}{2qE}-\frac 12} \\
&=&\left({\cal{W}}^{+}_{\epsilon} (U,V)\right)^* \label{Wirep}\qquad,
\end{eqnarray} where we have reexpressed  the Rindler coordinates $\xi$
and
$\tau$ in terms of  light-like Minkowski coordinates, through the
relations $\xi=\frac 1{2a}
\ln(-\epsilon a^2UV)$ and $\tau= \frac 1{2a} \ln (-\epsilon \frac V U)$.\\
 Let us now consider a wave packet, denoted hereafter as $\Omega^{p \,
in}_\omega$, of inertial in-particles
$\phi^{p\, in}_\sigma$:
\begin{eqnarray}
\Omega_{p , in}^\omega=\int_{-\infty}^{\infty}  A(\frac
\sigma {\sqrt{qE}}, \frac \omega a)\phi^{p \, in}_ \sigma\,d(\frac \sigma
{\sqrt{qE}} )
\qquad.\label{Omegapin}
\end{eqnarray} Using the integral representation (\ref{Direp}), we can
rewrite this superposition as the convolution~:
\begin{equation}
\frac  {e^{\frac {\pi m^2}{8qE}}e^{\frac {i\pi} 8}e^{-i\frac {qE}4
V(V-U)}}{M\,\Gamma[\frac 12 -i\frac{m^2}{2qE}]}\int_0^\infty
e^{ix\sqrt{2qE}\frac {V-U}2 -i\frac {x^2}2}x^{-i\frac {m^2}{2qE}-\frac
12} F\left[ \frac \omega a;\sqrt{qE}(-V+ x \sqrt{\frac 2
{qE}})\right]\,dx\label{Spinm2}\\
\end{equation} where we introduced a two variable function, $F$, built
from the coefficients of the wavepacket (\ref{Omegapin})
 through the relation:
\begin{equation}  F\left[{\frac\omega
a};\beta\right]=\int_{-\infty}^{+\infty}  A(\tilde
\sigma ,{\frac\omega a}) e^{i  {\beta\tilde\sigma} }  e^{-i \frac
{{\tilde\sigma}^2} {2}}\, d\tilde\sigma  \qquad .\label{Fdef}
\end{equation}
If we choose the arbitrary function $F$ such as~:
\begin{equation}  F\left[\frac\omega a;\beta\right]=\left( \frac
{qE}{2a^2}\right)^{\frac14} e^{i \beta^2 /4} (\beta / \sqrt 2)^{i(\frac
{m^2}{2qE}-\frac\omega a )-\frac 12}\label{Fansatz}
\end{equation} (the choice of the coefficient in front of this expression
anticipating a subsequent normalization),
 eq.(\ref{Spinm2}) becomes an integral representation of Rindlerian $\cal
V$ $in$-modes on \TP and \TL , i.e.
${\cal W}^{-}_{\epsilon}$ functions (see eq. (\ref{RindFunW}))~:
\begin{equation}
\Omega^{p , in}_\omega=\left(\frac{8\pi^2a^2}{qE}\right)^{\frac
14}\frac{e^{i\frac\pi 8} e^{\frac{\pi m^2}{8qE}}}{M}
e^{-i\epsilon\frac\pi 4}e^{-\epsilon\frac{\pi\omega}{4a}}\,{\cal
W}^{-}_{\epsilon}\qquad .
\end{equation}
\par\noindent While general arguments of completeness insure the existence
of the function
$A(\sigma/\sqrt{qE}, \omega/a)$, we need its explicit expression to obtain
the continuation of
$\Omega^{p ,in}_\omega$ on the rest of  space-time (in the regions $V>0$).
Taking the inverse Fourier transform  (with respect to the variable $\beta$) of eq. (\ref{Fansatz}), we obtain~:
\begin{eqnarray}  A(\frac{\sigma}{\sqrt{qE}}, \frac{\omega}{a})&=&
\left(\frac{qE}{2a^2}\right)^{\frac14}
\frac {e^{i {\frac {\sigma^2} {{2qE}}}}}{\sqrt 2
\pi} \int_0^{\infty} e^{i\frac {p^2} 2}e^{-i\sqrt 2 {\frac \sigma
{\sqrt{qE}}} p}p^{-i \frac \omega a +i
\frac {m^2}{2qE}-\frac 12}\,
dp\label{Aint}\\&=&\left(\frac{qE}{2a^2}\right)^{\frac14}\frac
{e^{{\frac \pi 2}({\frac \omega {2a}}-\frac{
m^2}{4qE})}e^{\frac{i\pi}8}}{\sqrt 2 \pi}
\Gamma[i(\frac {m^2}{2qE}-{\frac \omega a}]+\frac 12) D_{i({\frac \omega
a}-\frac {m^2}{2qE})-\frac 12}(\sqrt 2 e^{\frac {3i\pi}4 }{\frac \sigma
{\sqrt{qE}}})\nonumber\qquad .
\end{eqnarray}

  Using twice the integral representations (\ref{Aint}) of
the parabolic cylinder functions,
  the integrals over
$\frac \sigma {\sqrt{qE}}$ and $x$ become straightforward. We obtain an
expression involving only
${\cal W}^{+}_{\epsilon}$ functions, i.e. modes ${\cal
U}_{in,R}$ and ${\cal U}_{out,F}$~:
\begin{equation}
\Omega^{p , in}_\omega=\left(\frac{8\pi^2 a^2}{qE}\right)^{\frac
14}\frac{e^{i\frac\pi 8} e^{\frac{\pi
m^2}{8qE}}\Gamma[i(\frac{m^2}{2qE}-\frac\omega a)+\frac
12]}{M\,\Gamma[\frac 12-i\frac{m^2}{2qE}]} e^{i\epsilon\frac\pi
4}e^{\epsilon\frac{\pi\omega}{4a}}\,{\cal W}^{+}_{\epsilon}\qquad .
\end{equation}

More explicitly, by introducing the characteristic functions
$\chi_{L,R,P,F}$ (which are equal to zero or to one according to that the
coordinates considered correspond to a point belonging to the sector
labeling the function or not), we may write~:
\begin{eqnarray}
\Omega^{p , in}_\omega&=&\chi_P\, {\cal V}^{\omega}_{in,P} -i\,
e^{-\frac{\pi m^2}{4qE}}\left({\frac{\cosh[\pi(\frac \omega
a-\frac{m^2}{2qE})]}{|\sinh(\frac{\pi\,\omega}a)|}}\right)^{\frac 12}
\left[\chi_L\,{{\cal V}^{\omega}_{in,L}} +i\,\chi_R\, e^{ \frac{\pi
\omega} {2a}}\frac {\Gamma[i(\frac {m^2}{2qE}-\frac \omega a )+\frac 12]}
{\Gamma[\frac 12 -i\frac{m^2}{2qE}]}\,{{\cal
U}^{\omega}_{in,R}}\right]\nonumber\\ &&-i \,\chi_F\, e^{-\frac{\pi
\omega}{2a}}\frac {\Gamma[i(\frac {m^2}{2qE}-\frac \omega a )+\frac 12]}
{\Gamma[\frac 12 -i\frac{m^2}{2qE}]}\,{{\cal U}^{\omega}_{out,F}}
\label{OmegapinRLFP}\qquad .
\end{eqnarray} The mode (\ref{OmegapinRLFP}) so constructed defines an
Unruh mode. Indeed it is  a (superposition of) positive frequency modes,
algebraically related to Rindler modes. Similarly, we may define other
orthonormalized $in$-Unruh modes through wave superpositions built as
in eq. (\ref{Omegapin})~:
\begin{eqnarray}
\varpi_{p \, in}^\omega&=&\int_{-\infty}^{\infty}  A(-\frac
\sigma {\sqrt{qE}}, \frac \omega a)\phi^{p \, in}_ \sigma\,d(\frac \sigma
{\sqrt{qE}} )=e^{i\psi}\sgn (\omega)\left[\chi_R {{\cal
V}^{\omega}_{in,R}}+i\chi_F e^{\frac{\pi\omega}{2a}}{{\cal
V}^{\omega}_{in,F}}\right]\qquad,\label{varpipin}\\
\Omega_{a \, in }^{\omega}&=&\int_{-\infty}^{\infty}  A(-\frac
\sigma {\sqrt{qE}}, \frac \omega a)\phi^{a \, in}_ \sigma\,d(\frac \sigma
{\sqrt{qE}} )\nonumber\\ &=&\chi_P\,{{\cal U}^{\omega\,
*}_{in,P}}-i\,e^{-\frac{\pi m^2}{4qE}}
\left({\frac{\cosh[\pi(\frac \omega
a-\frac{m^2}{2qE})]}{|\sinh(\frac{\pi\,\omega}a)|}}\right)^{\frac
12}\left[\chi_R{{\cal U}^{\omega\, *}_{in,R}} +i\,e^{\frac{\pi
\omega}{2a}}\frac{\Gamma[i(\frac{m^2}{2qE}-\frac\omega a )+\frac 12]}
{\Gamma[\frac 12 -i\frac{m^2}{2qE}]}\,\chi_L\,{{\cal V}^{\omega\,
*}_{in,L}}\right]\nonumber\\ &&+i\,e^{-\frac{\pi
\omega}{2a}}\frac{\Gamma[i(\frac{m^2}{2qE}-\frac\omega a)+\frac 12]}
{\Gamma[\frac 12 -i\frac{m^2}{2qE}]}{{\cal V}^{\omega\, *}_{out,F}}
\qquad,\label{Omegaain}\\
\varpi_{a \, in }^{\omega}&=&\int_{-\infty}^{\infty}  A(\frac
\sigma {\sqrt{qE}}, \frac \omega a)\phi^{a \, in}_ \sigma\,d(\frac \sigma
{\sqrt{qE}} ) =e^{i\psi}\sgn (\omega)\left[\chi_L {{\cal U}^{\omega\,
*}_{in,L}}+i\chi_F e^{\frac{\pi\omega}{2a}}{{\cal U}^{\omega\,
*}_{in,F}}\right]
\qquad,\label{varpiain}
\end{eqnarray} where the phase $\psi$ is given by~:
\begin{equation}
\psi= \arg
\left\{ \frac{\Gamma[i\frac \omega a]}{\Gamma[ i(\frac{m^2}{2qE}+\frac
\omega a)+\frac 12]}\right\}\qquad .
\end{equation}  Let us emphasize some properties of these Unruh modes. By
considering on $\Scrim$ the supports  of the Rindler modes that define them, it
is obvious that these Unruh modes are orthogonal to each other and normalized. This
also results from the relation~:
\begin{eqnarray}
\int^{\infty}_{-\infty}d(\frac \sigma {\sqrt{qE}})A(\sigma,
\omega)A^*(\varepsilon\,\sigma,
\omega^{\prime})&=&\theta(\varepsilon)\delta({\omega^\prime} -\omega )\qquad \varepsilon=\pm 1\qquad.
\label{AAint}
\end{eqnarray} that can be  easily proved using the  integral
form (\ref{Aint}) of the coefficients $A(\sigma,\omega)$. On the other hand, once we
realize that these modes have to correspond to Minkowskian particle or
antiparticle $in$-modes, starting from $\ScrimR$ or $\ScrimL$, they may
be built from purely algebraic considerations. Indeed it suffices to fix
their behavior on $\Scrim$ and continue them across the horizon by
requiring continuity (always in term of wave packets). For instance,
suppose we want to start with modes that born on $\ScrimPR$, we have to
take on \TP ,  ${\cal V}_{in,P}$ modes (see fig. ({\bf
3})). Then in order to maintain zero Cauchy data on
$\ScrimRP$ and
$\ScrimL$ we have to paste them respectively to ${\cal U}_{in,R}$ and
${\cal V}_{in,L}$ modes. Then it remains to fix the linear combination
of modes defined on
\TF that fit our construction on $\HorpR$ and $\HorpL$.
So, using the relation (\ref{Worig}) we may fix the relative weights
(and phases) between these modes on each quadrant. And an a priori
unexpected
 property appears: the interferences are such that the particle Unruh mode
$\Omega^{p\,in}$ vanishes on
$\ScripFL$ and the antiparticle Unruh mode $\Omega^{a\,in}$ on
$\ScripFR$. This  finds its mathematical origin in an analyticity
property: the  fact that the
$\varpi$ Unruh modes can be expressed as linear combination of only
functions
${\cal M}^{\pm}_{\epsilon}$  and the modes $\Omega$ as pure combination of
${\cal W}^{\pm}_{\epsilon}$ functions.

 Just as we have introduced an
$in$ Unruh basis (\ref{Omegapin}\ref{varpiain})  we may consider a
$out$ Unruh basis built out from
 superposition of $out$-particle and antiparticle modes with as
coefficient the complex conjugate of those used in the construction of
the $in$ Unruh basis. Actually these modes are obtained by performing a
space-time inversion [$(\tau_F,\xi_F)\leftrightarrow(\tau_P,\xi_P),\
(\tau_R,\xi_R)\leftrightarrow(\tau_L,\xi_L)$] on the previous modes and
taking their complex conjugate. They are given by~:
\begin{eqnarray}
\Omega_{p , out}^\omega&=&\int_{-\infty}^{\infty}  A(\frac
\sigma {\sqrt{qE}}, \frac \omega a)^*\phi^{p \, out}_ \sigma\,d(\frac
\sigma {\sqrt{qE}} )\label{Omegapout}\qquad,\\
\varpi_{p \, out}^\omega&=&\int_{-\infty}^{\infty}  A(-\frac
\sigma {\sqrt{qE}}, \frac \omega a)^*\phi^{p \, out}_ \sigma\,d(\frac
\sigma {\sqrt{qE}} )
\label{varpipout}\qquad,\\
\Omega_{a \, out }^{\omega}&=&\int_{-\infty}^{\infty}  A(-\frac
\sigma {\sqrt{qE}}, \frac \omega a)^*\phi^{a \, out}_ \sigma\,d(\frac
\sigma {\sqrt{qE}} )
\label{Omegaaout}\qquad,\\
\varpi_{a \, out }^{\omega}&=&\int_{-\infty}^{\infty}  A(\frac
\sigma {\sqrt{qE}}, \frac \omega a)^*\phi^{a \, out}_ \sigma\,d(\frac
\sigma {\sqrt{qE}} )\qquad.
\label{varpiaout}
\end{eqnarray} These modes are related to Rindler $out$-modes by a
Bogoljubov transformation given in Appendix {\bf {B}} and schematically
represented on fig. ({\bf 4}).\\
To compute the coefficient of the Bogoljubov transformation between the
$in$ and $out$ Unruh modes, we use the Bogoljubov transformation
(\ref{bogomink}) to transfer the relation between modes into relation between
connectors linking Rindlerian and Minkowskian modes. For instance we
obtain for the modes
$\Omega^{p \, out}_\omega$~:
\begin{eqnarray}
\Omega^{p \, out}_\omega=\int_{\infty}^{\infty}
A(\frac\sigma{\sqrt{qE}},\omega)^*(\gamma\phi^{p\,in}_{\sigma}-\delta^*\phi^
{a\,in\,*}_{\sigma})d(\frac\sigma{\sqrt{qE}})\qquad.\label{poutin}
\end{eqnarray}  Fortunately, the $\gamma$ and $\delta$ coefficients are
$\sigma$ independent and factorize out of the integral. The contribution
of the antiparticle modes $\phi_{\sigma}^{a\, in}$ sums directly
(see eq. (\ref{varpiain})) and gives~:
\begin{equation}
 e^{i\frac \pi 2}e^{-\frac{\pi m^2}{2qE}}\varpi^{a\,in\,*}_\omega\qquad.
\end{equation} The summation of the particle modes needs a little bit
more work. Thanks to the linear relations connecting parabolic cylinder
functions, we get:
\begin{equation}
 A(\pm\sigma,\omega)^*=\frac{\Gamma[\frac12+i\frac \omega
a-i\frac{m^2}{2qE}]}{\sqrt{2\pi}}\left\{ e^{-i\frac \pi 2}e^{- \frac{\pi
m^2}{4qE}}e^{ \frac{\pi\omega}{2a}}A(\mp\sigma,\omega)+ e^{ \frac{\pi
m^2}{4qE}}e^{-
\frac{\pi\omega}{2a}}A(\pm\sigma,\omega)\right\}\label{AAintbis}
\end{equation} and the particle modes contribution to eq. (\ref{poutin}) can
be written as a sum of  two $in$ Unruh modes. So we get:
\begin{eqnarray}
\Omega^{p \, out}_\omega&=&i\sqrt{q_1}\Omega^{p \,
in}_\omega+e^{-\frac{\pi m^2}{2qE}}e^{\frac{\pi
\omega}a}\sqrt{q_1}e^{i\psi'}\varpi^{p
\, in}_\omega +i e^{-\frac{\pi m^2}{2qE}}\varpi^{a \, in \, *}_\omega\qquad,
\end{eqnarray} where $q_1$ is defined by eq. (\ref{QUin}) and $\psi'$ is
the phase:
\begin{equation}
\psi'=\arg \left\{\frac{\Gamma[i(\frac\omega a-\frac{m^2}{2qE})+\frac
12]}{\Gamma[\frac 12 + i\frac{m^2}{2qE}]}\right\}
\qquad .
\end{equation}
 In the same way, we obtain:
\begin{eqnarray}
\varpi^{p \, out}_\omega&=&ie^{i\psi'}\sqrt{q_1}\varpi^{p
\, in}_\omega +e^{-\frac{\pi m^2}{2qE}}e^{\frac{\pi
\omega}a}\sqrt{q_1}e^{i\psi'}\Omega^{p \, in}_\omega+i e^{-\frac{\pi
m^2}{2qE}}\Omega^{a \, in \, *}_\omega\qquad ,\nonumber \\
\Omega^{a \, out }_\omega&=&ie^{i\psi'}\sqrt{q_1}\Omega^{a
\, in}_\omega +ie^{-\frac{\pi m^2}{2qE}}\varpi^{p \, in\,
*}_\omega+e^{i\psi'}e^{-\frac{\pi
m^2}{2qE}}e^{\frac{\pi\omega}{a}}\sqrt{q_1}\varpi^{a \, in }_\omega\qquad
,
\nonumber \\
\varpi^{a \, out }_\omega&=&ie^{i\psi'}\sqrt{q_1}\varpi^{a\, in}_\omega
+e^{i\psi'}e^{-\frac{\pi m^2}{2qE}}
e^{\frac{\pi\omega}{a}}\sqrt{q_1}\Omega^{a \, in }_\omega + ie^{-\frac{\pi
m^2}{2qE}}\Omega^{p \,in}_\omega\qquad .
\end{eqnarray}

We found instructive to recover, using these basis, the decay rate
(\ref{Drate}) between $in$ and $out$ Minkowskian vacua. Formally we obtain~:
\begin{equation}
\vert\braket {0,out}{0,in}\vert^2=\exp\left[-2
\frac{{\cal{T}}}{2\pi}\int_{-\infty}^{\infty}d\omega\
\ln\left(1+e^{-\frac{\pi m^2}{qE}}\right)\right]\nonumber
 \label{Drate2}\quad.
\end{equation} Let us emphasize that $2
({{\cal{T}}}/{2\pi})\int_{-\infty}^{\infty}d\omega=2(\sum_{\omega>0}+\sum_{\omega<0})$ gives the expected total space-time volume factor $LT/2\pi$, in accordance with our interpretation of $\sum_{\omega>0}$
and
$\sum_{\omega<0}$ as the space-time volume of the various quadrant.\\ More
interesting is the computation of  Rindlerian particle  population in
Minkowski vacua. Expressing the field as superposition of Unruh modes and
Rindler we obtain, thanks to eqs (\ref{OmegapinRLFP}-\ref{varpiain}), we
obtain the Bogoljubov transformation between Unruh and Rindler creation
and annihilation operators~:
\begin{eqnarray}  a_{{\cal{U}}_R}^{in}(\omega)&=&\alpha_{\Omega
{\cal{U}}_R}^{in}(\omega)a_{\Omega}^{in}(\omega)+\gamma_{\Omega
{\cal{U}}_R}^{in}(\omega)b_{\Omega}^{in\,\dagger}(\omega)\qquad\qquad
\mbox{\rm with}\ \omega>0\qquad ,\nonumber\\
a_{{\cal{V}}_L}^{in}(\omega)&=&\alpha_{\Omega
{\cal{V}}_L}^{in}(\omega)a_{\Omega}^{in}(\omega)+\gamma_{\Omega
{\cal{V}}_L}^{in}(\omega)b_{\Omega}^{in\,\dagger}(\omega)\qquad\qquad
\mbox{\rm with}\ \omega<0\qquad ,\nonumber\\
b_{{\cal{U}}_R}^{in\,\dagger}(\omega)&=&\beta_{\Omega
{\cal{U}}_R}^{in}(\omega)a_{\Omega}^{in}(\omega)+\epsilon_{\Omega
{\cal{U}}_R}^{in}(\omega)b_{\Omega}^{in\,\dagger}(\omega)\qquad\qquad
\mbox{\rm with}\ \omega<0\qquad ,\nonumber\\
 b_{{\cal{V}}_L}^{in\,\dagger}(\omega)&=&\beta_{\Omega{\cal{V}}_L}^{in}
(\omega)a_{\Omega}^{in}(\omega)+
\epsilon_{\Omega{\cal{V}}_L}^{in}(\omega)b_{\Omega}^{in\,\dagger}(\omega)
\qquad\qquad \mbox{\rm with}\ \omega>0\qquad ,\nonumber\\
a_{{\cal{V}}_R}^{in}(\omega)&=&\alpha_{\varpi
{\cal{V}}_R}^{in}(\omega)a_{\varpi}^{in}(\omega)\qquad ,\nonumber\\
b_{{\cal{U}}_L}^{in\,
\dagger}(\omega)&=&\epsilon_{\varpi
{\cal{U}}_L}^{in}(\omega)b_{\varpi}^{in\,\dagger}(\omega)\qquad ,
\end{eqnarray}  the various coefficient $\alpha,\ \beta,\ \gamma$ etc
\dots being defined implicitly by eqs (\ref{Omegapin}, \ref{varpipin}-\ref{varpiain}).
These relations allows us to express \cite{KaUm} the  $in$ Minkowski
vacuum
 as (see Appendix {\bf{B}}):
\begin{eqnarray}
\vert \, 0,\, Mink,\,in\rangle&=&\prod_{\omega>0}\frac 1{\vert
{\alpha^{in\,*}_{\Omega {\cal{U}}_R}\vert^2}}\prod_{\omega>0}\frac
1{\vert {{\alpha^{in\,*}_{\Omega
{\cal{V}}_L}}\vert^2}}\exp\left[\sum_{\omega>0}\frac{\beta^{in\,*}_{\Omega
{\cal{V}}_L}}{\alpha^{in\,*}_{\Omega
{\cal{U}}_R}}a_{{\cal{U}}_R}^{in\,\dagger}(\omega)b_{{\cal{V}}_L}^{in\,
\dagger}(\omega) \right]\nonumber\\ &&\exp\left[\sum_{\omega<0}
\frac{\beta^{in\,*}_{\Omega {\cal{U}}_R}}{\alpha^{in\,*}_{\Omega
{\cal{V}}_L}}a_{{\cal{V}}_L}^{in\,\dagger}(\omega)b_{{\cal{U}}_R}^{in\,
\dagger}(\omega)\right]
\kvac_{in}^{\R}\otimes\kvac_{in}^{\L}\nonumber\\
\end{eqnarray} and can then be interpreted as a superposition of Rindler's
pairs consisting in  Rindlerian particles and antiparticles and their
partners, anti particles and particles localized in the opposite sector.
Using the explicit form of the Bogoljubov coefficients given in Appendix
{\bf {B}}, we obtain the mean density number of Rindlerian $out$ particles
that an  accelerated observer (in the sector
$\tR$)  detects when the  quantum state is the
$in$ Minkowski vacuum~:
\begin{eqnarray}  n_{{\cal{U}}^{out}_{\omega  \,R}}&=&\langle 0, Mink ,\,
in\vert a ^{\dag \, out}_{{\cal{U}}_R}(\omega)a
^{out}_{{\cal{U}}_R}(\omega)\vert  0, Mink \, in\rangle
 \nonumber\\ &=&\theta(\omega)\vert
\alpha^R_{{\cal{U}}{\cal{U}}}\vert^2\vert \gamma^{in}_{\Omega \,
{\cal{U}}_R}(\omega)\vert^2+\theta(-\omega)\vert
\beta^R_{{\cal{U}}{\cal{U}}}\vert^2\vert \epsilon^{in}_{\Omega \,
{\cal{U}}_R}(\omega)\vert^2 \nonumber\\ &=&e^{-\frac {\pi m^2}{
qE}}\label{npUoutR}
\end{eqnarray} and:
\begin{eqnarray}  n_{{\cal{V}}^{out}_{\omega>0  \,R}}&=&\langle 0, Mink
,\, in\vert a ^{\dag \, out}_{{\cal{V}}_R}(\omega)a
^{out}_{{\cal{V}}_R}(\omega)\vert  0, Mink \, in\rangle
\nonumber\\ &=&\vert \alpha^R_{{\cal{V}}{\cal{U}}}\vert^2\vert
\gamma^{in}_{\Omega \, {\cal{U}}_R}(\omega)\vert^2 \nonumber\\
&=&\frac{(1+e^{-\frac {\pi m^2}{ qE}})}{e^{2\pi\frac
\omega a}-1}\label{npVoutR}
\end{eqnarray} for the various particles, while the density number of
antiparticles is~:
\begin{eqnarray}
 n_{{\cal{V}}^{out}_{\omega < 0 \,R}}&=&\langle 0, Mink ,\, in\vert b
^{\dag \, out}_{{\cal{V}}_R}(\omega)b ^{out}_{{\cal{V}}_R}(\omega)\vert
0, Mink ,\, ,in\rangle
 \nonumber\\ &=&\vert
\epsilon^R_{{\cal{V}}{\cal{U}}}\vert^2\vert \beta^{in}_{\Omega \,
{\cal{U}}_R}(\omega)\vert^2+\vert
\gamma^R_{{\cal{V}}{\cal{V}}}\vert^2\vert \alpha^{in}_{\varpi \,
{\cal{V}}_R}(\omega)\vert^2 \nonumber\\&=&\frac{(1+e^{-\frac {\pi m^2}{
qE}}e^{-2\pi\frac\omega a})}{e^{-2\pi\frac \omega a}-1}\qquad
.\label{naVoutR}
\end{eqnarray}
In the same way, an  accelerated observer in the $\tL$ sector will detect
the following populations:
\begin{eqnarray} & n_{{\cal{U}}^{out}_{\omega > 0 \,L}}
=n_{{\cal{V}}^{out}_{\omega > 0
\,R}}=\frac{1+e^{-\frac {\pi m^2}{ qE}}}{e^{2\pi\frac
\omega a}-1}&\qquad ,\label{naUoutL} \\
&n_{{\cal{V}}^{out}_{\omega \,L}}=n_{{\cal{U}}^{out}_{\omega
\,R}}=e^{-\frac{\pi m^2}{qE}}\label{naVoutL}
\end{eqnarray} of antiparticles and:
\begin{equation} n_{{\cal{U}}^{out}_{\omega < 0
\,L}}=n_{{\cal{V}}^{out}_{\omega < 0 \,R}}=
\frac{(1+e^{-\frac {\pi m^2}{ qE}}e^{-2\pi\frac\omega a})}{e^{-2\pi\frac
\omega a}-1}\qquad .\label{npUoutL}
\end{equation} of particles.\\ All these populations are given by Bose
factor and corrective terms proportional to the Schwinger factor. On the
other hand, an  accelerated observer in the right sector  measures a
total charge given by:
\begin{eqnarray}  Q_R&=&\int_0^\infty d\omega
N_{{\cal{V}}^{out}_{\omega\,R}}+\int_{-\infty}^\infty d\omega
N_{{\cal{U}}^{out}_{\omega\,R}}-\int_{-\infty}^0 d\omega
N_{{\cal{V}}^{out}_{\omega  \,R}}\nonumber\\
 &=&\int_0^\infty d\omega e^{-\pi\frac{m^2}{qE}}\qquad .\label{QRIGHT}
\end{eqnarray} Charge conservation implies that a left observer sees:
\begin{eqnarray}  Q_L=-\int_0^\infty d\omega e^{-\pi\frac{m^2}{qE}}\label{QLEFT}\qquad
.
\end{eqnarray} Note  (a check of all this algebra) that, always due to
charge conservation,
 the same expressions of $Q_R$ and $Q_L$ are obtained if we compute them
using $in$ density number of particles, but the individual contributions
of $\cal U$ and $\cal V$ type of (anti)particles are different.\\ Of
course the total charges $Q_F$ and $Q_P$ are zero.\\
 Moreover, when $\frac{qE}{m^2}\rightarrow 0$ (i.e. for a weak electric
field or  massive particles), some populations  (\ref{npUoutR},
\ref{naVoutL}) vanish, and the other (\ref{npVoutR},\ref{naVoutR},
\ref{naUoutL},\ref{npUoutL}) become Boltzmanian in
character $\approx 1/ {e^{\frac {2\pi |\omega |}a}-1}$. This can be
understood as follows.
Let us consider, for instance, the ${\cal{U}}^{out}_{\omega  \,R}$ modes.
In the limit
where $qE$ goes to zero, these modes become localized near $i^0_R$. Their
charges (see eqs
(\ref{QUoutR}) ) on the horizon components  vanishes, while on the past and
future horizon
they become equal to unity. Indeed these modes tend to Bessel-I functions,
i.e. function which grows
exponentially when
$\rho$ goes to infinity, and that do not contribute to  the Hilbert space of
the neutral quantum field (see for instance ref. \cite{Sp}). Classically,
they correspond
to hyperbolic trajectories pushed away to infinity, hyperbolic trajectories
whose radius
$(m/qE)$ become
infinite and whose
asymptotic points on ${\cal I}^-_R$ and ${\cal I}^+_R$ slip to $i^0$. On
the contrary,
${\cal{V}}^{out}_{\omega  \,R}$ correspond classically to hyperbolic motion
whose past and
future asymptotic points tend respectively near $i^-$ and $i^+$ i.e.
inertial trajectories.
When such trajectories cross the \TR quadrant, they enter in and leave it
out across the
horizon components ${\cal H}^-_R$ and ${\cal H}^+_R$. This is confirmed by
the fact that
for such modes their charge contents (eqs (\ref{QVoutR})) vanishes at
infinity but become
equal to unity on the horizon components.
\section{Uniformly accelerated charged detector} In a previous work
\cite{GMPS} we have computed transition amplitudes (see hereafter, formulas
\ref{Akkk} to \ref{Bcharged}) between charged particles of masses and charges
 that we denote
here
$M,\ Q$ and
$M',\ Q'$ (instead of $M,\; Q$ and $m,\; q$ as in ref.\cite{GMPS})
interacting by  exchange of a third kind of particles of mass $m$ and
charge
$q$ (that was called $\mu$ and $\alpha$ in ref.\cite{GMPS}). Their
interactions were described by a three  field interaction Hamiltonian and
first order perturbation calculations were performed. Our purpose now is
first to clarify
some aspect of the physics of the accelerated detector and to make an
explicit contact
between it and the model built on the three interacting fields.\\ The
detector can be
seen as a ``two-level ion" propagating in a constant electric field. The
ion levels are
supposed to have rest masses
$M$ and $M'$ (resp. charges $Q$ and $Q'$), both much greater than the mass
$m$ (resp. charge $q$) of the exchanged quanta. This ion  makes transition,
without recoil, between its two levels, i.e. whatever is its internal
configuration, it always moves with constant acceleration
$a=Q\;E/M=Q'\;E/M'$. Hereafter we show how this model appears as a
limiting case of the three field model, by comparing amplitude of
transition obtained in both cases.
In the following, all the calculations are done at first order of the
interaction coupling constant.
The physics of the accelerated two level detector  is described through the
effective
hamiltonian~:
\begin{eqnarray} {\cal H}^{int}(\tau)=a\tilde g\left[ Ae^{-i \Delta M \tau}
e^{iq\int A_\mu dx^\mu} \hat\phi^{\dag}(\tau,\xi=0)+ A^{\dag}e^{i \Delta
M \tau} e^{-iq\int A_\mu dx^\mu}
\hat\phi(\tau,\xi=0)\right],\label{HamTau}
\end{eqnarray}
 where we have taken into account the prescribed trajectory of the
detector
$\xi =0,\;(\rho = a^{-1})$. The operators $A$ and $A^{\dag}$ are
annihilation and
creation operator acting on the two dimensional Hilbert space generated by
the two
quantum states $up$ and $down$ of the ion~:
\begin{equation}
A\vert up\rangle=\vert down\rangle\qquad,
\qquad A\vert down \rangle=0 \qquad,\qquad
A^{\dag}\vert down\rangle=\vert up\rangle\qquad,\qquad
A^{\dag}\vert up\rangle=0\qquad .
\end{equation}
The energy difference between these two levels  is given by the mass gap
$\Delta M>0$. Indeed, for the uniformly accelerated detector following the
trajectory $\xi = 0$, the local rest frame time ( the detector proper
time ) coincides with the Rindler $\tau$ time,
fixing the conjugate energy to be the rest frame energy difference between
 the two levels.
 The integral phase factor $\int
A_\mu\;dx^\mu$  insures  gauge invariance of the interaction. It has to be
computed along the trajectory of the detector, from an arbitrary origin to
the position of the detector at   interaction proper time
$\tau$. For the model we consider here, it reduces to
$\frac{qE\tau}{2a}$, up to an arbitrary constant of integration.

 Before discussing this model, we would like to recall the reader a few
results
\cite{PaMa} concerning the simpler situation where the exchanged agent is
massless and chargeless. In this case the field can be expressed, on \tR, as a
superposition of Unruh $U$ and $V$ modes (see ref. \cite{BMPS} for details)~:
\begin{equation}
\hat\phi_R=\hat \phi^U_R+\hat\phi^V_R
=\int_0^{\infty}d\omega\;\left(a^U_\omega\;\varphi^U_\omega+a^V_\omega
\varphi^V_\omega\right) + \mbox{\rm{herm.\ conj.}}
\end{equation} The detector transitions, when the field state is the vacuum,
can only occur by  emission of Minkowskian neutral quanta. At first order of
perturbation, the probability amplitudes of such transitions from the
lower mass state of the detector to the higher one (denoted by $B$) or the
converse
(denoted by $A$) via the emission of one quantum associated to an Unruh $U$
mode are
given by~:
\begin{equation}
\begin{array}{lccl} B(\omega;\Delta M)&=&\langle O_M | \langle + |a_\omega
{\cal T}_\tau e^{-i \int d\tau {\cal H}^int} |-\rangle|0_M \rangle&= -i
\,\tilde g
\sqrt{\frac {\pi}{\vert\omega\vert}}\alpha_\omega  \delta(\frac {\Delta M
}a +\frac \omega a)\\ &&&\nonumber\\ A(\omega;\Delta M)&=&\langle O_M|
\langle - |a_\omega {\cal T}_\tau e^{-i \int d\tau {\cal H}^int}
|+\rangle|0_M \rangle&=-i\, \tilde g
\sqrt{\frac {\pi}{\vert\omega\vert}}\alpha_\omega  \delta(\frac {\Delta M}
a -\frac \omega a)
\end{array}
\end{equation} where $\alpha_\omega = 1/\sqrt{(1-e^{-\frac
{2\pi\;\omega}a})}$ and ${\cal T}_\tau$ denotes the time ordering
operator with respect to the detector proper time. Its is interesting to
compare these formula with their analog where the emitted quanta are
those associated to Minkowski modes of energy $k$ (eq. (2.48) of ref.
\cite{BMPS}). In the latter case the instant of emission is given by
$\tilde\tau(k)=a^{-1}
\ln k/\Delta M $ with a width $1/\sqrt{\Delta M}$. When the  emitted
quanta are in pure Unruh's states, their ``energy" are fixed ($\omega =-\Delta
M$), but the process becomes completely delocalized in time. In other words the
Rindler ``energy'' balance is actually a ``boost'' constant of motion
conservation relation.
When recoils effects are taken into account but the classical picture of
the detector still
valid, the ion jumps from one hyperbola to another whose centers are such
that their
difference in localization remains in accord with  eq.(\ref{Omegacentre})
for the different
values of
$\omega$.
Summing the square of these amplitudes and interpreting as
usual the $\delta(0)$, we obtain the  detector excitation probability per
unit of proper time~:
\begin{equation} PB(\Delta M)=\int d\omega \vert B(\omega;\Delta
M)\vert^2={\tilde g}^2  \frac{\pi}{\Delta
M}\frac 1{({ e^{\frac {2\pi\;\Delta M}a}-1})}
\end{equation} and similarly for the desexcitation probability per unit
of proper time~:
\begin{equation} PA(\Delta M)= {\tilde g}^2  \frac{\pi}{\Delta
M}\frac 1{{ (1-e^{-\frac {2\pi\;\Delta M}a})}}
\end{equation} The ratio of these  probabilities is:
\begin{eqnarray}
\frac {PB(\Delta M)}{PA(\Delta M)}=e^{-\frac{2\pi}a \Delta M}
\end{eqnarray} which reflects the thermal equilibrium of the detector and
the radiation, at Unruh temperature $a/2\pi$.

Now let us repeat the same calculation, but in the framework of a
charged exchanged agent. First we compute the probability amplitude of
spontaneous excitation of the detector by emission of an
$\Omega$ (resp $\varpi$) antiparticle of quantum number
$\omega$. Expressing the quantum field $\hat \phi$ as a superposition of
Unruh modes, we obtain at first order in the coupling constant~:
\begin{eqnarray} B({\{{\Omega_\omega}\atop{\varpi_\omega}}; \Delta M)&=&
-i\tilde g a\langle 0,Mink,out\vert b^{out}_{\left\{ {\Omega}\atop
{\varpi}
\right.}(\omega)\int {\cal H}^{int}(\tau)d\tau
\vert 0,Mink,in\rangle \\ \nonumber. &=&-i\frac{{\tilde g}\, a}{N}\int d\tau
e^{i\tau(\Delta M-\frac{qE}{2a})}\left\{ {\Omega^{a \,out
\,*}_{\omega}}\atop{\varpi^{a
\,out
\,*}_\omega}\right. - i\frac{{\tilde g} \,a}{NqE}\frac \delta \gamma\int d\tau
e^{i\tau (\Delta M-\frac{qE}{2a})}\nonumber \\
&&\int_{-\infty}^{\infty}d\sigma
 \int_{-\infty}^{\infty}d\tilde{\omega}\, \left\{ \alpha^{out\,* }_{\Omega
\varphi}(\sigma,{\tilde{\omega}})
\epsilon^{out}_{\left\{ \Omega  \atop \varpi
\right.\varphi}(\sigma,{\omega})\Omega^{p \, out}_{\tilde{\omega}}+
\alpha^{out\,* }_{\varpi
\varphi}(\sigma,{\tilde{\omega}})
\epsilon^{out}_{\left\{ \Omega \atop \varpi
\right.\varphi}(\sigma,{\omega})\varpi^{p \,
out}_{\tilde{\omega}}\right\}\label{Bcom}\qquad .
\end{eqnarray} Here $N=\braket{0,Mink,out}{0,Mink,in}$ is the
normalization factor (\ref{videoutin}), $\delta/\gamma$ is defined by eq;
(\ref{gamdel}) and
$\alpha^{out}_{\left\{
\Omega
\atop
\varpi
\right.\varphi}$ and $\epsilon^{out}_{\left\{ \Omega  \atop \varpi
\right.\varphi}$ are the coefficients of the (non frequency mixing)
 Bogoljubov transformation between Unruh and Minkowski modes
(\ref{Omegapout}
\ref{varpiaout}).  Finally, expressing the  Unruh modes in terms of
Rindler modes (eqs
\ref{outoutRL}), the $\tau$ integration in eq. (\ref{Bcom}) becomes
trivial and gives $\delta$ functions. So we obtain the excitation
amplitude probability $B({\Omega_{\omega}};\Delta M)$ of the detector by
emission of an
${\Omega_{\omega}}$ quantum  as the sum of
\begin{equation} -i\frac{{\tilde
g}\sqrt{2\pi}}{N}\epsilon_{\Omega\,{\cal{V}}_R}^{out}{\cal{N}}({\cal{V}}_{
\omega\,R}^{out})W_{-i({\omega\over
2 a} - {m^2 \over 2 qE}), -i {\omega\over 2a}} \left[{i qE \over2a^2} \right]
\delta (\frac{\Delta M}a-\frac{qE}{2a^2}-\frac\omega a).
\end{equation} and the integral~:
\begin{eqnarray}
 &&-i\frac{{\tilde g}\sqrt{2\pi}}{NqE}\frac \delta \gamma
\int_{-\infty}^{\infty}d\tilde{\omega}\int_{-\infty}^{\infty}d\sigma\,\delta
(\frac{\Delta M}a-\frac{qE}{2a^2}-\frac {\tilde{\omega}}a)\nonumber\\
&&\left\{\alpha^{out\,* }_{\Omega
\varphi}(\sigma,{\tilde{\omega}})\epsilon^{out }_{\Omega
\varphi}(\sigma,\omega)\alpha^{out}_{\Omega{\cal{V}}_R}({\tilde{\omega}})
{\cal{N}}({\cal{V}}_{{\tilde{\omega}},R}^{out})W_{-i({{\tilde{\omega}}\over
2 a} - {m^2 \over 2 qE}), -i {{\tilde{\omega}}\over 2a}}
\left[{i qE
\over2a^2} \right]+\right.\nonumber \\ &&\left.\alpha^{out\,* }_{\varpi
\varphi}(\sigma,{\tilde{\omega}})\epsilon^{out }_{\Omega
\varphi}(\sigma,\omega)\alpha^{out}_{\varpi{\cal{U}}_R}({\tilde{\omega}})
{\cal{N}}({\cal{U}}_{{\tilde{\omega}}\,R}^{out})M_{i({{\tilde{\omega}}\over
2 a} - {m^2 \over 2 qE}), i {{\tilde{\omega}}\over 2a}} \left[-{i qE
\over2a^2} \right]\right\}\qquad .
\end{eqnarray} Using the relations (\ref{AAint}) and (\ref{AAintbis}) for integrals of
products of
$A(\sigma,\omega)$ functions, it easy to show that the
$\sigma$ integral gives a
$\delta(\omega-\tilde\omega)$ function; as a consequence  the  integral over
$\tilde\omega$ is trivial to carry out. So we get for the second term
in the expression (\ref{Bcom}) of $B({\Omega_{\omega}};
\Delta M)$~:
\begin{eqnarray} &&-i\frac{{\tilde g}\sqrt{2\pi}}{N}\frac \delta
\gamma\delta (\frac{\Delta M}a-\frac{qE}{2a^2}-\frac
{\omega}a)\Gamma(\frac 12-i\frac\omega
a+i\frac{m^2}{2qE})\frac1{\sqrt{2\pi}}\nonumber\\ &&\left\{e^{i\frac \pi
2}e^{\frac{\pi \omega}{2a}}e^{-\frac{\pi
m^2}{4qE}}\alpha^{out}_{\Omega{\cal{V}}_R}({{\omega}}){\cal{N}}({\cal{V}}_{{
{\omega}},R}^{out})W_{-i({\omega\over 2 a} - {m^2 \over 2 qE}), -i
{\omega\over 2a}} \left[{i qE \over2a^2} \right]+\right.\nonumber \\
&&\left.e^{-i\psi}e^{-\frac{\pi \omega}{2a}}e^{\frac{\pi
m^2}{4qE}}\alpha^{out}_{\varpi{\cal{U}}_R}({{\omega}}){\cal{N}}({\cal{U}}_{{
{\omega}}\,R}^{out})M_{i({{{\omega}}\over 2 a} - {m^2
\over 2 qE}), i {{{\omega}}\over 2a}} \left[-{i qE \over2a^2}
\right]\right\}\qquad ,\nonumber\\
\end{eqnarray} which, once the explicit expressions of the coefficients
$\epsilon_{\Omega\,{\cal{V}}_R}^{out}$ and
$\alpha^{out}_{\Omega{\cal{V}}_R}$ (eqs (\ref{coeffoutout})) are
explicited, lead to:
\begin{eqnarray} B({\Omega_{\omega}}; \Delta M)&=&-i\frac{{\tilde
g}\sqrt{2\pi}}{2N}\delta (\frac{\Delta M}a-\frac{qE}{2a^2}-\frac
{\omega}a)(\frac{qE}a)^{-\frac12}\frac{e^{-\frac{3\pi
\omega}{4a}}e^{\frac{\pi m^2}{4qE}}}{\cosh\pi(\frac\omega
a-\frac{m^2}{2qE})}\nonumber\\ &&\frac{\Gamma(\frac
12+i\frac{m^2}{2qE})}{\Gamma(\frac 12+i\frac\omega
a-i\frac{m^2}{2qE})}W_{i({{{\omega}}\over 2 a} - {m^2 \over 2 qE}), i
{{{\omega}}\over 2a}} \left[-{i qE \over2a^2} \right].\label{BOmega}
\end{eqnarray} In the same way we obtain the excitation amplitude via
emission of a
$\varpi_\omega$ antiparticle~:
\begin{eqnarray} B({\varpi_\omega}; \Delta M)&=&-i\frac{{\tilde
g}\sqrt{2\pi}}{NqE}\frac
\delta \gamma
 \int_{-\infty}^{\infty}d\tilde{\omega}\int_{-\infty}^{\infty}d\sigma\,
\delta (\frac{\Delta M}a-\frac{qE}{2a^2}-\frac
{\tilde{\omega}}a)\nonumber\\ &&\left\{\alpha^{out\,* }_{\Omega
\varphi}(\sigma,{\tilde{\omega}})\epsilon^{out
}_{\varpi\varphi}(\sigma,\omega)\alpha^{out}_{\Omega{\cal{V}}_R}({
\tilde{\omega}}){\cal{N}}({\cal{V}}_{{\tilde{\omega}},R}^{out})W_{-i({{\tilde{\o
mega}}
\over 2 a} - {m^2 \over 2 qE}), -i {{\tilde{\omega}}\over 2a}} \left[{i qE
\over2a^2} \right]+\right.\nonumber \\ &&\left.\alpha^{out\,* }_{\varpi
\varphi}(\sigma,{\tilde{\omega}})\epsilon^{out }_{\varpi
\varphi}(\sigma,\omega)\alpha^{out}_{\varpi{\cal{U}}_R}({\tilde{\omega}})
{\cal{N}}({\cal{U}}_{{\tilde{\omega}}\,R}^{out})M_{i({{\tilde{\omega}}\over
2 a} - {m^2 \over 2 qE}), i {{\tilde{\omega}}\over 2a}} \left[-{i qE
\over2a^2} \right]\right\}\nonumber\\
\end{eqnarray} which, again using the relations (\ref{AAint}, \ref{AAintbis}), lead to~:
\begin{eqnarray} &&B({\varpi_\omega}; \Delta M)=-i\frac{{\tilde
g}\sqrt{2\pi}}{2N}\delta (\frac{\Delta M}a-\frac{qE}{2a^2}-\frac
{\omega}a)(\frac{qE}a)^{-\frac12}e^{-i\psi}\nonumber\\ &&\frac{e^{i\frac
\pi 2}e^{\frac{\pi \omega}{4a}}e^{-\frac{\pi
m^2}{4qE}}}{\cosh\pi(\frac\omega
a-\frac{m^2}{2qE})}\frac{\Gamma(\frac12+i\frac{m^2}{2qE})}{\Gamma(\frac12-i\frac{m^2}{2qE}+i\frac\omega a)} W_{i({{{\omega}}\over 2 a} - {m^2 \over 2
qE}), i {{{\omega}}\over 2a}} \left[-{i qE \over2a^2} \right].
\end{eqnarray}

Similar calculations lead the  desexcitation amplitude of probability of
the detector, by emission of a particle of type $\Omega$ or $\varpi$~:
\begin{eqnarray} A({\Omega_\omega}; \Delta
M)&=&-i\frac{{\tilde g}\sqrt{2\pi}}{2N}\delta(\frac {\Delta M} a -
\frac{qE} {2a^2}+\frac
\omega a)(\frac{qE}a)^{-\frac12}
\frac{\Gamma(\frac12+i\frac{m^2}{2qE})}{\Gamma(\frac12+i\frac\omega
a-i\frac{m^2}{2qE})}\nonumber\\ &&\frac{e^{\frac{\pi
\omega}{4a}}e^{\frac{\pi m^2}{4qE}}}{\cosh\pi(\frac\omega
a-\frac{m^2}{2qE})}
 W_{i({{{\omega}}\over 2 a} - {m^2 \over 2 qE}), i {{{\omega}}\over 2a}}
\left[-{i qE \over2a^2} \right],\label{norecoilAO}
\end{eqnarray} and:
{\par\noindent\footnotesize
\begin{eqnarray*}
&&A({\varpi_\omega}; \Delta M)=-i\frac{{\tilde
g}\sqrt{2\pi}}N\delta(\frac {\Delta M} a - \frac{qE} {2a^2}+\frac \omega
a)(\frac{qE}a)^{-\frac12}\left(\frac{\vert\sinh\frac{\pi
\omega}a\vert}{{\cosh\pi(\frac\omega
a-\frac{m^2}{2qE})}}\right)^{\frac12}\varepsilon(\omega)\vert\frac\omega
a\vert^{-\frac12}\Gamma[1-i\frac\omega a]\nonumber\\
&&\left\{\frac
1{\Gamma(\frac12-i\frac{m^2}{2qE})}\frac{e^{\frac{3\pi
m^2}{4qE}}e^{-\frac{3\pi \omega}{4a}}}{2\cosh\frac{\pi
m^2}{2qE}}W_{i({\omega\over 2 a} - {m^2 \over 2 qE}), i {\omega\over 2a}}
\left[-{i qE \over 2a^2}
\right]+\right.\nonumber\left.\frac 1{\Gamma(\frac12-i\frac\omega
a+i\frac{m^2}{2qE})}{e^{i\frac \pi 2}e^{\frac{\pi m^2}{4qE}}e^{-\frac{\pi
\omega}{4a}}}W_{-i({\omega\over 2 a} - {m^2 \over 2 qE}), -i {\omega\over
2a}} \left[{i qE \over 2a^2}  \right]\right\}\ \ .\label{norecoilAv}
\end{eqnarray*}}
\par\noindent\normalsize On the other hand let us notice that the
transition amplitudes due to absorption of quanta~:
\begin{eqnarray} B^{\prime}({{\Omega_\omega}\atop{\varpi_\omega}}; \Delta
M)&=&\langle+\vert\langle 0,Mink,out\vert {\cal{T}}e^{-i\int d\tau \,{\cal
H}^{int}(\tau)}a^{in\,\dag}_{\left\{ {\Omega}\atop {\varpi}
\right.}(\omega)\vert 0,Mink,in\rangle\vert-\rangle\nonumber\\
\\ A^{\prime}({{\Omega_\omega}\atop{\varpi_\omega}}; \Delta
M)&=&\langle-\vert\langle 0,Mink,out\vert {\cal{T}}e^{-i\int d\tau
\,{\cal H}^{int}(\tau)}b^{in\,\dag}_{\left\{ {\Omega}\atop {\varpi}
\right.}(\omega)\vert 0,Mink,in\rangle\vert+\rangle\nonumber\\
\end{eqnarray} are directly related to the previous ones by~:
\begin{equation}
B^{\prime}({{\Omega_\omega}\atop{\varpi_\omega}}; \Delta
M)=A({{\Omega_\omega}\atop{\varpi_\omega}}; \Delta M)\qquad \mbox{\rm and}
\qquad
A^{\prime}({{\Omega_\omega}\atop{\varpi_\omega}}; \Delta
M)=({{\Omega_\omega}\atop{\varpi_\omega}}; \Delta M)\qquad .
\end{equation}
These last relation can be verified by repeating the previous calculation,
or more
directly using {\bf CPT} transformation arguments \cite{PaMa}.\\
The occurrence of the Dirac distribution $\delta(\frac {\Delta M} a -
\frac{qE} {2a^2}+\frac
\omega a)$ in all these amplitudes  makes trivial the estimation of
probabilities of transition per unit of proper time. With obvious
notation, $T$ beeing a large detector proper time interval, we obtain ~:
\begin{eqnarray}
PB(\Omega,\Delta M) & = & T^{-1}\int \vert B(\Omega_\omega;\Delta
M)\vert^2\, d\omega
\quad =\quad \vert
{\cal{C}}\vert ^2 \left\vert e^{-\frac{3\pi{\omega_\star}}{2a}}\vert
W_{i({{\omega_\star}\over 2 a} - {m^2 \over 2 qE}), i {{\omega_\star}\over 2a}}
\left[-{i qE
\over 2a^2}\right]\right\vert^2 \\
PB(\varpi ,\Delta M)&=&\vert
{\cal{C}}\vert^2e^{\frac{\pi{\omega_\star}}{2a}}e^{-\pi\frac{m^2}{qE}}\left\vert
W_{i({{\omega_\star}\over 2 a} - {m^2 \over 2 qE}), i {{\omega_\star}\over 2a}}
\left[-{i qE \over 2a^2}\right]\right\vert^2\label{PBvarpi}\\ PA(\Omega ,\Delta M)&=&\vert
{\cal{C}}\vert^2e^{\frac{\pi{\omega_\star}}{2a}}\left\vert
W_{i({{\omega_\star}\over 2
a} - {m^2 \over 2 qE}), i {{\omega_\star}\over 2a}} \left[-{i qE \over
2a^2}\right]\right\vert^2\\ PA(\varpi ,\Delta M)&=&\vert {\cal{C}}\vert^2
e^{-\frac{\pi{\omega_\star}}{2a}}\frac
\pi{\cosh\pi\frac{m^2}{2qE}}
\left|W_{i({{\omega_\star}\over 2 a} - {m^2 \over 2 qE}), i
{{\omega_\star}\over 2a}}
\left[-{i qE \over 2a^2}  \right]\right|^2
\nonumber\\ &&\left|\frac{1}{\Gamma(1-i\frac{\omega_\star}
a)}\frac{M_{-i({{\omega_\star}\over 2 a} - {m^2 \over 2 qE}), -i
{{\omega_\star}\over
2a}} \left[{i qE \over 2a^2}
\right]}{W_{i({{\omega_\star}\over 2 a} - {m^2 \over 2 qE}), i
{{\omega_\star}\over
2a}}
\left[-{i qE \over 2a^2}  \right]}+\frac{e^{-i\frac\pi
2}e^{-\pi\frac{m^2}{2qE}}}{\Gamma(\frac12-i\frac{\omega_\star}
a+i\frac{m^2}{2qE})}\frac{W_{-i({{\omega_\star}\over 2 a} - {m^2 \over 2
qE}), -i
{{\omega_\star}\over 2a}} \left[{i qE \over 2a^2}
\right]}{W_{i({{\omega_\star}\over
2 a} - {m^2 \over 2 qE}), i {{\omega_\star}\over 2a}} \left[-{i qE \over 2a^2}
\right]}\right|^2\label{modAvarpi}\nonumber\\
\label{PAvarpi}
\end{eqnarray} with ${\omega_\star}\equiv \Delta M-qE/2a$  fixed by the mass
shell condition dictated by the delta functions and where we have set
$\vert{\cal{C}}\vert^2= {\tilde g}^2
\frac{{\pi}}{2N^2}(\frac{qE}a)^{- 1} e^{\pi\frac{m^2}{2qE}}/\left(\cosh
\pi(\frac{\omega_\star} a-\frac {m^2}{2qE})\cosh\pi\frac{m^2}{2qE}\right)$\\ In
the limit of small charge exchange (
${{qE} \over {m^2}} \mapsto 0$), approximating, in the limit $(qE\mapsto 0)$,
Whittaker functions  by Bessel functions (see ref. \cite{AbSt}, section
{\bf 13.3})~:
\begin{eqnarray*}
\Gamma(\frac12+i\frac{\omega_\star}
a-i\frac{m^2}{2qE})W_{-i({{\omega_\star}\over 2
a} - {m^2 \over 2 qE}), -i {{\omega_\star}\over 2a}} \left[{i qE \over 2a^2}
\right]&\approx& 2(\frac m{2a})^{i\frac{\omega_\star}
a}(\frac{iqE}{2a^2})^{\frac12-i\frac{\omega_\star}{2a}}\;K_{-i\frac{\omega_\star
}
a}(\frac m a)
\\
\frac{1}{\Gamma(\frac12-i\frac{\omega_\star} a)}M_{-i({{\omega_\star}\over
2 a} - {m^2
\over 2 qE}), -i {{\omega_\star}\over 2a}} \left[{i qE \over 2a^2}
\right]&\approx&(\frac m{2a})^{i\frac{\omega_\star}
a}(\frac{iqE}{2a^2})^{\frac12-i\frac{\omega_\star}{2a}}\;I_{-i\frac{\omega_\star
}
a}(\frac m a)
\end{eqnarray*}
it is easy to see that the probabilities
$PA(\Omega ,\Delta M)$ and
$PB(\Omega ,\Delta M)$ are of order one compared to the probabilities
$PA(\varpi ,\Delta M)$ and
$PB(\varpi ,\Delta M)$  which of order $e^{-\pi{m^2}/{qE}}$.
Indeed an amplitude such as $B(\varpi_\omega;\Delta M)$ corresponds to the
probability amplitude of excitation of  the detector by absorption of a quantum
described by a
$\varpi_\omega$ mode, i.e. a rindlerian mode ${\cal U}^{out}_R$ on the \TR
quadrant.
But, as discussed at the end of the previous section (see eq.
(\ref{npUoutR})), such
modes corresponds to fluctuations produced by the Schwinger mechanism, and
in the
limit of vanishing charge, their populations become exponentially dampen (here
we suppose that $qE/m^2<<1$) compared to the populations of fluctuations
due to the
Unruh mechanism. Accordingly, the probabilities of transition involving
such modes
are also exponentially small, compared to those refereeing to process
involving the
modes created by the Unruh mechanism. This illustrate how the two kinds of
creation
process cooperate  to the physics of the detector. This situation is
similar to the
physics of the Reissner-Nordstr\"om black-holes which  start to loose their
charge  by
the Schwinger mechanism, and after by  Hawking evaporation
\cite{Gi}.
In the limit of small
charges we obtain as ratio of the probabilities of transition the
thermodynamical
equilibrium expression~:
\begin{eqnarray}
 \frac{PB(\Delta M)}{PA(\Delta M)}=
\frac {[PB(\Omega;\Delta M)+PB(\varpi;\Delta M)]}{[PA(\Omega;\Delta
M)+PA(\varpi;\Delta M)]}
\approx e^{-\frac{2\pi}a(\Delta M -\frac {qE}{2a})}
\end{eqnarray} whose meaning is discussed at length in \cite{GMPS}. On fig.
({\bf 5})
we have plot this ratio of total probabilities
divided by the Boltzmanian factor $\exp[-(2\,\pi(\Delta M/a-qe/2a^2)\equiv
\exp[-2\,\pi\,\omega_\star]$. This figure illustrate the role played by the
Schwinger
factor which control the instability of the quantum state of exchanged
quanta. For
large mass or, equivalently small charge, we recover the usual Boltzman
equilibrium
formula, otherwise new  equilibrium relations  have to be considered. Once
again the
same physics occurs when we consider  a quantum state in a strong gravitational
field. For instance if a black hole is enclosed in a box, it may reach an
equilibrium configuration with surrounding radiation. However the radiation
alone will
not obey a local equation of state linking energy density and pressure, but
it is
only the whole system that will satisfy an equilibrium relation involving
its total
mass energy,  the geometry of the box, etc.

In ref \cite{GMPS} we also obtained expressions for the transition
amplitudes of an accelerated ion, coupled to a massive and charged scalar
field $\Phi_m$ of mass $m$ and charge $q=Q-Q'$, when its recoil is taken
into account. The two level ion was modeled by introducing two scalar
fields
$\Psi_M$ and
$\Psi_M'$, of masses
$M$ and $M'$ and charges $Q$ and $Q'$. The transitions were dictated
through the interaction hamiltonian~:
\begin{eqnarray} H^{int}=g\int dx\left( \Psi _M^{\dagger }\Psi _{M'}\Phi
_{m}+\Psi _M\Psi _{M'}^{\dagger }\Phi _{m}^{\dagger }\right)
\label{hamint2}\qquad .
\end{eqnarray} We get for the desexcitation  amplitude~:
\begin{eqnarray}  {\cal{A}}(k|k',k'')&\equiv&\langle 0,out\mid a_{m}
^{out}(k^{\prime \prime })\ a_{M'}^{out}(k^{\prime }){\cal T} e^{(-i)\int
dt H^{int}} a_M^{in\dagger }(k)\mid 0,in\rangle \nonumber\\ &=&{\mbox{\rm
C}}\,
\delta (k-k^{\prime }-k^{{\prime }{\prime }}) \Ar\label{Akkk}
\end{eqnarray} which at first order of a perturbation expansion reduces to
\begin{eqnarray}
\Ar & =&\frac {\sqrt{2\pi} e^{-\frac{i\pi }4}e^{-\frac \pi 2
(\epsilon_{M'}+\epsilon_{m})} e^{\frac{\pi{\cal{E}}}4} e^{\frac
i2(\frac{k^2}{QE}\,-\frac{k^{\prime\,2}}{Q'E}-\frac{k^{\prime\prime}\,^2}{qE
})} e^{\frac{i{\cal Q} ^2}2}} { (2QE)^{1/4}(2Q'E)^{1/2}(2 qE)^{1/2}
\Gamma (i\epsilon _m+1/2)\Gamma (i\epsilon _\mu+1/2)}\nonumber\\
&&\Gamma(i{\cal{E}}+\frac 12)(\frac Qq)^{\frac i2\epsilon_{\mu}} (\frac Q
\alpha)^{\frac i2\epsilon_{M'}}{\cal{I}}\label{Acharged}
\end{eqnarray} which $\cal I$ denoting the complicated expression:
\footnotesize
\begin{eqnarray} {\cal{I}} &=& \left\{
\left(\frac{\sqrt{qE}}{\sqrt{Q'E}}\right)^{\frac12+i\epsilon_M}
D_{-i{\cal{E}}-\frac 12}\left[+ \sqrt{2}e^{i\frac{\pi} 4}{\cal Q} \right]
B(i\epsilon _{m} +\frac 12,-i{\cal{E}}+\frac 12)  _2F_1(i\epsilon
_M+\frac 12,i\epsilon _{m} +\frac 12,1+i\epsilon _M-i\epsilon
_{M'};-\frac{qE }{Q'E})   \right.  \nonumber \\ && +
\left(\frac{\sqrt{Q'E}}{\sqrt{qE}}\right)^{\frac12+i\epsilon_M}\left.
D_{-i{\cal{E}}-\frac 12}\left[
\sqrt{2}e^{-i\frac{\pi} 4}{\cal Q} \right]  B(i\epsilon _{M'}+\frac
12,-i{\cal{E}}+\frac 12)\
  _2F_1(i\epsilon _M+\frac 12,i\epsilon _{M'}+\frac 12,1+i\epsilon
_M-i\epsilon _{m} ;-\frac{Q'E}{qE })\right\}\nonumber\\
\label{AI}
\end{eqnarray}
\normalsize and where we have set:
\begin{eqnarray} {\cal Q}&=&\frac{Q'E \,k-QE\, k^{\prime}}
{\sqrt{QEQ'EqE}}\qquad,\\
{\cal{E}}&=&\epsilon_{M'}+\epsilon_{m}-\epsilon_M
\end{eqnarray} while the constant ${\mbox{\rm C}}$ is related to the
normalization factors and Bogoljubov coefficients through~:
\begin{equation} {\mbox{\rm C}} = {-ig \over \sqrt{ 2 \pi}}
\frac 1{{\alpha _M{\cal{N}}_M} {\alpha _{M'}{\cal{N}}_{M'}} {\alpha _{m}
{\cal{N}}_{m}}}\qquad .
\end{equation} Similarly, the amplitude of spontaneous excitation for the
two level ion, defined as:
\begin{eqnarray} {\cal{B}}(k^{\prime }\vert k,-k^{{\prime }{\prime
}})&=&\langle 0,out\mid a_{m} ^{out}(k^{\prime \prime })\ b_M^{out}(-k)\
{\bf {\tau}}e^{-i\int dt H^{int}} b_{M'}^{in\dagger }(-k^{\prime})\mid
0,in\rangle  \nonumber \\
\label{GenB22}
\end{eqnarray} was obtained into the form:
\begin{eqnarray} {\cal{B}}(k^{\prime }\vert k,-k^{{\prime }{\prime
}})&=&{\mbox{\rm C}}\,
\delta (k-k^{\prime }-k^{{\prime }{\prime }}) \Br
\end{eqnarray} with at first order:
\footnotesize
\begin{eqnarray}
\Br&=& (\frac qQ)^{\frac i2(\epsilon_{m}-i\epsilon_M)} (\frac
{\alpha}Q)^{\frac i2(\epsilon_{M'}-i\epsilon_M)}
\frac {e^{-\frac \pi 2 (\epsilon_{M'}+\epsilon_{m})} e^{\frac{i\pi
}4}\sqrt{2\pi} e^{\frac i 2(\frac{k^2}{QE}\,-\frac{k^{\prime
\,2}}{Q'E}+\frac{k{\prime\prime\,^2}}{qE })}}
{{(2QE)^{3/4}(2Q'E)^{1/4}(2qE)^{1/4}} \Gamma (i\epsilon _m+1/2)\Gamma
(i\epsilon _{M'} +1/2)}\nonumber \\  &&\Gamma (i{\cal{E}}+\frac 12)
e^{\frac{i{\cal Q} ^2}2} e^{\frac{\pi {\cal{E}}}8}e^{-\frac{i\pi }8}
D_{-i{\cal{E}}-\frac 12}\left[ \sqrt{2}e^{i\frac {5\pi} 4}{\cal Q}
\right]\nonumber\\  &&\left\{ {e^{-\frac{3i\pi}8}}B(i\epsilon _{m} +\frac
12,-i\epsilon_M+\frac 12)\;\/_2F_1(i{\cal{E}} +\frac 12,\frac
12-i\epsilon _M ,1+i\epsilon _{m}-i\epsilon _M;\frac{Q'E}{Q E }) \right.
\nonumber \\
 & & +\left. {e^{-{\pi}
\epsilon_M}e^{\frac{i\pi}8}}B(i\epsilon _{M'} +\frac
12,-i\epsilon_M+\frac 12)\;
\/ _2F_1(i{\cal{E}} +\frac 12,\frac 12-i\epsilon _M ,1+i\epsilon
_{M'}-i\epsilon _M;\frac{qE }{QE})\right\}\quad.\nonumber\\
\label{Bcharged}
\end{eqnarray}
\normalsize

These amplitudes were obtained by using modes solving the filed equation
by separation of variables in the gauge $A=-Etdz$, modes such that their
spatial dependence are of the form $e^{ikz}$. To make contact between
these amplitudes and those obtain in the framework of the detector model
we first have to change our Fock space basis in order to use, in both
situations, the same quantum numbers to label the states. These $k$ modes,
gauge transformed so as to solve eq. (\ref{feqxt}),  are related to the
modes Minkowski modes introduced in eq. (\ref{Dmodein}) by the (trivial)
Bogoljubov transformation:
\begin{eqnarray}
\varphi_k^{p\,in\;}&=&\frac{e^{i\frac{3\pi}8}}{\sqrt{2\pi Q'E}}\int
d\sigma e^{-i\frac{\sigma\;k}{Q'E}} \varphi_\sigma^{p\,
in}\nonumber\;\;\;\qquad ,\\
\varphi_{-k}^{a\,in\,*}&=&\frac{e^{-i\frac{3\pi}8}}{\sqrt{2\pi Q'E}}\int
d\sigma e^{-i\frac{\sigma\;k}{Q'E}} \varphi_\sigma^{a\,
in\,*\;}\nonumber\qquad ,\\\
\varphi_k^{p\,out}&=&\frac{e^{i\frac{-3\pi}8}}{\sqrt{2\pi Q'E}}\int
d\sigma e^{-i\frac{\sigma\;k}{Q'E}} \varphi_\sigma^{p\,
out}\nonumber\;\;\qquad ,\\\
\varphi_{-k}^{a\,out\,*}&=&\frac{e^{i\frac{3\pi}8}}{\sqrt{2\pi Q'E}}\int
d\sigma e^{-i\frac{\sigma\;k}{Q'E}} \varphi_\sigma^{a\, out\,*}\qquad .
\end{eqnarray}

In the following, we  only discuss the first order expansion of the
${\cal{A}}(k\vert k^\prime,k^{\prime\prime})$ amplitude~:
\begin{eqnarray} {\cal{A}}(k\vert k^\prime,k^{\prime\prime})&=&-i\langle
0,out\mid a_{m} ^{out}(k^{\prime \prime })\ a_{M'}^{out}(k^{\prime })
\int dt H^{int} a_M^{in\dagger }(k)\mid 0,in\rangle \nonumber\\ &=&-i\int
d\sigma d\sigma^\prime
d\sigma^{\prime\prime}a(\sigma^{\prime\prime},-k^{\prime\prime})
a(\sigma^{\prime},-k^{\prime})a(\sigma,k) \langle 0,out\mid a_{m}
^{out}(\sigma^{\prime \prime })\ a_{M'}^{out}(\sigma^{\prime })
\int dt H^{int} a_M^{in\dagger }(\sigma)\mid 0,in\rangle \nonumber\\
\end{eqnarray} and show how from this amplitude we may recover the
transition amplitude of the two level detector.

We shall evaluate  the limit of this ``3-field" amplitudes of transition
for $M,M'\to \infty$ with
$\Delta M/M\equiv (M-M')/M\ll 1$ and $\Delta m/a\ll 1$,  which corresponds
to the situation where the three field model mimics the heavy two level
detector, without recoil.   The detector states are given by sharply
localized states built out of the vacua $\ket{0_M}$ and $\ket{0_{M'}}$.
They are
obtained as~:
\begin{eqnarray}
\ket{+}={\int f_+(k) dk a_{M}^{\dagger}(k)\ket{0_M}\ket{0_{M'}}}\quad
,\quad \ket{-}={\int f_-(k^\prime) dk^\prime
a_{M^\prime}^{\dagger}(k^\prime)
\ket{0_M}\ket{0_M'}}
\end{eqnarray} and the operator $A^\dagger$, acting on this two
dimensional state space, is:
\begin{eqnarray} A^\dagger=\int f_+(k)a_{M}^{\dagger}(k)\;dk\int
f_-^*(k^\prime)a_{M'}^{\dagger}(k') dk^\prime
\end{eqnarray}
 These states are supposed to be normed ($\braket{+}{+}=\int f_+\;f_+^*\;
dk=1,\ \braket{-}{-}=\int f_-\;f_-^*\; dk^\prime=1,\ \braket{+}{-}=0$)
and, as in the large  mass limit, the Schwinger process vanishes we have
not to distinguish between $in$ or $out$ vacua $\ket{0_M}$ and
$\ket{0_{M'}}$. At first order of perturbation, the amplitude of
transition between these states becomes
\begin{eqnarray} {\cal{A}}(k^{\prime\prime})& =&{\int f_+(k) dk\int
f_-(k^\prime) dk^\prime {\cal{A}}(k\vert
k^\prime,k^{\prime\prime})}\label{Aintegrated}\\ &\simeq &-ig\bra{0,
m,out}a_m^{out}(k'')\int
dt\;dz\;\phi_M(t,z)\hat\Phi_m^*(t,z)\phi_{M^\prime}^*(t,z)\ket{0,m,in}
\end{eqnarray} where ${\cal{H}}_{int}(\tau)$ is given by eq.
(\ref{HamTau}) and where
$\phi_{M}(t,z)$ and $\phi_{M'}(t,z)$ are classical solution of the wave
equation, given by the superposition of modes weighted by the function
$f_{\pm}(k)$; for example: $\phi_{M}(t,z)=\int f_+(k)
\phi^{p}_k(t,z)\;dk$. Note that at this first order of the perturbation
expansion, a
necessary condition for
${\cal{A}}(k^{\prime\prime})$ to be non zero is that the wave packets
$\phi_{M}(t,z)$ and $\phi_{M'}(t,z)$ overlap, but at higher order virtual
detector states allow ''tunneling" transitions between non overlapping
configurations of the detector. The heavy detector corresponds to the
limit where these wave packets have their supports blend into a single
classical trajectory, here a hyperbola whose center is at the origin of the
$(t,z)$ coordinates. Such configurations can be obtained by using, in the
large mass limit, the maximally localized packets introduce in
\cite{BMPPS} or more directly just by considering suitable superpositions
of  W.K.B. approximate solutions of the wave equation~:
\begin{equation}
\phi^p_k(t,z)\approx \left(\frac{QEt}{M}\right)^{(i\frac{M^2}{2QE}-\frac
12)}e^{-ik(z-t)}e^{i\frac{QEt(t-z)}2}\qquad .
\end{equation}
 In the limit considered, we thus obtain:
\begin{eqnarray}
\phi_{M}^*(t,z)\; \phi_{M'}(t,z)&\propto &e^{\frac {\Delta M}{2a}\ln
at}\delta (\rho-a^{-1})
\nonumber\\ &=& e^{i \Delta M \;\tau } e^{-i\int A_\mu\;dx^\mu}\;\delta
(\rho-a^{-1})
\end{eqnarray} because $\Delta M \;a =(Q-Q')E=qE$. So we recover, from
the field description, the energy difference between the two detector
levels as predicted by the ''equivalence" principle. The amplitude  reduces
to~:
\begin{eqnarray} {\cal{A}}(k^{\prime\prime})&\simeq&-i \int
d\sigma^{\prime\prime}a(\sigma^{\prime\prime},-k^{\prime\prime})\langle
-\vert \langle 0,m,out\vert a_m^{out}(\sigma^{\prime\prime})\int d\tau
{\cal{H}}_{int}(\tau)\vert 0,m,in\rangle\vert+\rangle \qquad ,
\end{eqnarray} and using the  Bogoljubov transformation (\ref{outoutRL})
between Unruh and Rindler modes, we get :
\begin{eqnarray} {\cal{A}}(k^{\prime\prime})&\simeq&e^{-i\frac \pi 8}\int
\frac{e^{i\frac{\sigma k''}{qE}}}{\sqrt{2\pi qE}}
\left\{\alpha_{{\Omega}
\varphi}^{out}(\sigma,\omega)A({\Omega}_\omega;\Delta M)+\alpha_{\varpi
\varphi}^{out}(\sigma,\omega)A(\varpi_\omega;\Delta
M)\right\}\;\frac{d\sigma}{\sqrt{qE}}\;d\omega  \nonumber \\
\end{eqnarray} which, can be explicitly integrate (thanks to the integral
representations of Whittaker and parabolic cylinder functions) to give at
the end:
\begin{eqnarray} {\cal{A}}(k^{\prime\prime})&\simeq&- \frac{{ g}}{
N}\frac { a}{{(2q E)}^{3/4}} e^{\frac{\pi m^2}{8q E}}\frac{\Gamma[\frac
12+i\frac{m^2}{2q E}]}{\cosh [\pi(\frac{m^2}{2Q'E}-\frac{\mu}
a)]}\nonumber\\ &&\left\{D_{i\frac {\mu} a-i\frac{m^2}{2q E}-\frac
12}\left[-e^{i\frac
\pi 4}k^{\prime\prime}\sqrt{\frac 2{q E}}\right]\;
\frac{W_{-i(\frac {\mu}{2a}-\frac{m^2}{2q E}),
-i\frac{\mu}{2a}}\left[\frac{iq
E}{2a^2}\right]}{\Gamma[\frac12+i\frac{m^2}{2q E}]}+
\right.\nonumber\\ &&\left.i\;D_{i\frac {\mu} a-i\frac{m^2}{2q E}-\frac
12}\left[e^{i\frac \pi 4}k^{\prime\prime}\sqrt{\frac 2{q
E}}\right]e^{-\frac{\pi {\mu}}{2a}}\;\frac{M_{i(\frac
{\mu}{2a}-\frac{m^2}{2q E}), i\frac{m}{2a}}\left[-\frac{iq
E}{2a^2}\right]}{\Gamma[1+i\frac{\mu} a]}\right\}\label{Alim1}\qquad ,
\end{eqnarray} where ${\mu}=\Delta\;M-Q'E/2a$.

This result can also directly be obtained  from the expression
(\ref{Acharged}) of  the amplitude ${\cal A}(k|k',k'')$ by taking its
limit for infinite masses and charges. Indeed, in this
limit, the  amplitude factorizes into a term involving only the $k$ and
$k'$ momenta and a term that tends to eq. (\ref{Alim1}), the effective
coupling constant $\tilde g$ being then implicitly defined as the product
of $g$ times a double integral involving the functions $f_+^*(k)$ and
$f_-(k')$ and the $k$, $k'$ term that comes out from the full
amplitude.\\  First let us discuss the factor
$\cal I$ (eq.
\ref{AI}). The limits of the parabolic functions
$D$ are simply obtained by direct substitution: ${\cal Q}\mapsto
\frac{k^{\prime\prime}}{\sqrt{qE}}$ and
${\cal{E}}\mapsto -i\frac{\cal Q} a +i\frac{{m}^2}{2qE}$. For the
hypergeometric functions occurring in $\cal I$, we use its series
expansion and obtain~:
\begin{equation} {\lim_{M,m\to\infty} }\,_2F_1(i\epsilon _M+\frac
12,i\epsilon _{m} +\frac 12,1+i\epsilon _M-i\epsilon_{M'};-\frac{qE
}{Q'E})=e^{-i\frac{qE}{4a^2}}(-i\frac{qE}{2a^2})^{-\frac 12-i\frac{\cal
Q}{2a}}M_{i\frac{\cal Q}{2a}-i\frac{{m}^2}{2qE},i\frac{\cal
Q}{2a}}(-\frac{iqE}{2a^2})\qquad ,
\end{equation} while for the second, we first have to use the inversion
relation to pass from the argument $Q'/q$ to $q/Q'$ and the Stirling
formula to finally get:
\begin{eqnarray} &&{\lim_{M,m\to\infty} }\,_2F_1(i\epsilon_M+\frac
12,i\epsilon _m+\frac 12,1+i\epsilon _M-i\epsilon _{m} ;-\frac{Q'E}{qE
})={\lim_{M,m\to\infty} }\,e^{i\frac\pi 4}e^{\pi\frac{{m}^2}{4qE}}(\frac
{M^2}{2QE})^{\frac 12-i\frac{{m}^2}{2qE}}\nonumber\\
&&\left\{\frac{\Gamma(-i\frac {\cal Q} a)}{\Gamma(\frac
12-i\epsilon_{m})}(\frac{Q'E}{qE})^{-\frac12-i\epsilon_M}e^{-\frac{\pi
{\cal Q}}{2a}}(\frac {M^2}{2QE})^{i\frac {\cal Q}
a}e^{-i\frac{qE}{4a^2}}(-i\frac{qE}{2a^2})^{-\frac 12-i\frac{\cal
Q}{2a}}M_{i\frac{\cal Q}{2a}-i\frac{{m}^2}{2qE},i\frac{\cal
Q}{2a}}(-\frac{iqE}{2a^2})\right.\nonumber\\ &&\left.+\frac{\Gamma(i\frac
{\cal Q} a)}{\Gamma(\frac 12-i\epsilon_{m}+i\frac {\cal Q}
a)}(\frac{Q'E}{qE})^{-\frac12-i\epsilon_{M'}}e^{-i\frac{qE}{4a^2}}(-i\frac{q
E}{2a^2})^{-\frac 12+i\frac{\cal Q}{2a}}M_{i\frac{\cal
Q}{2a}-i\frac{{m}^2}{2qE},-i\frac{\cal Q}{2a}}(-\frac{iqE}{2a^2})\right\}
\end{eqnarray} Moreover this can further simplified  using eqs (\ref{A23})
and the limit of the
${\cal{A}}$ amplitude reads:
\begin{eqnarray} &&{\lim_{M,m\to\infty} }\,{-ig \over \sqrt{ 2 \pi}}
\frac 1{{\alpha _M{\cal{N}}_M} {\alpha _{M'}{\cal{N}}_{M'}} {\alpha _{m}
{\cal{N}}_{m}}}
\frac {\sqrt{2\pi} e^{-\frac{i\pi }4}e^{-\frac \pi 2
(\epsilon_{M'}+\epsilon_{m})} e^{\frac{\pi{\cal{E}}}4} e^{-\frac
{i}2(\frac{k^{\prime\prime}\,^2}{qE })} e^{\frac{i{\cal Q} ^2}2}} {
(2QE)^{1/4}(2Q'E)^{1/2}(2 qE)^{1/2}
\Gamma (i\epsilon _m+1/2)\Gamma (i\epsilon _{m}+1/2)}\nonumber \\
&&\Gamma(\frac 12+i\frac{{m}^2}{2qE}-i\frac {\cal Q} a)(\frac Qq)^{\frac
i2\epsilon_{{m}}} (\frac Q \alpha)^{\frac
i2\epsilon_{M'}}{\cal{I}}\nonumber\\ &=&{\lim_{M,m\to\infty} }\,{-g a
\over { 2 \pi}} (\mbox{\rm{phase}})\,\frac 1{{{\cal{N}}_{m}}}
\frac { 1} {(QE)^{1/4}(Q'E)^{3/4}(2 qE)^{3/4} }\Gamma(\frac
12+i\frac{{m}^2}{2qE}-i\frac {\cal Q} a)\nonumber \\ &&\Gamma(\frac
12-i\frac{{m}^2}{2qE}+i\frac {\cal Q} a)\Gamma(\frac
12+i\frac{{m}^2}{2qE})e^{\frac{\pi {m}^2}{8qE}} \nonumber\\
&&\left\{\frac{1}{\Gamma(1+i\frac {\cal Q} a)} e^{-\frac{\pi {\cal
Q}}{2a}}e^{i\frac \pi 2}D_{-i{\cal{E}}-\frac 12}\left[+
\sqrt{2}e^{i\frac{\pi} 4}\frac{k^{\prime\prime}}{\sqrt{qE}}
\right]M_{i\frac{\cal Q}{2a}-i\frac{{m}^2}{2qE},i\frac{\cal
Q}{2a}}(-\frac{iqE}{2a^2})\right.\nonumber\\ &&\left.+\frac 1
{\Gamma(\frac 12+i\frac{{m}^2}{2qE})}D_{-i{\cal{E}}-\frac 12}\left[
\sqrt{2}e^{-i\frac{\pi} 4}\frac{k^{\prime\prime}}{\sqrt{qE}} \right]
W_{-i\frac{\cal Q}{2a}+i\frac{{m}^2}{2qE},i\frac{\cal
Q}{2a}}(\frac{iqE}{2a^2})
 \right\}
\end{eqnarray} This is in perfect agreement with eq. (\ref{Alim1}) if the
coupling constants  $\tilde g$  and $g$ are connected by a
$(q,m)$-independent relation, dependent of the precise form of the weight
functions characterizing the detector.

So we may relate the terms of the amplitude of transition
(\ref{Acharged}) to Unruh modes, and we see that the physical reason that favor
one kind of modes with respect to the other found its root in that their
own existence
is a manifestation of the Schwinger mechanism and that their probabilities
of interaction is weighted by a Schwinger factor.
\section*{Conclusion}
In this paper, we have studied, in Rindler coordinates, the quantization of
a charged field interacting with a constant electric field. The main
characteristic of this problem is that it involves
 two acceleration parameters: the acceleration of the Rindler
observer (the detector of section {\bf {7}}) and the natural
acceleration of the charged quanta ($qE/m$). So we obtain a toy (but
exactly solvable) model for the quantization of a charged field in a
Reissner-Nordtr\"om black hole geometry, in the same way as the Unruh detector
mimics the physics around a Schwarschild black hole. It is that similarity
that constitutes the main motivation of our work.\\
Now, let us summarize the main points of our analysis:
\begin{itemize}
\item The quantization of a charged field in Rindler coordinates illustrates
 the ``symmetry breaking" between particles and antiparticles in the {\bf R}
and {\bf L} quadrants. Eqs (\ref{QRIGHT}) and (\ref{QLEFT}) show that the
Minkowskian vacuum state carries a positive charge on the {\bf R} and (of
course) the opposite charge on the {\bf L} quadrant. This was
expected, as antiparticles ({\em resp.} particles) are always obliged to
leave the  {\bf R} ({\em resp.} {\bf L}) quadrant, which is not necessarily the
case for particles ({\em resp.} antiparticles) that may stay inside for ever.
Nevertheless, although obvious, we find interesting to see how this property
emerges from the rules of quantum mechanics and is encoded in the wave
function of the various modes used. Let us note that when the quantization is
performed on the full Minkowski space, no such charge polarization effect
comes into evidence because in this framework operator expectation values
are obtained by averaging over the complete space instead of over just
 one quadrant.

\item The discussion of the classical trajectories that leads to interpret
the conserved quantum number $\omega$ as the invariant distance $\Delta $
 from the
center of the hyperbolic trajectories to the common vertex of the four
Rindler quadrants shows that (once more in terms of wave packets), it is
near the center of the trajectories that the pair production mechanism
occurs. Indeed $\Delta$ depends crucially on the sign of $\omega$, the extra
term $1/2$ being only the reflect of the quantum indeterminacy of the
position. For instance, it is only for
$\omega<0$ (eqs (\ref{inoutR}, \ref{vacioR}) that the Bogoljubov
transformation is non trivial (i.e. mixes particles and antiparticles) on
the {\bf R} quadrant.
\item On the Rindler quadrants  {\bf R} or {\bf L} the persistence of the
Rindler  vacua is given by
the usual Schwinger result modified by a surface term (\ref{surfacedilog}).
The latter becomes negligible in the case of constant electric field in the
large volume limit but it plays nevertheless an important r\^ole.
In the framework of
black hole physics it becomes the vacuum polarization term that cancels the
Hawking radiation flux in the Boulware vacuum.
\item The limitations of the W.K.B. evaluation of the Feynman propagator are
exemplified by evaluating it on the one hand as a mode superposition and on
the other hand using the Pauli-Van Vleck approximation of the Schwinger
kernel. Here the Jacobi fields were an essential simplifying ingredient of
the calculation, which shows that such approximation is nothing else than
a quadratic expansion of the classical potential around its minimum (eqs
\ref{Vquad}, \ref{Vquadbis}).
\item The interpretation of the usual Minkowskian propagator as a sum over
winding Rindlerian propagators is reinforced. We have shown that the rate
of particle production calculated from the zero term of the Minkowskian
propagator winding number expansion coincides with the rate obtained from
the Rindler propagator.
\item But on the other hand the calculation of the
Rindlerian population of the Minkowski vacuum leads to mode densities
 that are not equilibrium distributions in character. This is not surprising
as in a constant electric field there are no physical reasons to expect an
equilibrium distribution.\\
Moreover some of the distributions (\ref{npUoutR},\ref{naVoutL})  obtained vanish when the
electric field goes to zero. This reflects the fact that for a charged field we
obtain twice the number of modes encountered when we quantize a neutral
field. Indeed for the charged field the two linearly independent solutions of
the radial equation (\ref{FFweq}) have to be taken into account while for the
uncharged field only one has to be considered, the other blowing up
exponentially at infinity. Here again the analysis of the classical
trajectories helps to understand what happens. The modes that have to be
rejected correspond to particles pushed to spatial infinity, i.e. quantum
mechanically they describe fluctuations engendered by the Schwinger
mechanism and disappear when $E\mapsto \infty$.\\
This indicates for the accelerated detector as well as
for the charged black hole, that the Schwinger process first quickly
switches off the external field, by emission of preferentially charged
particles that will neutralize the plates of the condenser
 producing the electric field, or  the black hole.
Only then will the system reach a thermal
(quasi)-equilibrium state at the Unruh/Hawking temperature.
\item The ratio of population of a heavy two level charged
detector accelerated and in interaction with the charged scalar field are
 in accordance with thermodynamics, as far as the charge difference
  between the two levels of the detector is small compared to their
difference in masses. Moreover the amplitude of transition of such a detector
can be obtained as the large mass limit of a model built on
three interacting fields, a model where the levels of the detector
are quantized. This
shows that it is only in the limit where the recoil effects and pair
creation {\em \`a la Schwinger} of the quanta describing the detector levels
 become negligible that the
thermodynamical limit makes sense and is recovered. It is also noteworthy
that, as expected, in the limit of  electric field going to zero, the
probabilities of transition by absorption or emission (\ref{PBvarpi},\ref{PAvarpi})
 of the fluctuating modes whose distributions become null (\ref{npUoutR})
 also vanish.
\end{itemize}
\section*{Acknowledgments}
We would like to thank ours collegues Marianne Rooman, Serge Massar, Renaud Parentani and,
especially,
Robert Brout for numerous enlightening and fruitful discussions. We also
thank the Fonds National de la
Recherche  Scientifique (F.N.R.S.) for financial support.

\newpage
\appendix
\section{Asymptotic behavior of Whittaker's functions}\label{A}
 The Rindler's modes are built from the solutions of eq.(\ref{FFweq}),
which are given by Whittaker's functions. For completeness, we recall in
this appendix the few properties of these functions used in the main text.
\\ Defining $\bwp =\exp (a\xi)\cF$ as a function of the imaginary variable
$z=-i\epsilon
\frac{qE}{2a^2}\exp(2a\xi)$, it is immediate to verify that $\bwp [z]$ has
to satisfy the Whittaker's equation:
\begin{equation} \left\{\frac{d^2}{dz^2}+\left[-\frac14 +
\frac{\kappa}{z} +\frac{(1-4\mu^2)}{4z^2}\right]\right\}\bwp =0
\label{Whiteq}\qquad .
\end{equation}
 with
\begin{equation}
\kappa=\frac{i}2\left(\frac{\omega}a-\frac{m^2}{qE}\right)\qquad ,\qquad
\mu=i\frac{\omega}{2a}
\end{equation}
 The Whittaker's functions are related to the confluent hypergeometric
function by
\begin{equation} \bwp [z]=\pxe^{-z/2}z^{1/2+\mu}\mbox{\rm
F}[\kappa,\mu,z]\qquad ,\label{hypergeo}
\end{equation}
 where $\mbox{\rm F}[\kappa,\mu,z]$ is a solution of the Kummer's
equation:
\begin{equation}
\left\{z\frac{d^2}{dz^2}+(1+2\mu-z)\frac{d}{dz}-(\frac12+\mu-\kappa)\right\}
\mbox{\rm F}[\kappa,\mu,z]=0 \qquad .\label{Kumeq}
\end{equation}
 In momentum space ($p=-\partial/\partial z$), this equation becomes first
order. Solving it by an elementary quadrature and returning to the $z$
representation, we obtain an integral representation of the Whittaker's
function:
\begin{eqnarray}
\bwp[z]&=&\pxe^{-z/2}z^{\frac12+\mu}\frac{1}{\Gamma[\frac12+\mu-\kappa]}
\int_0^\infty\pxe^{-pz}p^{-\frac12+\mu-\kappa}(1+p)^{-\frac12+\mu+\kappa}dp
\nonumber
\\
&=&\pxe^{-z/2}z^{\frac12+\mu}\frac{z^{-2\mu}}{\Gamma[\frac12+\mu-\kappa]}
\int_0^\infty\pxe^{-q}q^{-\frac12+\mu-\kappa}(z+q)^{-\frac12+\mu+\kappa}dq
\qquad .\label{Wintrep1}
\end{eqnarray} The second  of these equations results from the change of
variable $q=pz$ and the Jordan's lemma allowing to integrate on the real
axis instead of the imaginary one. When $z$ goes to infinity,
eq.(\ref{Wintrep1}) gives immediately the asymptotic behavior of the
function:
\begin{equation}
\bwp[z]\mettresous{z\rightarrow\infty}\sous{\simeq}\pxe^{-z/2}z^{\kappa}
\qquad ,
\label{Winfty}
\end{equation}
 while near $z=0$, the Whittaker's function behaves like
\begin{equation} \bwp[z]\mettresous{z\rightarrow
0}\sous{\simeq}\left[\frac{\Gamma[2\mu]}{\Gamma[\frac12+\mu-\kappa]}z^{1/2-\mu}+
\frac{\Gamma[-2\mu]}{\Gamma[\frac12-\mu-\kappa]}z^{1/2+\mu}\right]
\qquad .\label{Worig}
\end{equation}
 If, instead of the change of function (\ref{hypergeo}), we use the one
with the opposite sign of $\mu$, setting:
\begin{equation} \bwp [z]=\pxe^{-z/2}z^{1/2-\mu}\mbox{\rm
F}[\kappa,-\mu,z]
\end{equation}
 we obtain a different integral representation, but of the same function
(see eq.(\ref{Worig})):
\begin{equation}
\pxe^{-z/2}z^{\frac12-\mu}\frac{1}{\Gamma[\frac12-\mu-\kappa]}
\int_0^\infty\pxe^{-pz}p^{-\frac12-\mu-\kappa}(1+p)^{-\frac12-\mu+\kappa}dp=
\bwp[z]\qquad .\label{Wintrep2}
\end{equation}
 Indeed, both representations are solutions of a second order differential
equation and have the same asymptotic behaviors. In the mathematical
literature, this function is usually denoted by:
$\mbox{\rm W}_{\kappa,\mu}(z)$, and satisfies the relation :
\begin{equation} \bwp[z]=\mbox{\rm W}_{\kappa,+\mu}(z)=\mbox{\rm
W}_{\kappa,-\mu}(z)\qquad .
\end{equation}
 A second, independent, solution of eq.(\ref{Whiteq}), is given by the
complex conjugate of the first one :
\begin{equation}
\bwm[z]=\left[\bwp[z]\right]^* \qquad .
\end{equation}
 So, we obtain complete sets of unnormalized modes, solutions of the wave
equation (\ref{Rweq}):
\begin{eqnarray}
\cWpe(\tau,\xi) & =& \frac{ e^{-i \omega\tau}}{\sqrt{2\pi}}  e^{-a\, \xi}
W_{+ i ({\omega\over 2a} - {m^2 \over 2 q E}), i {\omega\over 2 a}}
\left[ -i\epsilon {q E \over 2 a} e^{2 a\, \xi} \right]\qquad ,\nonumber\\
\cWme(\tau,\xi) & =& \frac{ e^{-i \omega\tau}}{\sqrt{2\pi}}  e^{-a\, \xi}
W_{- i ({\omega\over 2a} - {m^2 \over 2 q E}), i {\omega\over 2 a}}
\left[ +i\epsilon {q E \over 2 a} e^{2 a\, \xi} \right]\qquad
.\label{RindFunW}
\end{eqnarray}
 From the asymptotic expansions of the function $\mbox{\bf W}^\pm$, we
read immediately the various coefficient that define the charges carried
by these modes:
\begin{eqnarray}  C_+(\cWpe)&=\left(C_-(\cWme
)\right)^*&=\sqrt{2a}\left(\frac{qE}{2a^2}\right)^{[\frac12 +
\frac{i}2(\frac{\omega}a-\frac{m^2}{qE})]}
\pxe^{\epsilon\frac{\pi}4(\frac{\omega}a-\frac{m^2}{qE})} \\
C_-(\cWpe)&=C_+(\cWme )&=0
\\ D_+(\cWpe)&=\left(D_-(\cWme )\right)^*&=\sqrt{2\vert\omega\vert}
\left(\frac{qE}{2a^2}\right)^{[\frac12 +i
\frac{\omega}{2a}]}\pxe^{-i\epsilon\frac{\pi}{4}}\pxe^{\epsilon\frac{\pi\omega}{
4a}}\;
\frac{\Gamma[-i\frac{\omega}{a}]}
{\Gamma[\frac12-i(\frac{\omega}{a}-\frac{m^2}{2qE})]}\\
D_-(\cWpe)&=\left(D_+(\cWme )\right)^*&=\sqrt{2\vert\omega\vert}
\left(\frac{qE}{2a^2}\right)^{[\frac12 -i
\frac{\omega}{2a}]}\pxe^{-i\epsilon\frac{\pi}{4}}\pxe^{-\epsilon\frac{\pi
\omega}{4a}}\;
\frac{\Gamma[i\frac{\omega}{a}]}{\Gamma[\frac12+\frac{im^2}{2qE}]}
\end{eqnarray} whose  squared modulus are
\begin{eqnarray}
\vert C_+(\cWpe)\vert ^2&=\vert C_-(\cWme )\vert
^2&=\left(\frac{qE}{a}\right)
\pxe^{\epsilon\frac{\pi}2(\frac{\omega}a-\frac{m^2}{qE})}\\
\vert D_+(\cWpe)\vert ^2 &=\vert D_-(\cWme )\vert
^2&=\left(\frac{qE}{a}\right)\pxe^{\epsilon\frac{\omega\pi}{2a}}\;
\frac{\cosh[\pi(\frac{\omega}a-\frac{m^2}{2qE})]}{\left\vert
\sinh[\pi\frac{\omega} {a}]\right\vert}\\
\vert D_-(\cWpe)\vert ^2 &=\vert D_+(\cWme )\vert
^2&=\left(\frac{qE}{a}\right)
\frac{\pxe^{-\epsilon\frac{\omega\pi}{2a}}\;\cosh\frac{\pi}2\frac{m^2}{qE}}{
\left\vert
\sinh[\pi\frac{\omega} {a}]\right\vert}
\end{eqnarray} and satisfied the Wronskian relation:
\begin{equation} \sw\left[\vert D_+(\cWpe)\vert ^2-\vert D_-(\cWpe)\vert
^2\right]=\epsilon
\vert C_+(\cWpe)\vert ^2 \qquad .
\end{equation}
 In the main text, we also make use of  linear combinations of the
previous Whittaker's function whose behaviors near
$\xi=-\infty$, $[z=0]$ are particularly simple. They   are the functions
$M_{\pm\kappa,\mu}(\pm z)$ and its complex conjugate
$M_{\mp\kappa,-\mu}(\mp z)$:
\begin{equation} M_{\kappa,\mu}(z)  =\epsilon
\left(\frac{\omega}{a}\right)\Gamma[i\frac{\omega}{a}]
\left[\frac{\pxe^{\epsilon\frac{\pi
m^2}{2qE}}}{\Gamma\left[\frac12+i(\frac{\omega}{a}-\frac{m^2}{2qE})\right]}\bwp
-\frac{\pxe^{-i\epsilon\frac{\pi}2}\pxe^{-\epsilon\frac{\pi}2(\frac{\omega}{
a}-\frac{ m^2}{qE})}}{\Gamma\left[\frac12+i\frac{m^2}{2qE})\right]}\bwm
\right]\qquad ,
\end{equation}
 obeying the relations
\begin{eqnarray}
M_{\kappa,\mu}(z)&=&\pxe^{-i\epsilon^\frac{\pi}{2}}\pxe^{+\epsilon\frac{\pi\
omega}{2a}} M_{-\kappa,\mu}(-z)\qquad ,\nonumber\\
M_{\kappa,-\mu}(z)&=&\pxe^{-i\epsilon^\frac{\pi}{2}}\pxe^{-\epsilon\frac{\pi
\omega}{2a}} M_{-\kappa,-\mu}(-z)\qquad ,\label{A23}
\end{eqnarray} from which we have defined the (unnormalized) modes:
\begin{eqnarray}
\cMpe(\tau,\xi)   &=& \frac{ e^{-i \omega\tau}}{\sqrt{2\pi}}  e^{-a\, \xi}
M_{+ i ({\omega\over 2a} - {m^2 \over 2 q E}), i {\omega\over 2 a}}
\left[ -i\epsilon {q E \over 2 a} e^{2 a\, \xi} \right]\qquad ,\\
\cMme (\tau,\xi)  &=&\frac{ e^{-i \omega\tau}}{\sqrt{2\pi}}  e^{-a\, \xi}
M_{- i ({\omega\over 2a} - {m^2 \over 2 q E}), -i {\omega\over 2 a}}
\left[ +i\epsilon {q E \over 2 a} e^{2 a\, \xi} \right] \qquad ,
\end{eqnarray} whose charge content is given by the coefficients:
\begin{eqnarray}
\vert D_+(\cMpe  )\vert ^2&=\vert D_-(\cMme  )\vert
^2&=\left\vert{\omega}\right\vert\left(\frac{qE}{2a^2}\right)
\pxe^{\epsilon\frac{\pi\omega}{2a}}\\
\vert D_-(\cMpe  )\vert ^2&=\vert D_+(\cMme  )\vert ^2&=0\\
\vert C_+(\cMpe  )\vert ^2 &=\vert C_-(\cMme  )\vert
^2&=\left(\frac{\omega}{a}\right)\left(\frac{qE}a\right)
\frac{\pxe^{\epsilon\frac{\pi}2\left(\frac{\omega}{a}+\frac{m^2}{qE}\right)}\;
{\cosh\left[\pi\left(\frac{\omega}a-\frac{m^2}{2qE}\right)\right]}}{{\sinh
\left[\frac{\pi\omega}a\right]}}\nonumber\\ &&\\
\vert C_-(\cMpe  )\vert ^2 &=\vert C_+(\cMme  )\vert
^2&=\left(\frac{\omega}{a}\right)\left(\frac{qE}a\right)
\frac{\pxe^{-\epsilon\frac{\pi}2\left(\frac{\omega}{a}-\frac{m^2}{qE}\right)}\;
{\cosh\left[\pi\frac{m^2}{2qE}\right]}}{{\sinh\left[\frac{\pi\omega}a\right]
}}\nonumber\\ &&
\end{eqnarray}  verifying the Wronskian relation:
\begin{equation}
\sw\vert D_+(\cMpe  )\vert ^2=\epsilon\left(\vert C_+(\cMpe  )\vert
^2-\vert C_-(\cMpe  )\vert ^2\right)
\qquad .
\end{equation}
 Let us remark that ${\cal W}^{+}_{\epsilon}(\tau,\xi)=\left[{\cal
W}^{+}_{\epsilon}(-\tau,\xi)\right]^{*}$ and ${\cal
M}^{+}_{\epsilon}(\tau,\xi)=\left[{\cal
M}^{+}_{\epsilon}(-\tau,\xi)\right]^{*}$, i.e.
$in$ and $out$ classes of modes are related by a $\tau$ inversion
followed by a complex conjugation.
\section{Explicit forms of some Bogoljubov coefficients}\label{C} For
sake of completeness we recall here the basic relations between Fock
basis defined trough Bogoljubov transformations and give here the
explicit expressions of the various Bogoljubov coefficients that we have
used in the main text. Suppose a quantum field operator $\hat \Psi$,
satisfying a second order field equation, is defined in terms of two
sets   of linearly independent solutions $\phi^p_k,\;\phi^{a\;*}_k$ and
$\Phi^p_K,\;\Phi^{a\;*}_K$, labeled by the indices
$k$ and $K$, of the corresponding classical equation. We have
\begin{eqnarray}
\hat \Psi&=&\sum_{k}a_i \phi^p_k+b_i^\dagger\phi^{a\;*}_k\\
&=&\sum_{K}A_K \Phi^p_K+B_K^\dagger\Phi^{a\;*}_K\qquad,
\end{eqnarray} and
\begin{equation}
\left\{
\begin{array}{ccc}
\Phi^p_K&=&\sum_j\alpha_K^{\ j}\phi_j^p+\beta_K^{\ j}\phi_j^{a\;*}\\
\Phi^{a\;*}_K&=&\sum_j\gamma_K^{\ j}\phi_j^p+\epsilon_K^{\ j}\phi_j^{a\;*}
\end{array}\right.\qquad\mbox{\rm and\ }\qquad
\left\{
\begin{array}{ccc}
\phi^p_k&=&\sum_j\alpha_K^{\ j\;*}\Phi_K^p-\gamma_K^{\
j\;*}\Phi_K^{a\;*}\\
\phi^{a\;*}_k&=&\sum_j\epsilon_K^{\ j\;*}\Phi_K^{\ a\;*}-\beta_K^{\
j\;*}\Phi_K^{p}
\end{array}\right.
\end{equation} If, as usual, we suppose these basis orthonormal,
unitarity implies~:
\begin{equation}
\left\{
\begin{array}{rcl}
\delta_{K\;K'}&=&\sum_j \alpha_{K'}^{\ j\;*}\alpha_K^{\ j}-\beta_{K'}^{\
j\;*}\beta_K^{\ j}\\
\delta_{K\;K'}&=&\sum_j \epsilon_{K'}^{\ j\;*}\epsilon_K^{\
j}-\gamma_{K'}^{\ j\;*}\gamma_K^{\ j}\\ 0&=&\sum_j \gamma_{K'}^{\
j\;*}\alpha_K^{\ j}-\epsilon_{K'}^{\ j\;*}\beta_K^{\ j}
\end{array}\right.\qquad \qquad
\left\{
\begin{array}{rcl}
\delta_{k\;k'}&=&\sum_J \alpha_{J}^{\ k\;*}\alpha_J^{\ k'}-\gamma_{J}^{\
k\;*}\gamma_J^{\ k'}\\
\delta_{K\;K'}&=&\sum_J \epsilon_{J}^{\ k'\;*}\epsilon_J^{\
k}-\beta_{J}^{\ k'\;*}\beta_J^{\ k}\\ 0&=&\sum_J \gamma_{J}^{\
k'\;*}\epsilon_J^{\ k}-\alpha_{J}^{\ k'\;*}\beta_J^{\ k}
\end{array}\right.
\end{equation} Combining these relations we obtain the links between the
various creation and annihilation operators~:
\begin{equation}
\left\{\begin{array}{ccc} a_j&=&\sum_K A_K\alpha_K^{\
j}+B_K^\dagger\gamma_K^{\ j}\\ b^\dagger_j&=&\sum_K A_K\beta_K^{\
j}+B_K^\dagger\epsilon_K^{\ j}
\end{array}
\right.\qquad \mbox{\rm and\ similarlly\ }\qquad
\left\{
\begin{array}{ccc} A_K&=&\sum_j a_j\alpha_K^{\
j\;*}-b_j^\dagger\beta_K^{\ j\;*}\qquad ,\\ B^\dagger_K&=&\sum_K
A_K\beta_K^{\ j}+B_K^\dagger\epsilon_K^{\ j}\qquad ,\end{array}\right.
\end{equation} For each set of operator is associate a ``vacuum '' state
$\ket{\Omega}$ and $\ket{\omega}$ such that for all quantum number $k$ and
$K$~:
\begin{equation} A_K\ket{\Omega}=B_K\ket{\Omega}=0\qquad \mbox{\rm
and}\qquad a_k\ket{\omega}=b_k\ket{\omega}=0\qquad .
\end{equation} A standard computation \cite{KaUm,BMPS} gives the link
between these vacuum states~:
\begin{equation}
\ket{\Omega}={\cal N}e^{\{\sum_{k l}m^{k l}a_k^\dagger
b_l^\dagger\}}\;\ket{\omega}\qquad ,\qquad
\ket{\omega}={\cal N}^* e^{ \{\sum_{K L}M^{K L}A_K^\dagger
B_L^\dagger\}}\;\ket{\Omega}
\end{equation} where the matrices $m^{k l}$ and $M^{K L}$ are given by~:
\begin{equation} m^{k l}=\sum_J(\alpha^{-1})_J^k\beta_J^l\qquad \mbox
{\rm and}\qquad M^{K L}= \sum_j\gamma_L^j(\alpha^{-1})_K^j\qquad .
\end{equation} The normalization coefficients are of particular physical
interest; they give the projection of one vacuum state on the other (a
fact that we anticipate in the notation${\cal N}^*$). In the main text we
only encounter ``diagonal" Bogoljubov transformations, i.e. such that the
two sets of indices $\{k\}$ and $\{K\}$ are identical and  the various
matrices $\alpha_K^k,\
\dots \ \epsilon _K^k\propto \delta _K^k$ i.e. diagonal. In such cases the
matrices $m^{k l}$ and $M^{K L}$ are also diagonal and the normalization
coefficient are particularly easy to evaluate. By factorizing the vacuum
state according to the quantum number $k$:
$\ket{\omega}=\prod_k\ket{\omega_k}$, we obtain~:
\begin{equation} 1=|{\cal N}|^2\prod _k \sum
_{n=0}^{\infty}\frac1{(n!)^2}\bra{\omega_k}
\left(m_{k k}^*a_kb_k\right)^n
\left(m_{k k}a_k^\dagger b_k^\dagger\right)^n\ket{\omega_k}=\prod_k
\left(1-|\m_{k k}|^2\right)^{-1} =\prod_k |\alpha^k_k|^2\qquad ,
\end{equation} the phase of $\cal N$ remaining arbitrary, the vacuum
state being actually ray in the Hilbert space.
\subsection*{Bogoljubov transformations between Rindler $in$ and $out$
modes} On quadrant \TR, the coefficients connecting Rindler $in$ and
$out$-modes, coefficients occurring in eq.(\ref{bogoRindR}), are given
by:
\begin{eqnarray}
\alpha^{R}_{{\cal{U}}{\cal{U}}}(\omega>0)&=&\beta^{R}_{{\cal{U}}{\cal{U}}}(\
omega<0)=\frac{\Gamma[\frac 12-i(\frac \omega
a-\frac{m^2}{2qE})]}{\Gamma[-i\frac{\omega}{a}]}\frac{\NUwo{R}}{\NUwi{R}}\qquad
,\nonumber \\
\alpha^{R}_{{\cal{U}}{\cal{V}}}(\omega)&=&e^{i\frac \pi 2}e^{-\frac{\pi
\omega}{2a}}\frac{\Gamma[i\frac\omega a]}{\Gamma[-i\frac\omega a]}
\frac{\Gamma[\frac12-i(\frac \omega
a-\frac{m^2}{2qE})]}{\Gamma[\frac12+i\frac{m^2}{2qE}]}
\frac{\NUwo{R}}{\NVwi{R}}\qquad ,\nonumber \\
\alpha^{R}_{{\cal{V}}{\cal{U}}}(\omega>0)&=&\epsilon^{R}_{{\cal{V}}{\cal{U}}
}(\omega<0)=e^{i\frac
\pi 2}e^{-\frac{\pi
\omega}{2a}}\frac{\Gamma[\frac 12-i(\frac \omega
a-\frac{m^2}{2qE})]}{\Gamma[\frac
12-i\frac{m^2}{2qE}]}\frac{\NVwo{R}}{\NUwi{R}}\qquad ,\nonumber
\\
\epsilon^{R}_{{\cal{V}}{\cal{V}}}(\omega<0)&=&\gamma^{R}_{{\cal{V}}{\cal{V}}
}(\omega>0)=\left(\frac a
\omega\right)e^{-\frac {\pi m^2}{2qE}}\frac{\Gamma[\frac 12-i(\frac
{\omega}{a}-\frac{m^2}{2qE})]}{\Gamma[-i\frac{\omega}{a}]}\frac{\NVwo{R}}{\NVwi{R}}
\nonumber \qquad .
\end{eqnarray}  On quadrant \TP, see eq.(\ref{UVoiP}) they are given by:
\begin{eqnarray}
\alpha^{P}_{{\cal{U}}{\cal{V}}}(\omega>0)&=&\gamma^{P}_{{\cal{U}}{\cal{V}}}(
\omega<0)=
\pxe^{\pi\left(\frac{\omega}{2a}-\frac{m^2}{2qE}\right)}\frac{
\Gamma[1+i\frac\omega
a]}{\Gamma[\frac12+i\frac{m^2}{2qE}]}\frac{\NUwo{P}}{\NVwi{P}}\qquad
,\nonumber \\ &=&\beta^{P\;*}_{{\cal{V}}{\cal{U}}}(\omega<0)
=\epsilon^{P\;*}_{{\cal{V}}{\cal{U}}}(\omega>0)
\\
\beta^{P}_{{\cal{U}}{\cal{U}}}(\omega>0)&=&\epsilon^{P}_{{\cal{U}}{\cal{U}}}
(\omega<0)=e^{i\frac
\pi 2} {\pxe^{-\pi\frac{m^2}{2qE}}}\frac{ \Gamma[1+i\frac\omega a]}
{\Gamma[\frac12+i(\frac{\omega}{a}-\frac{m^2}{2qE})]}\frac{\NUwo{P}}{\NUwi{P
}}\qquad ,\nonumber \\
&=&\alpha^{P\;*}_{{\cal{V}}{\cal{V}}}(\omega<0)=\gamma^{P\;*}_{{\cal{V}}
{\cal{V}}}(\omega>0)
\qquad .
\end{eqnarray}
\subsection*{Bogoljubov transformation between Unruh and Rindler $out$
modes} In the \R\, and \L\, sectors, Unruh $out$ modes are related to
Rindler modes as:
\begin{eqnarray}
\Omega^{p \, out}_{\omega}&=&\theta(-\omega)\left\{\beta^{out}_{\Omega
{\cal{V}}_R}(\omega){\cal{V}}^{out}_{\omega,R}+\alpha^{out}_{\Omega
{\cal{U}}_L}(\omega){\cal{U}}^{out}_{\omega,L}\right\}+\nonumber \\
&&\theta(\omega)\left\{\alpha^{out}_{\Omega
{\cal{V}}_R}(\omega){\cal{V}}^{out}_{\omega,R}+\beta^{out}_{\Omega
{\cal{U}}_L}(\omega){\cal{U}}^{out}_{\omega,L}\right\} \nonumber\\
\varpi^{p \, out}_{\omega}&=&\alpha^{out}_{\varpi
{\cal{U}}_R}(\omega){\cal{U}}^{out}_{\omega,R} \nonumber\\
\Omega^{a \, out
\,*}_{\omega}&=&\theta(-\omega)\left\{\epsilon^{out}_{\Omega
{\cal{V}}_R}(\omega){\cal{V}}^{out}_{\omega,R}+\gamma^{out}_{\Omega
{\cal{U}}_L}(\omega){\cal{U}}^{out}_{\omega,L}\right\} +\nonumber\\
&&\theta(\omega)\left\{\gamma^{out}_{\Omega
{\cal{V}}_R}(\omega){\cal{V}}^{out}_{\omega,R}+\epsilon^{out}_{\Omega
{\cal{U}}_L}(\omega){\cal{U}}^{out}_{\omega,L} \right\}\nonumber\\
\varpi^{a \, out \,*}_{\omega}&=&\epsilon^{out}_{\varpi
{\cal{V}}_L}(\omega){\cal{V}}^{out}_{\omega,L},
\label{outoutRL}
\end{eqnarray} with
\footnotesize
\begin{eqnarray}
\alpha^{out}_{\Omega {\cal{U}}_L}(\omega)&=&\beta^{out}_{\Omega
{\cal{U}}_L}(\omega)=e^{i\frac \pi 2}e^{-\frac {\pi
m^2}{4qE}}\left\{\frac{\cosh\pi(\frac{m^2}{2qE}-\frac \omega a)}{\vert
\sinh\frac{\pi \omega}a \vert}\right\}^{\frac12}\nonumber\\
\alpha^{out}_{\Omega {\cal{V}}_R}(\omega)&=&\beta^{out}_{\Omega
{\cal{V}}_R}(\omega)=e^{-\frac {\pi m^2}{4qE}}e^{\frac{\pi
\omega}{2a}}\left\{\frac{\cosh\pi(\frac{\pi m^2}{2qE}-\frac \omega
a)}{\vert \sinh\frac{\pi \omega}a
\vert}\right\}^{\frac12}\frac{\Gamma(\frac 12-i\frac{m^2}{2qE}+i\frac
\omega a)}{\Gamma(\frac 12+i\frac{m^2}{2qE})}  \nonumber\\
\alpha^{out}_{\varpi {\cal{U}}_R}(\omega)&=&\epsilon^{out}_{\varpi
{\cal{V}}_L}(\omega)=\varepsilon(\omega)\nonumber\\
\epsilon^{out}_{\Omega {\cal{V}}_R}(\omega)&=&\gamma^{out}_{\Omega
{\cal{V}}_R}(\omega)= \alpha^{out\,*}_{\Omega
{\cal{U}}_L}(\omega)=\beta^{out\,*}_{\Omega {\cal{U}}_L}(\omega)\nonumber
\\ \gamma^{out}_{\Omega {\cal{U}}_L}(\omega)&=&\epsilon^{out}_{\Omega
{\cal{U}}_L}(\omega)=\alpha^{out\,*}_{\Omega
{\cal{V}}_R}(\omega)=\beta^{out\,*}_{\Omega {\cal{V}}_R}(\omega).
\label{coeffoutout}
\end{eqnarray}
\normalsize In the same way, we
get the expression of the Unruh $out$ modes  in terms of Rindler \TP and \TF
modes
\begin{eqnarray}
\Omega^{p \, out}_\omega&=&\alpha^{out}_{\Omega
{\cal{U}}_F}(\omega){\cal{U}}^{out}_{\omega,F}+\theta(\omega)\left\{\alpha^{
out}_{\Omega
{\cal{U}}_P}(\omega){\cal{U}}^{out}_{\omega,P}+\beta^{out}_{\Omega
{\cal{V}}_P}(\omega){\cal{V}}^{out}_{\omega,P}\right\}+ \nonumber\\
&&\theta(-\omega)\left\{\beta^{out}_{\Omega
{\cal{U}}_P}(\omega){\cal{U}}^{out}_{\omega,P}+\alpha^{out}_{\Omega
{\cal{V}}_P}(\omega){\cal{V}}^{out}_{\omega,P}\right\}\\
\varpi^{p \,
out}_\omega&=&\theta(\omega)\beta^{out}_{\varpi{\cal{U}}_P}(\omega){\cal{U}}
^{out}_{\omega,P}+
\theta(-\omega)\alpha^{out}_{\varpi
{\cal{U}}_P}(\omega){\cal{U}}^{out}_{\omega,P}\\
\Omega^{a \, out\,*}_\omega&=&\epsilon^{out}_{\Omega
{\cal{V}}_F}(\omega){\cal{V}}^{out}_{\omega,F}+\theta(\omega)\left\{\epsilon
^{out}_{\Omega
{\cal{V}}_P}(\omega){\cal{V}}^{out}_{\omega,P}+\gamma^{out}_{\Omega
{\cal{U}}_P}(\omega){\cal{U}}^{out}_{\omega,P}\right\}+ \nonumber\\
&&\theta(-\omega)\left\{\gamma^{out}_{\Omega
{\cal{V}}_P}(\omega){\cal{V}}^{out}_{\omega,P}+\epsilon^{out}_{\Omega
{\cal{U}}_P}(\omega){\cal{U}}^{out}_{\omega,P}\right\}\\
\varpi^{a \, out\,*}_\omega&=&\theta(\omega)\gamma^{out}_{\varpi
{\cal{V}}_P}(\omega){\cal{V}}^{out}_{\omega,P}+
\theta(-\omega)\epsilon^{out}_{\varpi
{\cal{V}}_P}(\omega){\cal{V}}^{out}_{\omega,P},
\end{eqnarray} with coefficients:
\footnotesize
\begin{eqnarray}
\alpha^{out}_{\Omega {\cal{U}}_F}(\omega)&=&\epsilon^{out}_{\Omega
{\cal{V}}_F}(\omega)=1\nonumber\\
\alpha^{out}_{\Omega {\cal{U}}_P}(\omega>0)&=&\beta^{out}_{\Omega
{\cal{U}}_P}(\omega<0)=\epsilon^{out\,*}_{\Omega
{\cal{V}}_P}(\omega>0)=\gamma^{out\,*}_{\Omega
{\cal{V}}_P}(\omega<0)=\alpha^{in\,*}_{\Omega
{\cal{V}}_F}(\omega>0)\nonumber\\
\beta^{out}_{\Omega {\cal{V}}_P}(\omega>0)&=&\alpha^{out}_{\Omega
{\cal{V}}_P}(\omega<0)=\gamma^{out\,*}_{\Omega
{\cal{U}}_P}(\omega>0)=\epsilon^{out\,*}_{\Omega
{\cal{U}}_P}(\omega<0)=\alpha^{in\,*}_{\Omega
{\cal{U}}_F}(\omega<0)\nonumber\\
\beta^{out}_{\varpi{\cal{U}}_P}(\omega>0)&=&\alpha^{out}_{\varpi
{\cal{U}}_P}(\omega<0)=\epsilon^{out\,*}_{\varpi
{\cal{V}}_P}(\omega<0)=\gamma^{out\,*}_{\varpi
{\cal{V}}_P}(\omega>0)=\beta^{in\,*}_{\varpi{\cal{V}}_F}(\omega<0).\nonumber
\\
\end{eqnarray}
\normalsize

\section{Schwinger representation for wave function products}\label{D} In
subsection \ref{exactPersist} we made use of several integral
representations of products of modes. Hereafter, we briefly indicate how
they are obtained.\\
$\bullet${Inertial observer} Using  the relations (9.240) and (6.669.3)
of ref. \cite{GR}, we may express the product of modes:
$\varphi_\sigma^{p\,out}(x)\varphi_\sigma^{a\,out}(x)$ as:
\begin{eqnarray}
\varphi_\sigma^{p\,out}(x)\varphi_\sigma^{a\,out}(x)&=&-e^{-i\frac \pi
4}\sqrt{\frac \pi 2}\frac {2 \sqrt{qE}}{\vert
M\vert^2}2^{-i\frac{m^2}{2qE}}\frac 1{\Gamma(\frac
14+i\frac{m^2}{4qE})\Gamma(\frac 34+i\frac{m^2}{4qE})}\nonumber \\
&&\int_0^\infty \frac{d\xi}{(\sinh \xi)^{-\frac 12}} e^{iqE(z+\frac \sigma
{qE})(\cosh \xi-\sinh \xi)}(\mbox{\rm{coth}}\frac \xi
2)^{-i\frac{m^2}{2qE}}.
\end{eqnarray} Introducing the new variable $s$ defined by $\sinh
2qEs=(\sinh \xi)^{-1}$ in this integral, we get:
\begin{eqnarray}
\varphi_\sigma^{p\,out}(x)\varphi_\sigma^{a\,out}(x)&=&=-\frac 1{\sqrt{2i
\pi}}\frac{e^{-\frac{3 \pi m^2}{4qE}}}{\sqrt{2\pi}}\frac
{\sqrt{qE}}{\Gamma(\frac 12+i\frac{m^2}{2qE})}\int_0^\infty \frac
{ds}{(\sinh 2qEs)^{\frac 12}}e^{-im^2s}e^{iqE(z+\frac \sigma
{qE})^2(\mbox{\rm{coth}}2qEs-\frac 1{\sinh
2qEs})}.\label{RepIntProdIn}\nonumber\\
\end{eqnarray}
$\bullet${Accelerated observer} In the same way, the product of wave
functions
${\cal{U}}_\omega^{out}(x){\cal{V}}_\omega^{in\,*}(x)$ can be expressed as
the integral~:
\begin{eqnarray}
{\cal{U}}_\omega^{out}(x){\cal{V}}_\omega^{in\,*}(x)&=&-i\frac {qE}2\vert
{\cal{N}}({\cal{U}}^{out}_\omega)\vert^2\frac 1{2\pi a^2
}\frac{\left(\Gamma(1+i\frac \omega a)\right)^2}{\Gamma(\frac 12+i\frac
\omega a-i\frac {m^2}{2qE})\Gamma(\frac 12+i\frac {m^2}{2qE})}\nonumber \\
&&\int_{-\infty}^\infty \frac {d\xi}{\cosh \xi}e^{i(\frac \omega
a-\frac{m^2}{qE})\xi}e^{i\frac {qE}2\rho^2 \tanh \xi} I_{i\frac \omega
a}(\frac {qE\rho^2}{2\cosh \xi})\label{RepIntProdAc}
\end{eqnarray}  thanks to the eq. (6.669.6) of \cite{GR}.

\newpage
\section*{Figures captions}
\begin{description}
\item[Fig. 1]The four Rindler patches and their coordinates.
\item[Fig. 2]Typical trajectories of charged particles on Rindler patches:
plain curves correspond to
particles trajectories, the dashed one to an antiparticle trjectory. The
numbers labelling the different
trajectories refer to the discussion in the main text.
\item[Fig. 3]Schematic representation, in the various
Rindler quadrant, of the charged carried by the various modes defined in section {\bf \ref{RindQuant}}, and of typical wavepackets built out of them with the appropriate mean value of $\omega$. On the $\bf R$ and $\bf L$ quadrants, we consider packets centered around a positive mean $\omega$; on the $\bf P$ and $\bf F$ quadrants, the packets are built out of modes whose $\omega$ are essentially negative.
\item[Fig. 4]$In$ and $out$  Unruh modes represented as superposition of
Rindler modes and their charge contents
\item[Fig. 5]Ratio of the total probabilities of excitation and
deexcitation of a charged
detector of mass gap $\Delta M/a=0.1$,  normalized with the Boltzman factor
$\exp-[2\,\pi\,\omega_\star/a]$, as
function of the mass
$m/a$ and ``charge" $qE/a^2$ of the exchanged quantum.
\end{description}
\newpage
\section*{Figures}

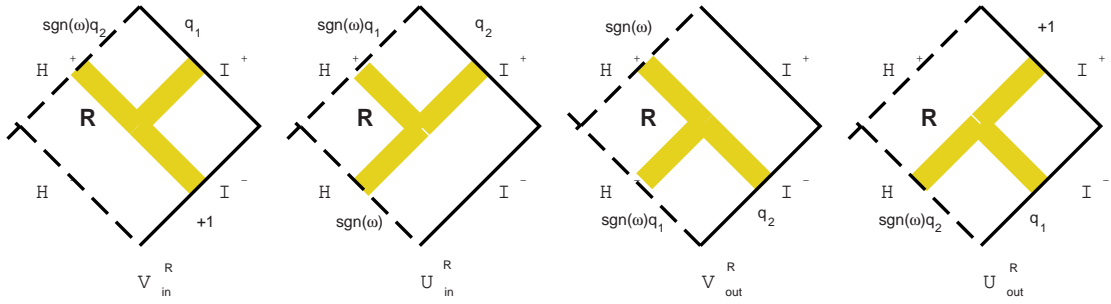
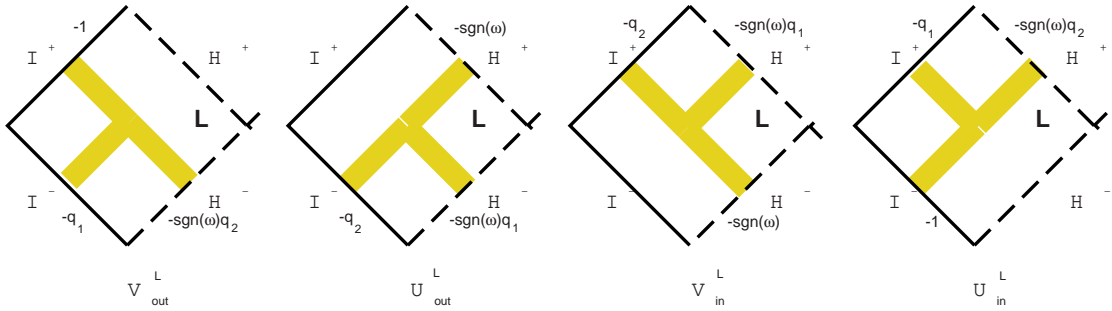
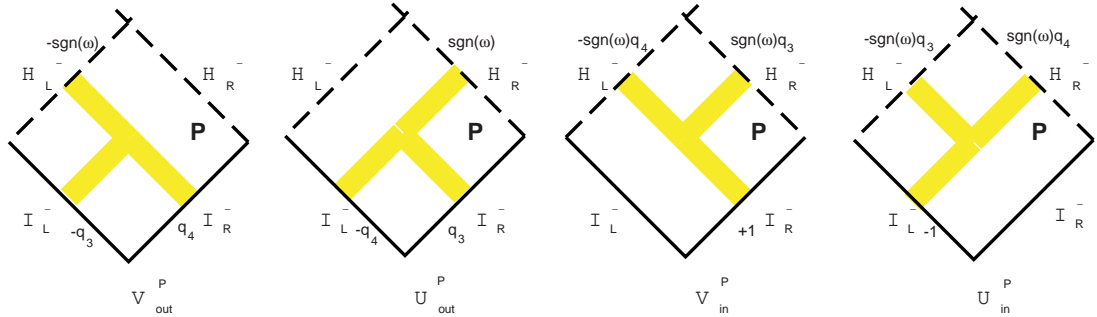
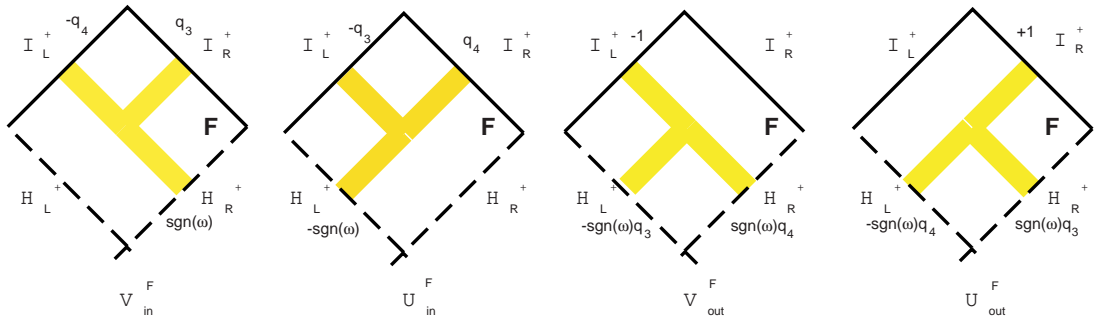
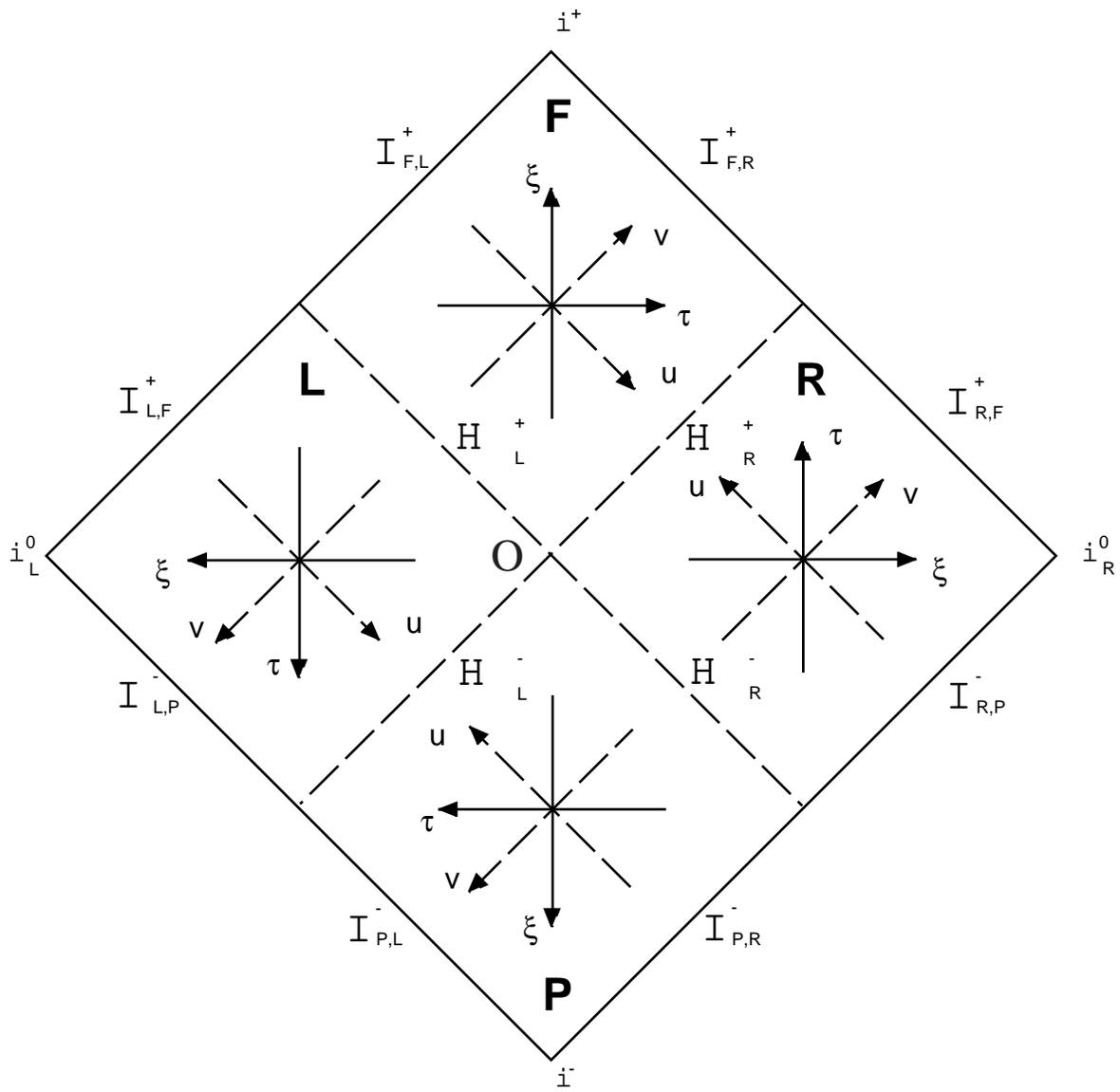
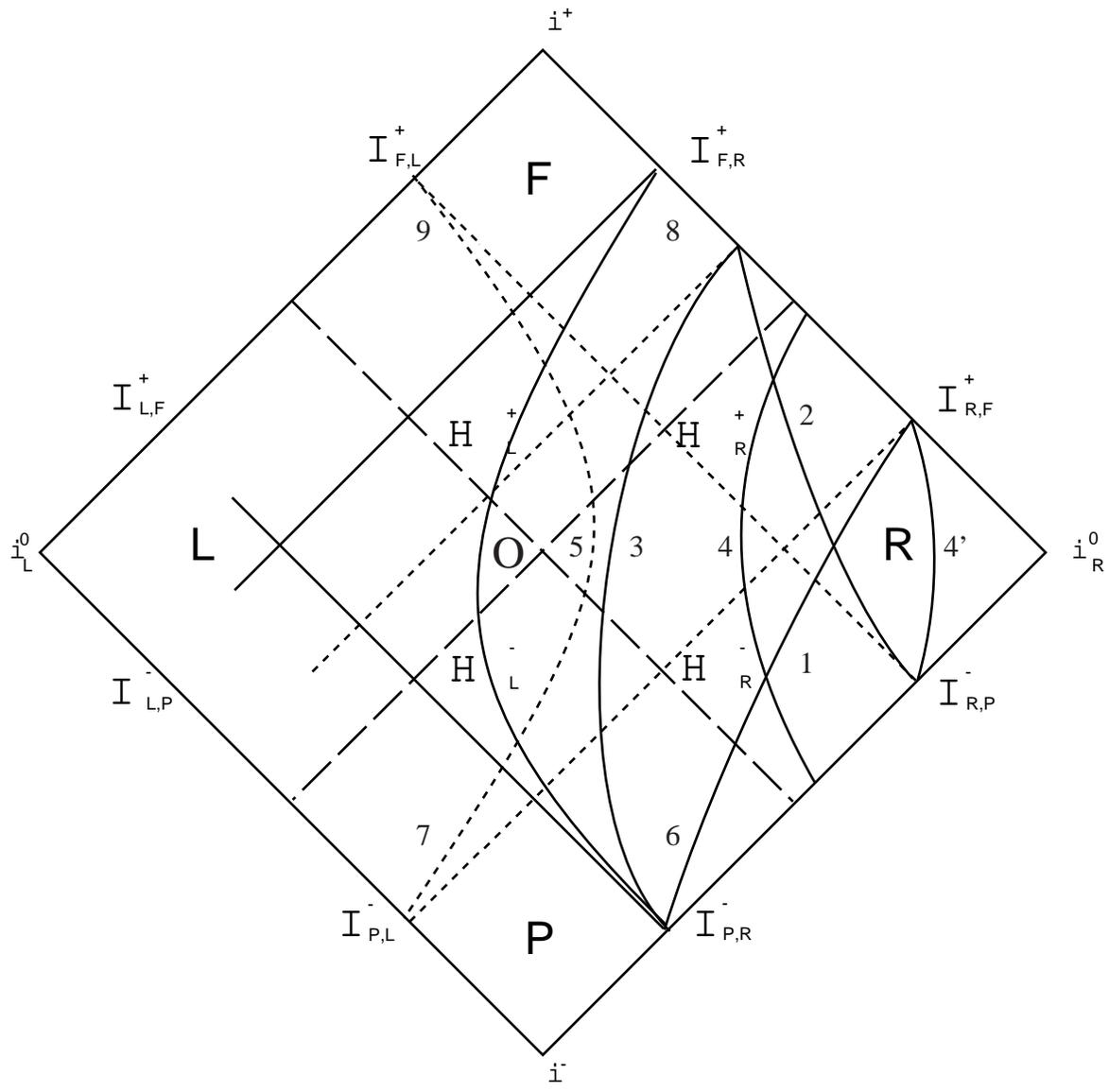
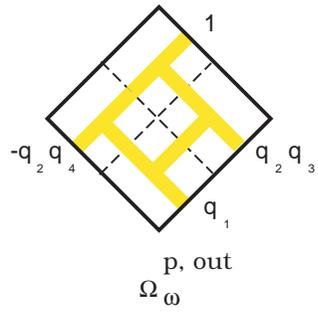
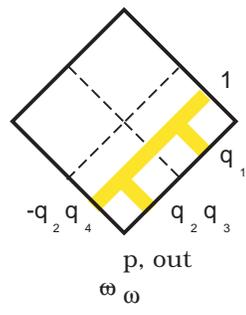
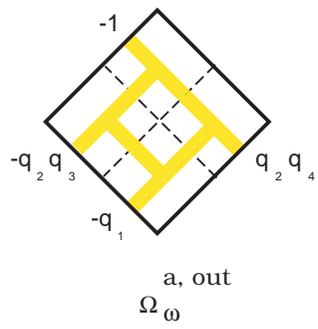
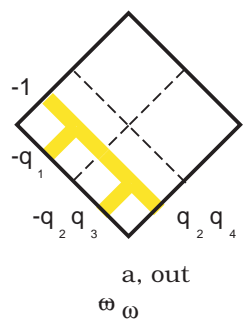
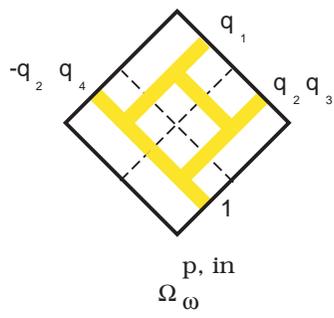
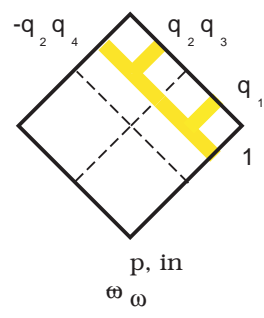
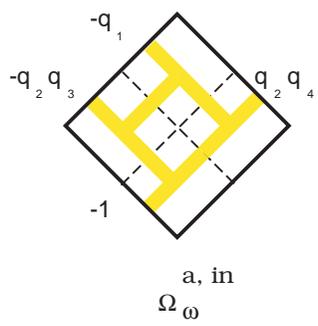
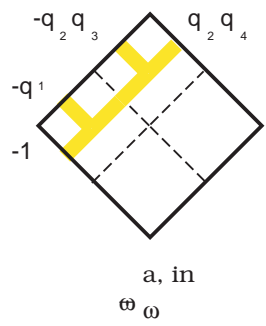
\end{document}